\documentclass[10pt,twocolumn,letterpaper]{article}

\usepackage[pagenumbers]{cvpr} 

\newcommand{\HazeMatching}{\mbox{\textsc{HazeMatching}}\xspace}

\newcommand{\HVAE}{\mbox{\textsc{LVAE}}\xspace}

\newcommand{\UNet}{\mbox{\textsc{U-Net}}\xspace}

\newcommand{\LPIPS}{\mbox{\textsc{LPIPS}}\xspace}
\newcommand{\FID}{\mbox{\textsc{FID}}\xspace}
\newcommand{\PSNR}{\mbox{\textsc{PSNR}}\xspace}
\newcommand{\MSE}{\mbox{\textsc{MSE}}\xspace}

\definecolor{cvprblue}{rgb}{0.21,0.49,0.74}
\usepackage[pagebackref,breaklinks,colorlinks,allcolors=cvprblue]{hyperref}
\usepackage{makecell} 
\usepackage{hhline}
\usepackage{lscape}
\usepackage{array}
\usepackage{ifthen}
\usepackage{colortbl}
\usepackage{multirow}
\usepackage{siunitx}
\usepackage{xspace}
\usepackage{algorithm}
\usepackage{algpseudocode}
\usepackage{graphicx} 
\usepackage{dsfont}
\usepackage{caption} 
\usepackage{wrapfig}
\usepackage{xcolor}
\usepackage{pifont} 
\usepackage{algpseudocode}
\usepackage{float}
\usepackage{graphicx}
\usepackage{placeins}
\usepackage{import}
\usepackage{etoolbox}
\def\paperID{****} 
\def\confName{CVPR}
\def\confYear{2026}

\title{\HazeMatching: Dehazing Light Microscopy Images with \\ Guided Conditional Flow Matching}

\usepackage{orcidlink}
\author{
\textbf{Anirban Ray}\orcidlink{0000-0002-7285-6727}$^{1,2}$ \quad
\textbf{Ashesh Ashesh}\orcidlink{0000-0003-3778-0576}$^{1,2}$ \quad
\textbf{Florian Jug}\orcidlink{0000-0002-8499-5812}$^{1}$ \\
$^{1}$Human Technopole, Milan, Italy \quad
$^{2}$Technische Universität Dresden, Germany \\
{\tt\small \{anirban.ray, ashesh.ashesh, florian.jug\}@fht.org}
}

\newcommand{\figTeaserMain}{
\begin{figure}[t]
  \centering
    \includegraphics[width=1.0\linewidth]{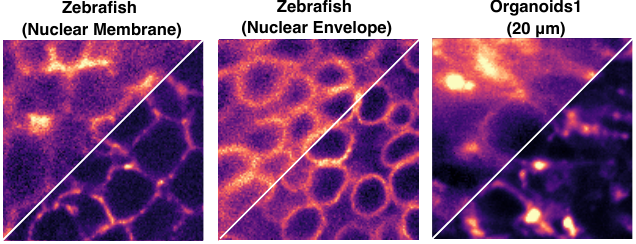} 
    \caption{\textbf{\HazeMatching} proposes a way to utilize Conditional Flow Matching to remove out-of-focus light (haze) from fluorescence microscopy data.}
  \label{fig:teaser_main}
\end{figure}
}

\newcommand{\figTraining}{
\begin{figure*}[h]
  \centering
    \includegraphics[width=1.0\textwidth]{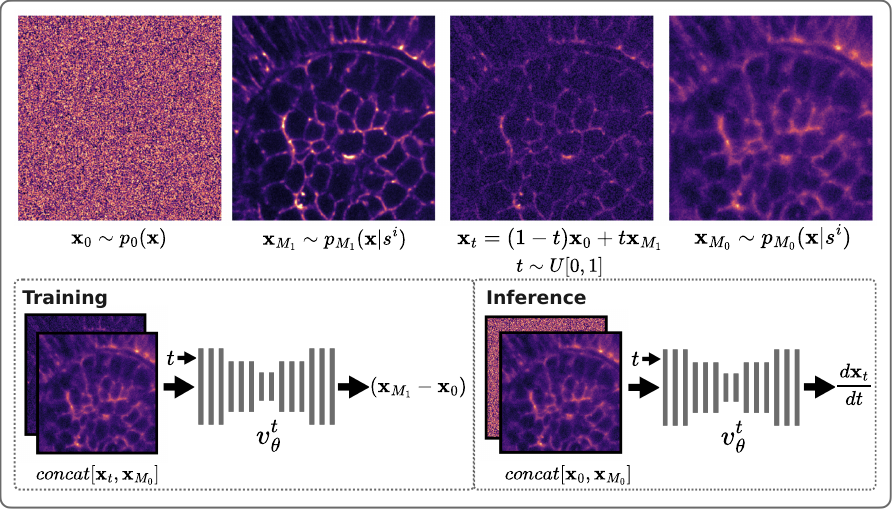}
    \caption{\textbf{Overview of our training and inference.} The figure illustrates the different distributions involved in our approach. The top panel shows how samples are drawn from the base noisy distribution $p_0(\mathbf{x})$, the target non-hazy image distribution $p_{M_1}(\mathbf{x}|s^i)$, and the hazy source image distribution $p_{M_0}(\mathbf{x}|s^i)$, and the computing of $\mathbf{x}_t$ with sampling time as $t \sim U[0,1]$. The lower left panel presents the training scheme, where the model is trained to learn a mapping between these distributions using conditional flow matching with inputs $concat[\mathbf{x}_t,\mathbf{x}_{M_0}]$ and $t$. The lower right panel depicts the inference process, where given a degraded observation $\mathbf{x}_{M_0}$, the trained model takes as inputs $concat[\mathbf{x}_0,\mathbf{x}_{M_0}]$ ($\mathbf{x}_0 \sim p_0(\mathbf{x})$) and $t$ and predicts the time dependent velocity field per pixel ($d\mathbf{x}_t/dt$) which is integrated using an (Euler) ODE solver iteratively using our proposed dehazing function $D$ to produce one prediction.}
  \label{fig:overview}
\end{figure*}
}

\newcommand{\figPlotsMain}{
\begin{figure*}[bt]
\centering
  \includegraphics[width=0.95\textwidth]{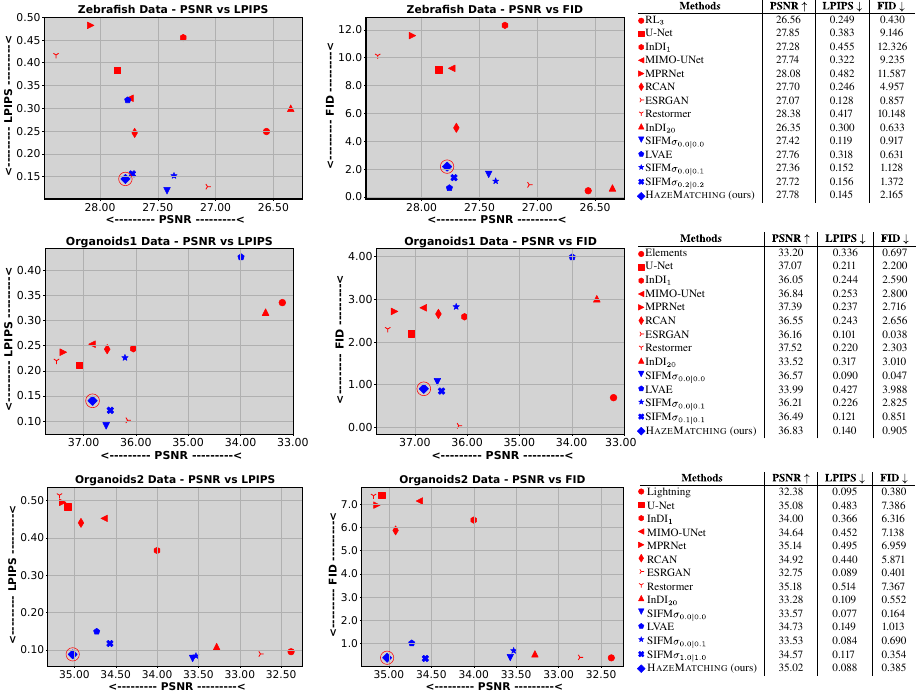}
\caption{
\textbf{Data fidelity vs.\ realism.} Each row corresponds to one dataset, \ie \textit{Zebrafish} (top), \textit{Organoids1} (middle), and \textit{Organoids2} (bottom). 
For each dataset, we show PSNR vs.\ LPIPS (left) and PSNR vs.\ FID (center), capturing the trade-off between pixel-level fidelity and perceptual quality. 
Our goal is to find a method that leads to high fidelity (high PSNR) while also leading to realistic looking predictions (low LPIPS/FID). 
\HazeMatching results are highlighted with an additional red circle around its blue marker.
Results tables (right) further quantify all results. 
Methods displayed in red are point-predicting baselines, while blue methods are generative posterior models (see main text).
Note that \HazeMatching consistently places itself close to the bottom left corner of all plots, \ie achieving the desired balanced performance across metrics and datasets. 
Additional results can be found in Supplementary Section~\ref{sup:results}.
}
  \label{fig:plot_main}
\end{figure*}
}

\newcommand{\figCalibMain}{
\begin{figure*}[tb]
  \centering
  \includegraphics[width=0.8\textwidth]{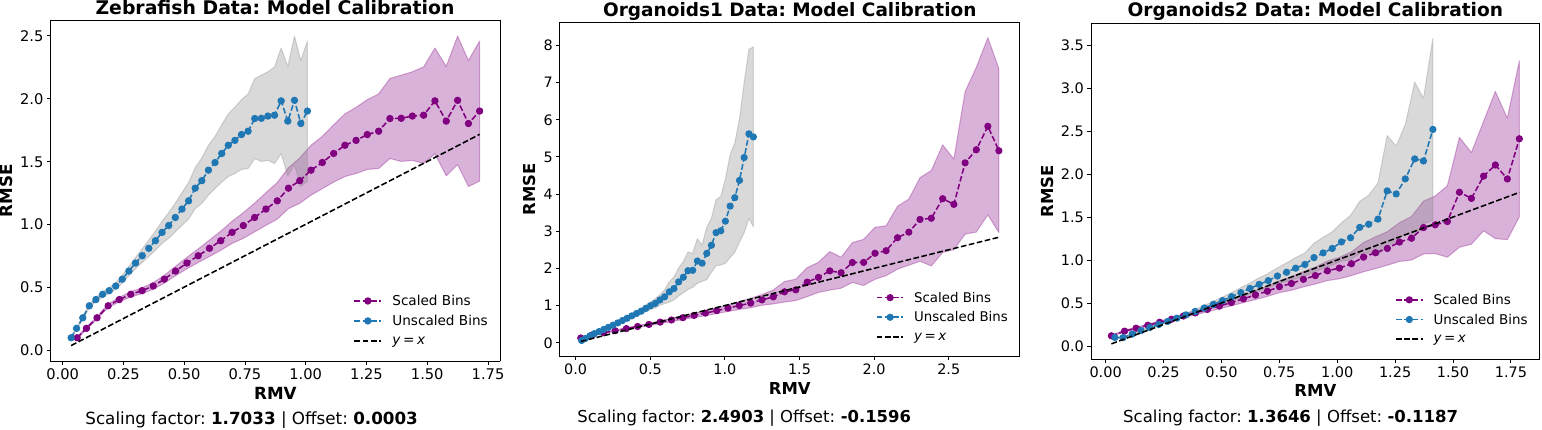} 
  \caption{\textbf{\HazeMatching models are well calibrated.} 
  RMSE vs.\ predicted RMV is shown for \textit{Zebrafish}, \textit{Organoids1}, and \textit{Organoids2} (left to right). 
  The dashed line is $y = x$. 
  Blue and purple circles show native calibration and calibrated results obtained by learning an additional calibration factor and offset, respectively.
  Shaded areas denoting standard error. 
  Learned calibration parameters (scaling and offset) are shown below each plot. 
  Additional results for \textit{Neuron} and \textit{Microtubule} are shown in Section~\ref{sup:results}.
  }
  \label{fig:calib_main}
\end{figure*}
}

\newcommand{\figQualitative}{
\begin{figure*}[tbp]
\centering
  \includegraphics[width=1.0\textwidth]{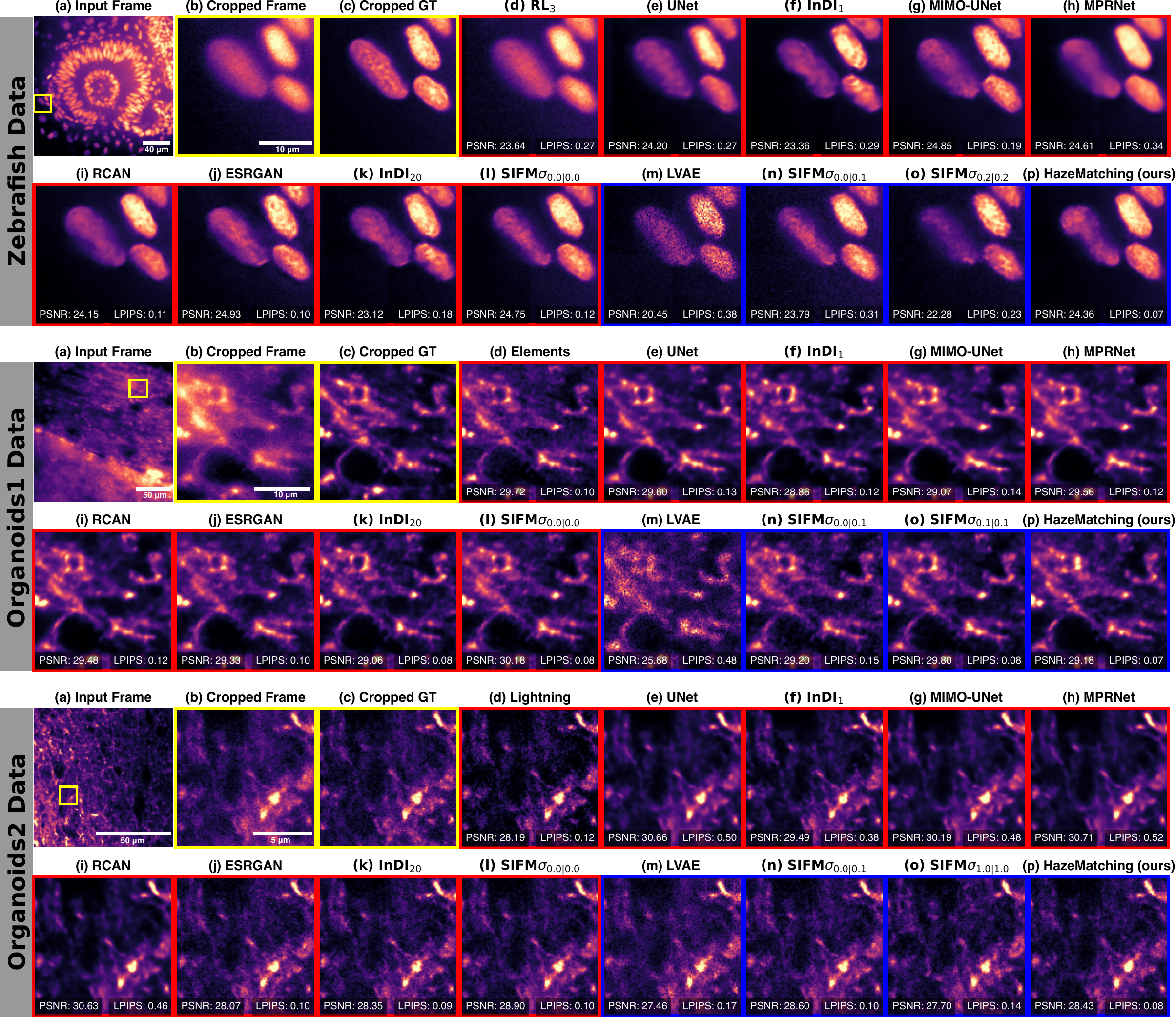} 
  \caption{\textbf{Qualitative results.} We show representative results for three datasets. 
  For each dataset we show 
  \textbf{(a)}~the full input ($1024 \times 1024$) and a selected crop ($128 \times 128$ yellow box), 
  \textbf{(b)}~the selected hazy crop, 
  \textbf{(c)}~the non-hazy ground truth crop, 
  \textbf{(d–o)}~predictions by all baseline methods (see Section~\ref{subsec:baselines}), and 
  \textbf{(p)}~results obtained with \HazeMatching. 
  Results with red borders are predictions by point-predictors, while methods with blue borders are results by generative posterior models (see also Figure~\ref{fig:plot_main} and main text). 
  We present additional results in the Supplementary Section~\ref{sup:results}.
}
  \label{fig:qualitative}
\end{figure*}
}

\newcommand{\figPlotsMainNeuron}{
\begin{figure*}[h]
  \includegraphics[width=1.0\textwidth]{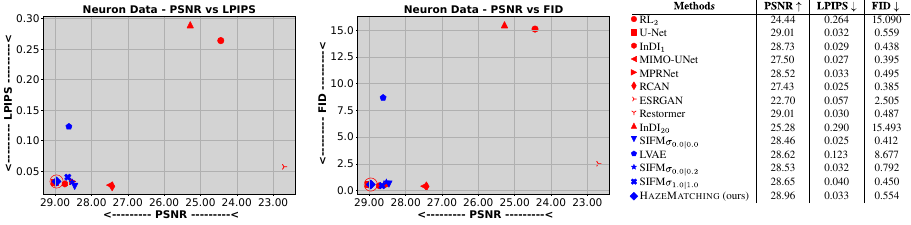}
\caption{
\textbf{Data fidelity vs.\ realism for the \textit{Neuron} dataset.} We show PSNR vs.\ LPIPS (left) and PSNR vs.\ FID (center), capturing the trade-off between pixel-level fidelity and perceptual quality. Our goal is to find a method that leads to high fidelity (high PSNR) while also leading to realistic looking predictions (low LPIPS/FID). (\HazeMatching is highlighted with an additional red circle.) Results tables (right) further summarize our results. Note that \HazeMatching consistently achieves balanced performance across all metrics and datasets. Methods displayed in red are point-predicting baselines, while blue methods are generative posterior models (see main text).
}
  \label{fig:plot_main_neuron}
\end{figure*}
}

\newcommand{\figCalibNeuron}{
\begin{figure*}[h]
  \centering
  \includegraphics[width=0.4\textwidth]{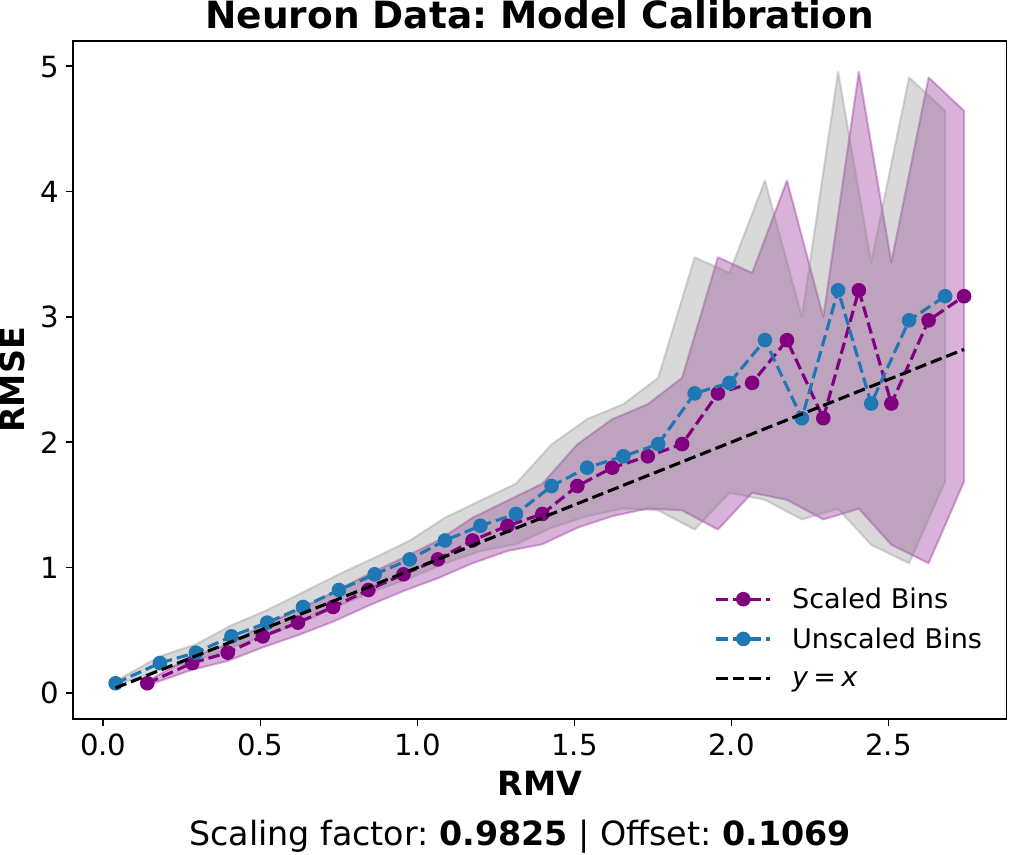}
  \caption{\textbf{Calibration of \HazeMatching.} RMSE vs.\ predicted RMV is shown for the \textit{Neuron} data. The dashed line indicates ideal calibration ($y = x$). Blue and purple circles show uncalibrated and calibrated results, respectively, with shaded areas denoting standard error. Calibration parameters (scaling and offset) are shown below.}
  \label{fig:calib_main_neuron}
\end{figure*}
}

\newcommand{\figPlotsMainMicrotubule}{
\begin{figure*}[h]
  \includegraphics[width=1.0\textwidth]{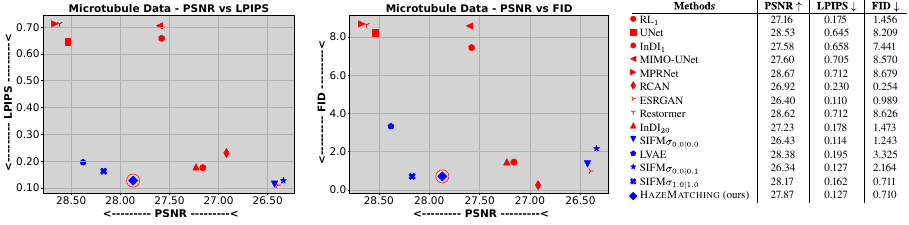}
\caption{
\textbf{Data fidelity vs.\ realism for the \textit{Microtubule} dataset.} We show PSNR vs.\ LPIPS (left) and PSNR vs.\ FID (center), capturing the trade-off between pixel-level fidelity and perceptual quality. Our goal is to find a method that leads to high fidelity (high PSNR) while also leading to realistic looking predictions (low LPIPS/FID). (\HazeMatching is highlighted with an additional red circle.) Results tables (right) further summarize our results. Note that \HazeMatching consistently achieves balanced performance across all metrics and datasets. Methods displayed in red are point-predicting baselines, while blue methods are generative posterior models (see main text).
}
  \label{fig:plot_main_microtubule}
\end{figure*}
}

\newcommand{\figCalibMicrotubule}{
\begin{figure*}[h]
  \centering
  \includegraphics[width=0.4\textwidth]{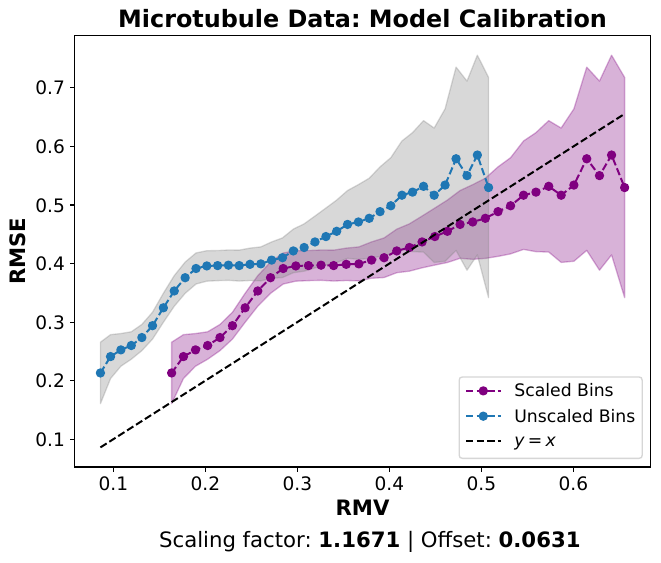}
  \caption{\textbf{Calibration of \HazeMatching.} RMSE vs.\ predicted RMV is shown for the \textit{Microtubule} data. The dashed line indicates ideal calibration ($y = x$). Blue and purple circles show uncalibrated and calibrated results, respectively, with shaded areas denoting standard error. Calibration parameters (scaling and offset) are shown below.}
  \label{fig:calib_main_microtubule}
\end{figure*}
}

\newcommand{\figCalibValVal}{
\begin{figure*}[h]
  \centering
  \includegraphics[width=1.0\textwidth]{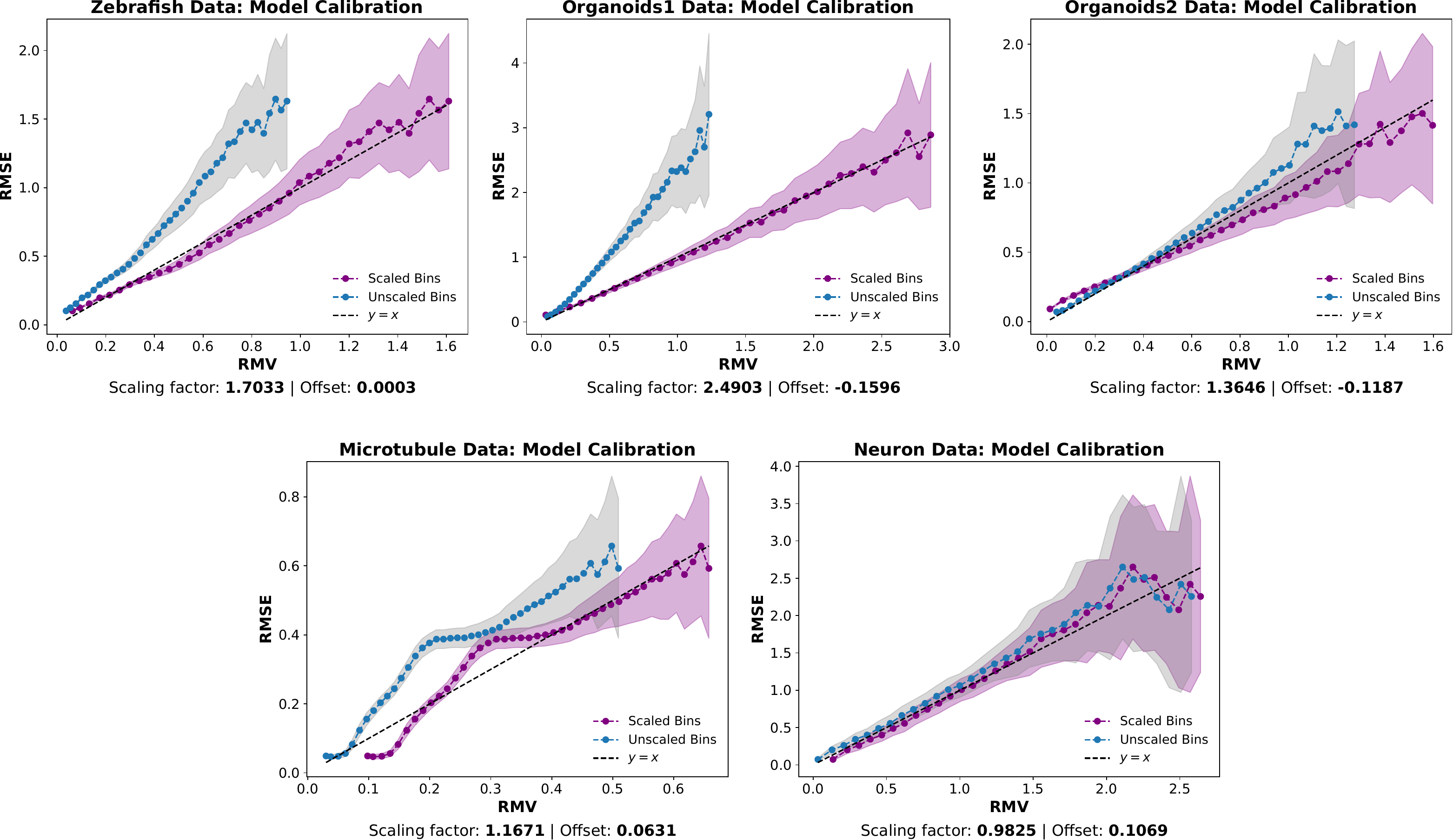}
  \caption{Calibration factors computed on validation data, applied on the validation data.}
  \label{fig:calib_val_val}
\end{figure*}
}

\newcommand{\figPlotsSSIM}{
\begin{figure*}[h]
\centering
  \includegraphics[width=0.8\textwidth]{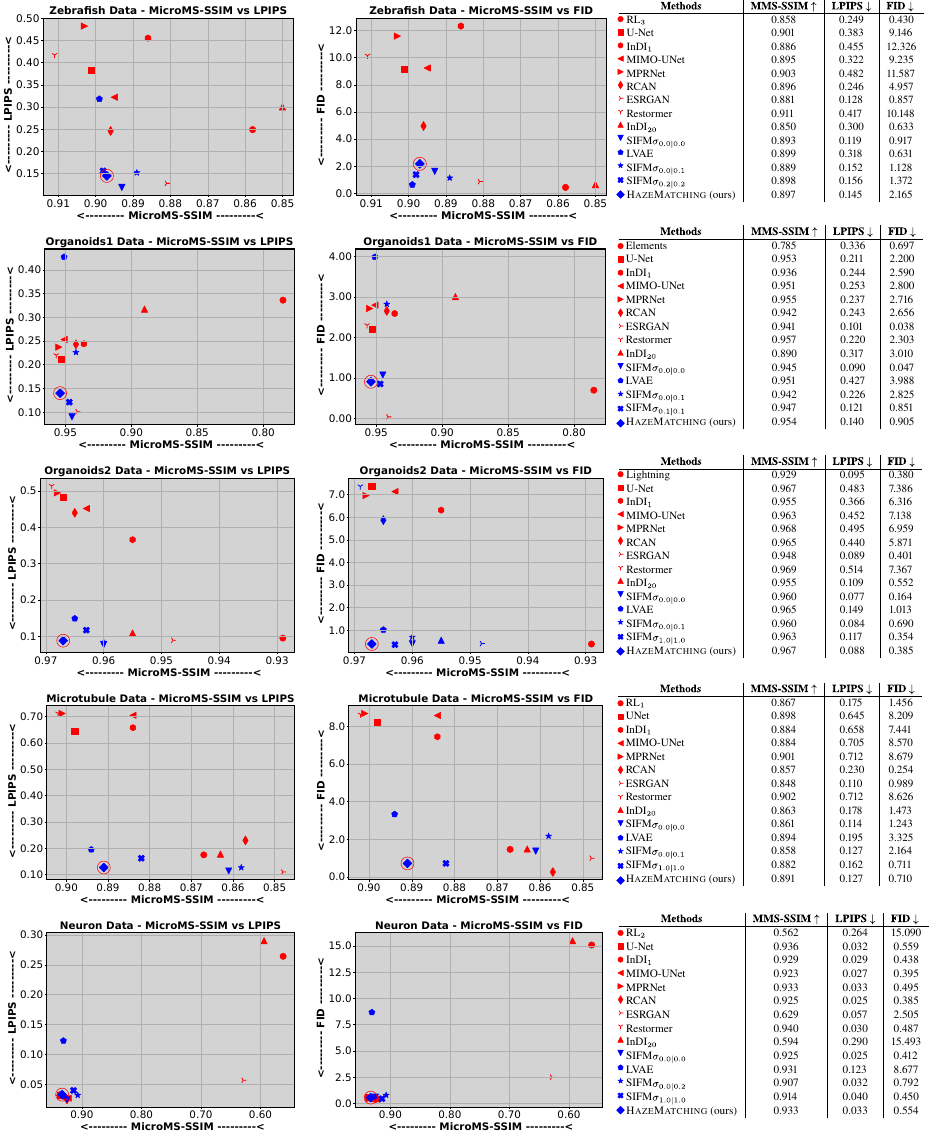}
\caption{
\textbf{Data fidelity (MicroMS-SSIM [MMS-SSIM]) vs.\ realism (LPIPS and FID).} Each row corresponds to one dataset, \ie \textit{Zebrafish} (row 1), \textit{Organoids1} (row 2), \textit{Organoids2} (row 3), \textit{Microtubule} (row 4), and \textit{Neuron} (row 5) datasets. 
For each dataset, we show MicroMS-SSIM vs.\ LPIPS (left) and MicroMS-SSIM vs.\ FID (center), capturing the trade-off between pixel-level fidelity and structural/perceptual quality. The MicroMS-SSIM metric highlights the preservation of local structural details and consistency in the restored images, complementing pixel-wise fidelity metrics like PSNR. 
Our goal is to identify methods that achieve both high fidelity (high MicroMS-SSIM) and strong perceptual realism (low LPIPS and FID). 
(\HazeMatching is highlighted with an additional red circle.)
Results tables (right) further summarize our results. 
Note that \HazeMatching consistently achieves a balanced trade-off across all metrics and datasets. 
Methods displayed in red are point-predicting baselines, while blue methods are generative posterior models (see main text).
}
  \label{fig:plot_main_ssim}
\end{figure*}
}

\newcommand{\figPlotsRL}{
\begin{figure*}[h]
\centering
  \includegraphics[width=1.0\textwidth]{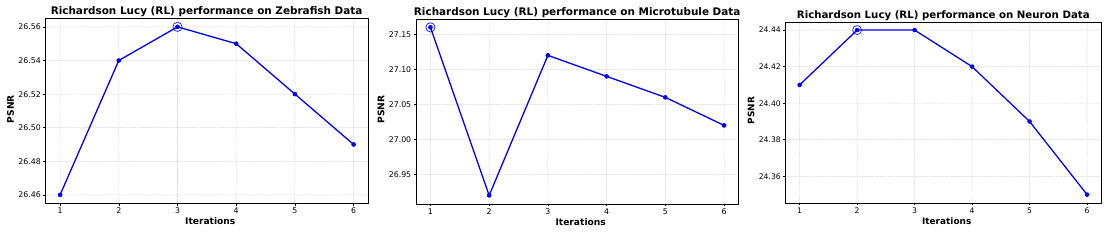}
\caption{\textbf{Selection of optimal Richardson--Lucy iterations across datasets.} For each dataset, we evaluate multiple RL iterations and select the one that yields the highest PSNR (indicated by a an additional circle marker). This ensures a fair comparison against learning-based methods. Note that RL performance varies across datasets, highlighting the importance of iteration selection based on quantitative fidelity.
}
  \label{fig:plot_rl_iters}
\end{figure*}
}

\newcommand{\figPlotsSIFM}{
\begin{figure*}[h]
\centering
  \includegraphics[width=0.95\textwidth]{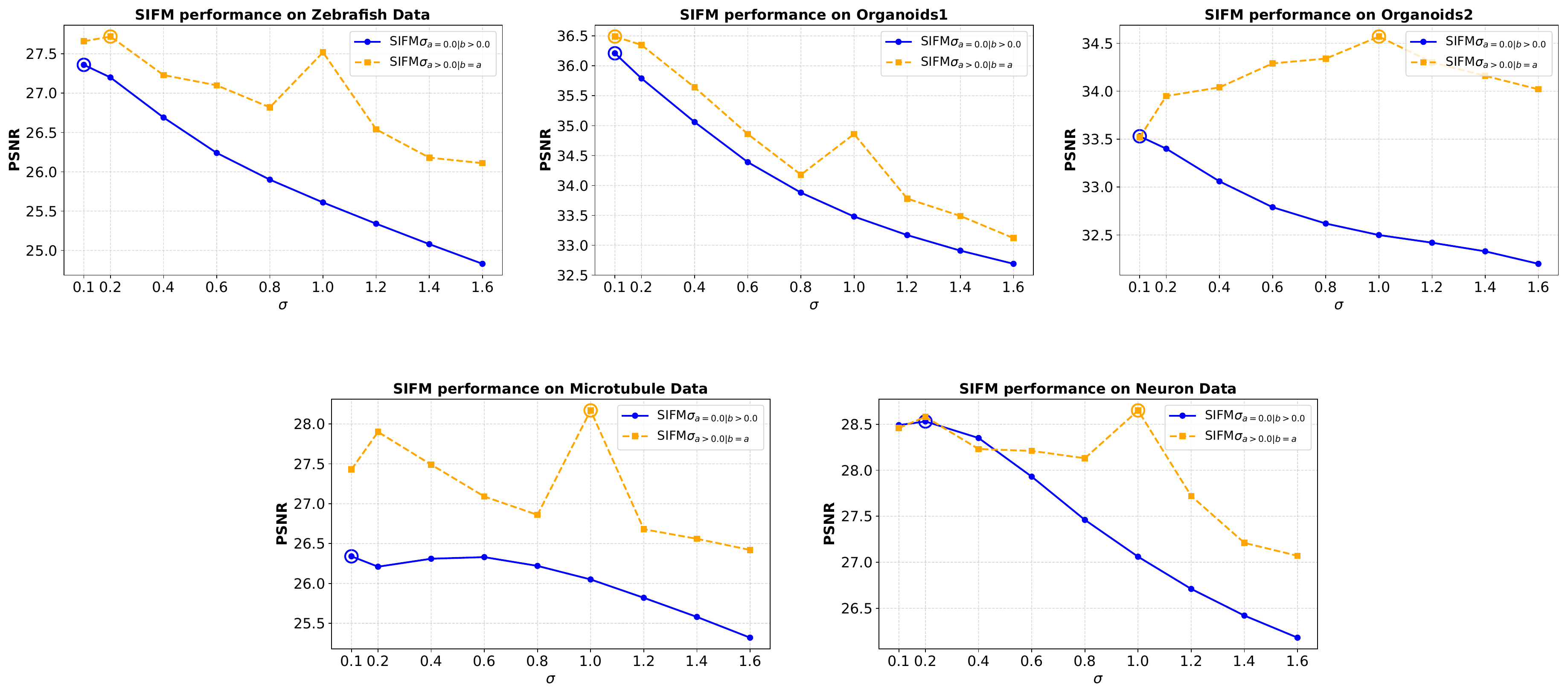}
\caption{ \textbf{Selection of optimal \texttt{SIFM}$_{a|b}$ configurations across datasets.} 
For each dataset, we evaluate multiple configurations of \texttt{SIFM} and chose the one for the main paper the one that achieves the highest PSNR (marked with a circle). This selection enables a fair comparison with other baselines. We experiment with two families of \texttt{SIFM} configurations: 
(1) \textbf{Blue solid lines:} SIFM{$\sigma_{a=0.0|b\geq0.0}$}, where the model is trained without any additive noise (i.e., $a=0.0$), and test-time noise of varying scale $b$ is applied to encourage sample diversity. (2) \textbf{Orange dashed lines:} SIFM{$\sigma_{a>0.0|b=a}$}, where the same level of Gaussian noise $\mathcal{N}(0, I)$ is added both during training and inference. Here, $a$ and $b$ represent the scaling of the noise used during training and inference, respectively. This setup allows us to investigate the impact of noise injection at different stages and assess the trade-off between sample diversity and fidelity. The optimal configuration varies across datasets, reflecting the sensitivity of \texttt{SIFM} to hyperparameter tuning and data-specific characteristics.
}
  \label{fig:plot_sifm_configs}
\end{figure*}
}

\newcommand{\figQualitativeSuppleZebrafishOne}{
\begin{figure*}[h]
\centering
  \includegraphics[width=1.0\textwidth]{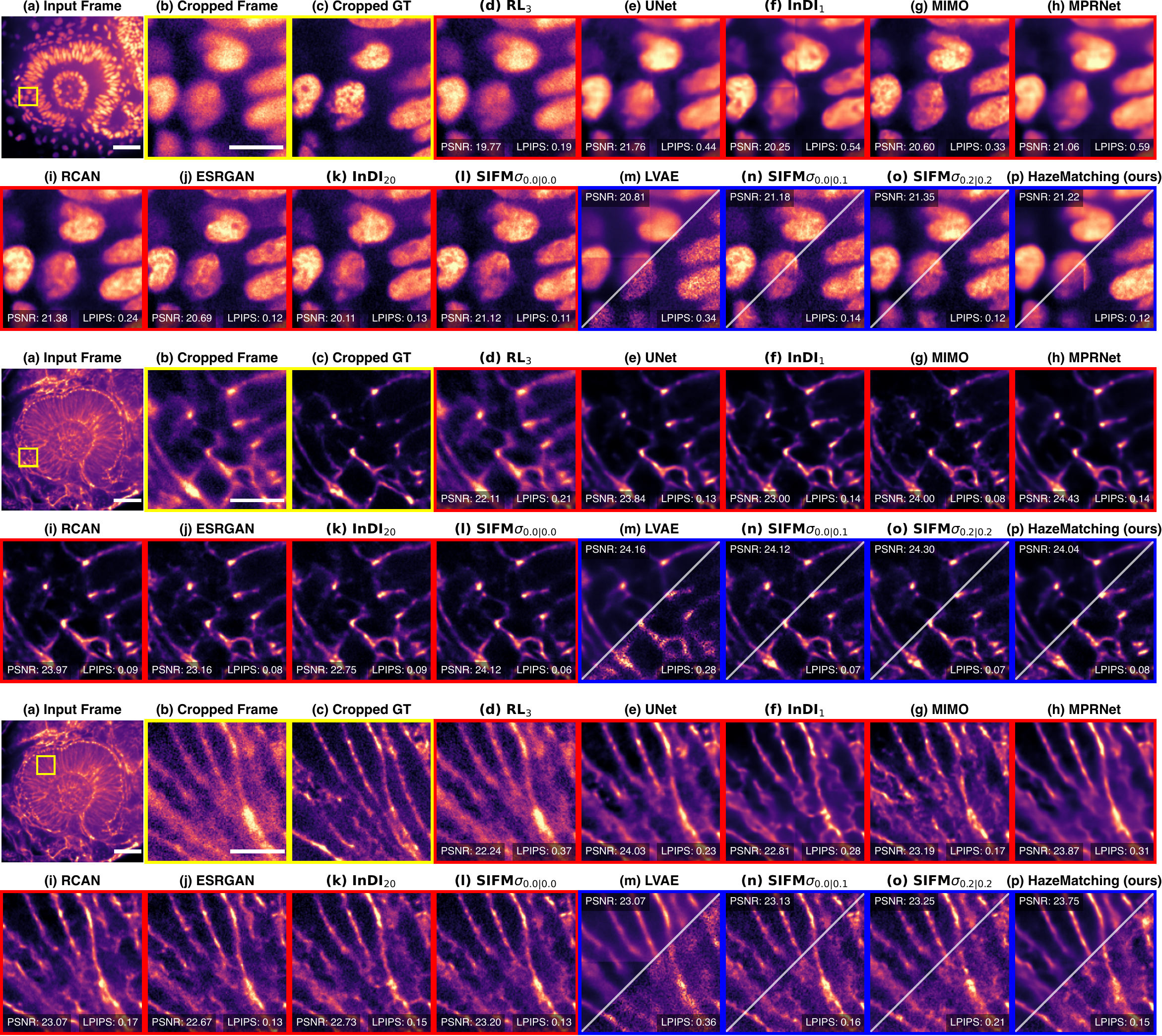}
  \caption{\textbf{Qualitative results on Zebrafish Data:} Here we present three examples, each illustrating two distinct structures: the Nuclei and the Nuclear Membrane. 
  \textbf{(a)}~the full input and a selected 128$\times$128 crop (yellow box); Scalebar: 40 $\mu m$, 
  \textbf{(b)}~the selected crop; Scalebar: 10 $\mu m$, 
  \textbf{(c)}~non-hazy ground truth, 
  \textbf{(d–o)}~predictions by all baseline methods (see Section~\ref{subsec:baselines}), and 
  \textbf{(p)}~results obtained with \HazeMatching. 
  Results with red borders are predictions by point-predictors, while methods with blue borders are results by generative posterior models (see also Figure~\ref{fig:plot_main} and main text). Note that \HazeMatching consistently produces sharper and more perceptually aligned predictions (lower LPIPS) compared to both deterministic and posterior-based baselines, while maintaining comparative fidelity (PSNR). For the point-prediction methods, PSNR and LPIPS are computed on the cropped region shown in yellow. For posterior models (blue borders), we plot the MMSE estimate in the upper triangle with its PSNR and one posterior sample in the lower-triangle with its LPIPS score.
}
  \label{fig:sup_qualitative1}
\end{figure*}
}

\newcommand{\figQualitativeSuppleZebrafishtwo}{
\begin{figure*}[h]
\centering
  \includegraphics[width=1.0\textwidth]{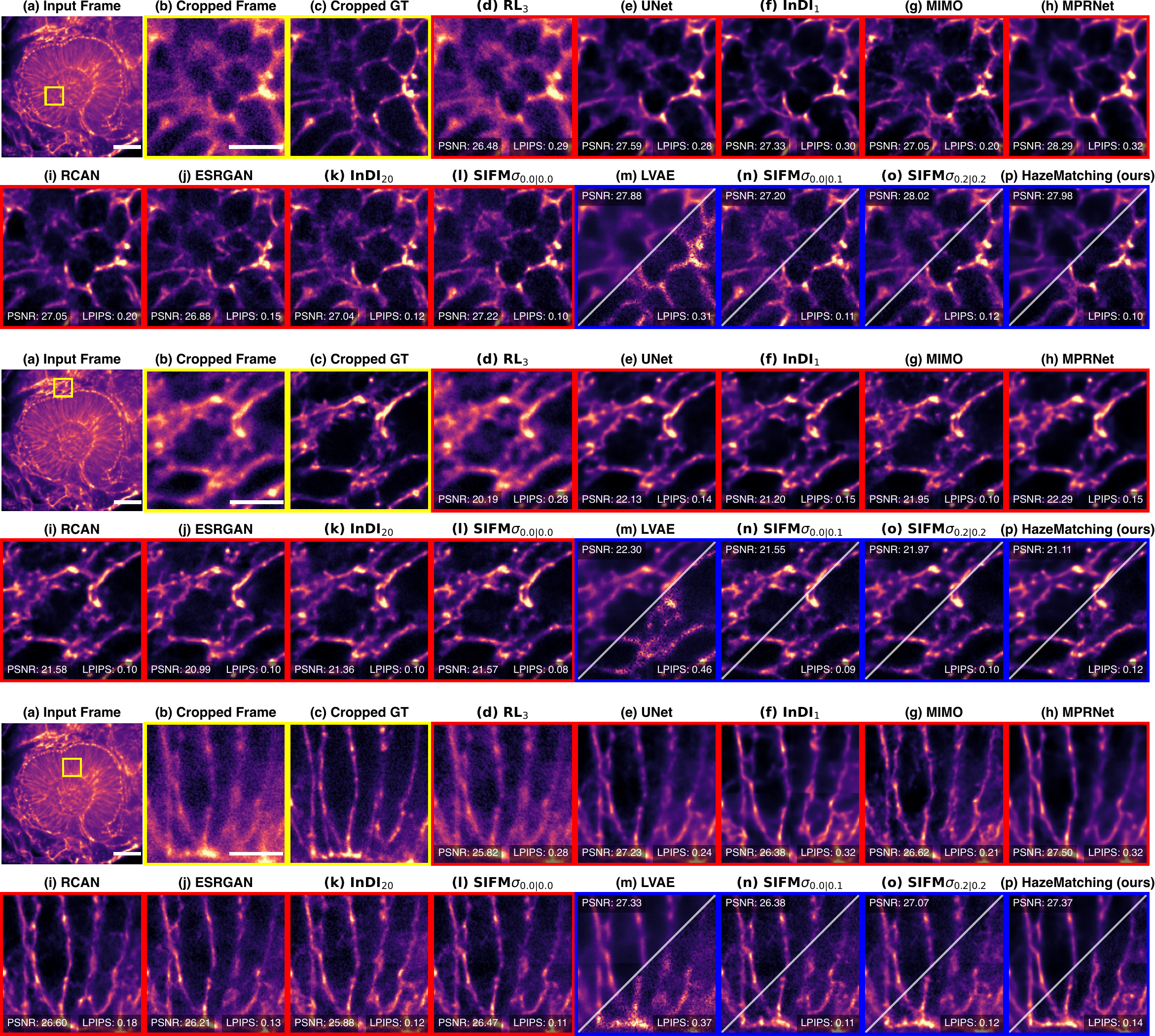}
  \caption{\textbf{Qualitative results on Zebrafish Data:} Here we present three examples showing Nuclear Membrane.
  \textbf{(a)}~the full input and a selected 128$\times$128 crop (yellow box); Scalebar: 40 $\mu m$, 
  \textbf{(b)}~the selected crop; Scalebar: 10 $\mu m$, 
  \textbf{(c)}~non-hazy ground truth, 
  \textbf{(d–o)}~predictions by all baseline methods (see Section~\ref{subsec:baselines}), and 
  \textbf{(p)}~results obtained with \HazeMatching. 
  Results with red borders are predictions by point-predictors, while methods with blue borders are results by generative posterior models (see also Figure~\ref{fig:plot_main} and main text). Note that \HazeMatching consistently produces sharper and more perceptually aligned predictions (lower LPIPS) compared to both deterministic and posterior-based baselines, while maintaining comparative fidelity (PSNR). For the point-prediction methods, PSNR and LPIPS are computed on the cropped region shown in yellow. For posterior models (blue borders), we plot the MMSE estimate in the upper triangle with its PSNR and one posterior sample in the lower-triangle with its LPIPS score.
}
  \label{fig:sup_qualitative2}
\end{figure*}
}

\newcommand{\figQualitativeSuppleZebrafishthree}{
\begin{figure*}[h]
\centering
  \includegraphics[width=1.0\textwidth]{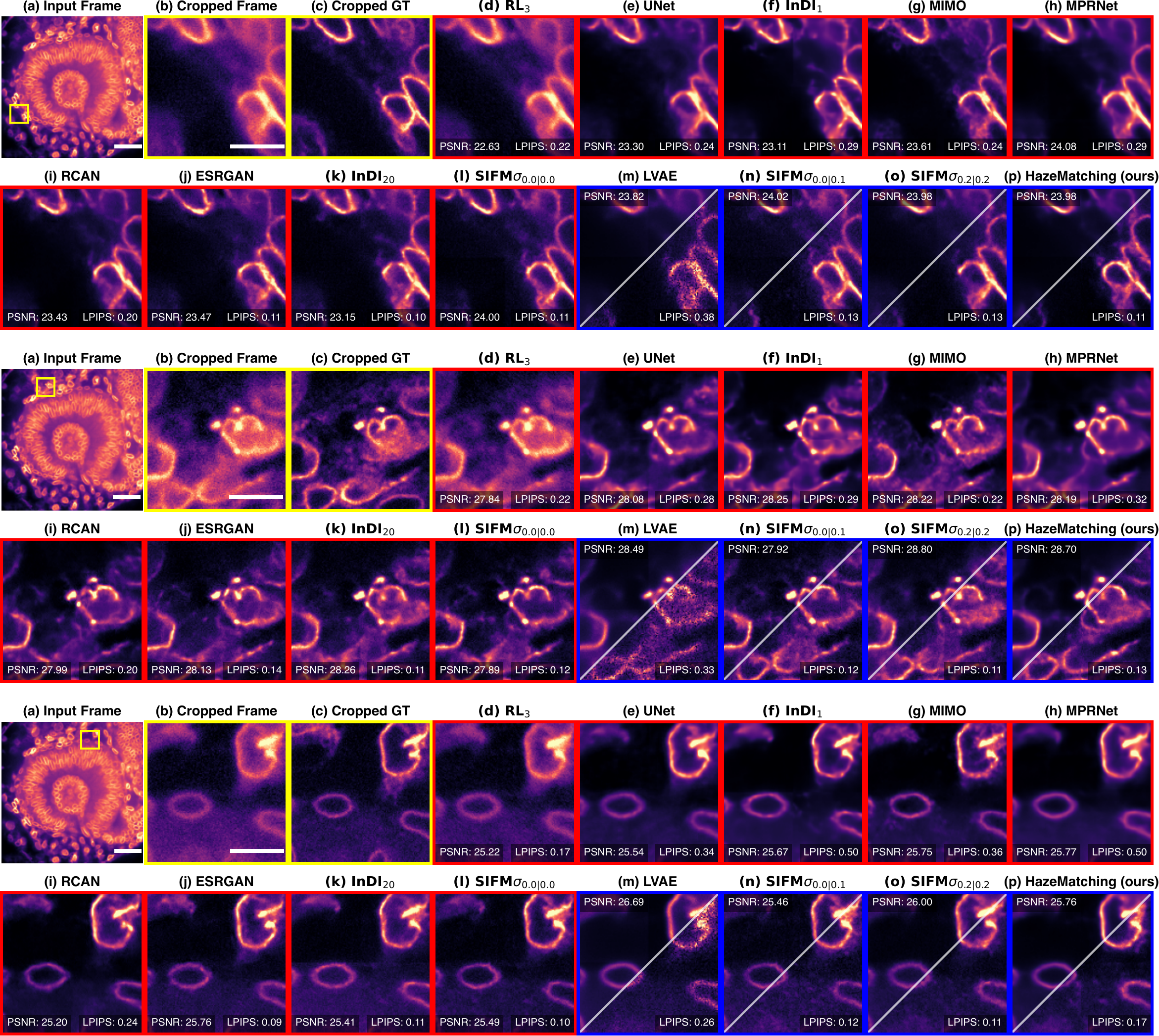}
  \caption{\textbf{Qualitative results on Zebrafish Data:} Here we present three examples showing Nuclear Envelope.  
  \textbf{(a)}~the full input and a selected 128$\times$128 crop (yellow box); Scalebar: 40 $\mu m$, 
  \textbf{(b)}~the selected crop; Scalebar: 10 $\mu m$, 
  \textbf{(c)}~non-hazy ground truth, 
  \textbf{(d–o)}~predictions by all baseline methods (see Section~\ref{subsec:baselines}), and 
  \textbf{(p)}~results obtained with \HazeMatching. 
  Results with red borders are predictions by point-predictors, while methods with blue borders are results by generative posterior models (see also Figure~\ref{fig:plot_main} and main text). Note that \HazeMatching consistently produces sharper and more perceptually aligned predictions (lower LPIPS) compared to both deterministic and posterior-based baselines, while maintaining comparative fidelity (PSNR). For the point-prediction methods, PSNR and LPIPS are computed on the cropped region shown in yellow. For posterior models (blue borders), we plot the MMSE estimate in the upper triangle with its PSNR and one posterior sample in the lower-triangle with its LPIPS score.
}
  \label{fig:sup_qualitative3}
\end{figure*}
}

\newcommand{\figQualitativeSuppleOrgOneOne}{
\begin{figure*}[h]
\centering
  \includegraphics[width=1.0\textwidth]{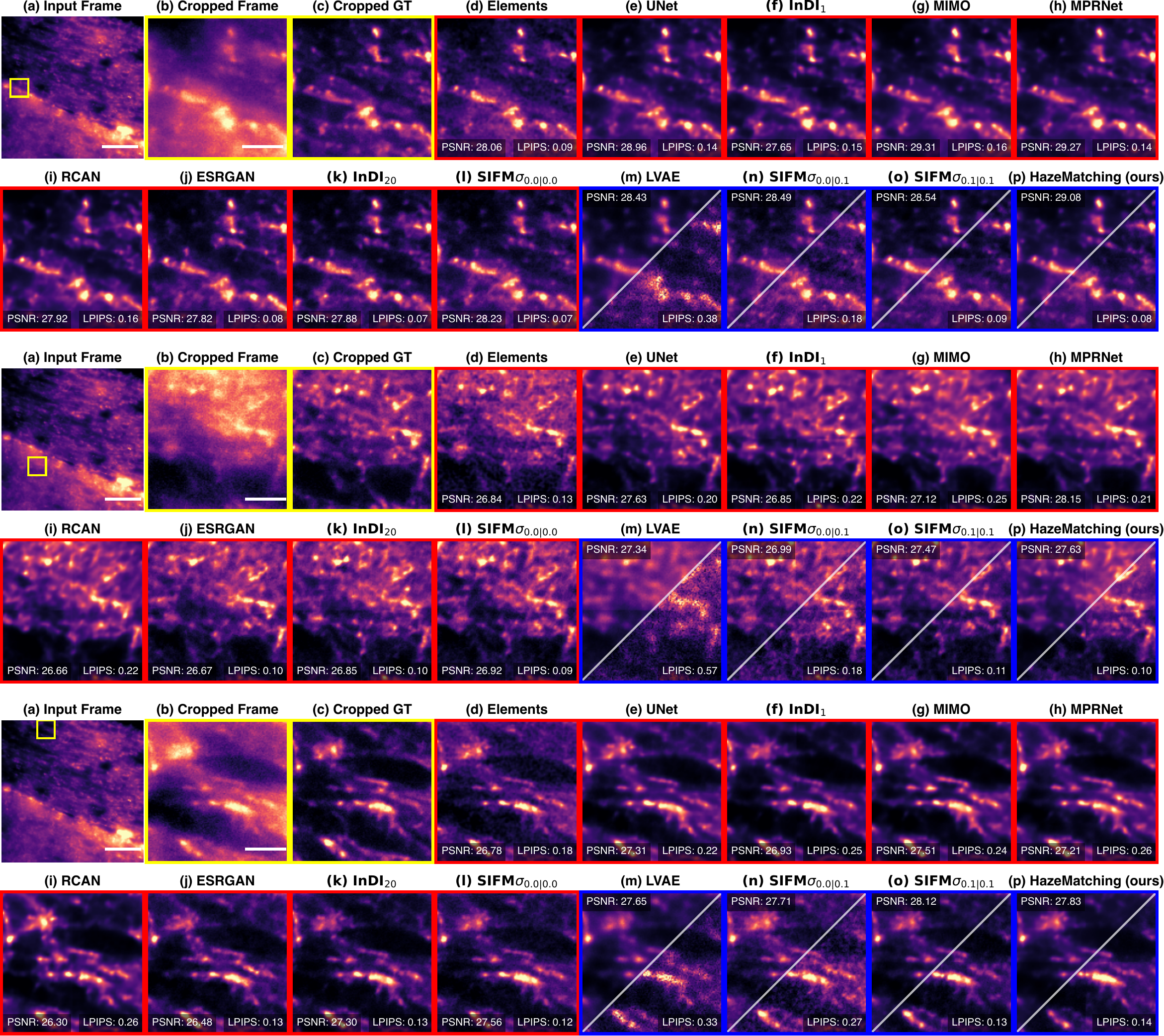}
  \caption{\textbf{Qualitative results on Organoids1 Data:} Here we present three examples showing Brain Organoids.  
  \textbf{(a)}~the full input and a selected 128$\times$128 crop (yellow box); Scalebar: 50 $\mu m$, 
  \textbf{(b)}~the selected crop; Scalebar: 10 $\mu m$, 
  \textbf{(c)}~non-hazy ground truth, 
  \textbf{(d–o)}~predictions by all baseline methods (see Section~\ref{subsec:baselines}), and 
  \textbf{(p)}~results obtained with \HazeMatching. 
  Results with red borders are predictions by point-predictors, while methods with blue borders are results by generative posterior models (see also Figure~\ref{fig:plot_main} and main text). Note that \HazeMatching consistently produces sharper and more perceptually aligned predictions (lower LPIPS) compared to both deterministic and posterior-based baselines, while maintaining comparative fidelity (PSNR). For the point-prediction methods, PSNR and LPIPS are computed on the cropped region shown in yellow. For posterior models (blue borders), we plot the MMSE estimate in the upper triangle with its PSNR and one posterior sample in the lower-triangle with its LPIPS score.
}
  \label{fig:sup_qualitative4}
\end{figure*}
}

\newcommand{\figQualitativeSuppleOrgOneTwo}{
\begin{figure*}[h]
\centering
  \includegraphics[width=1.0\textwidth]{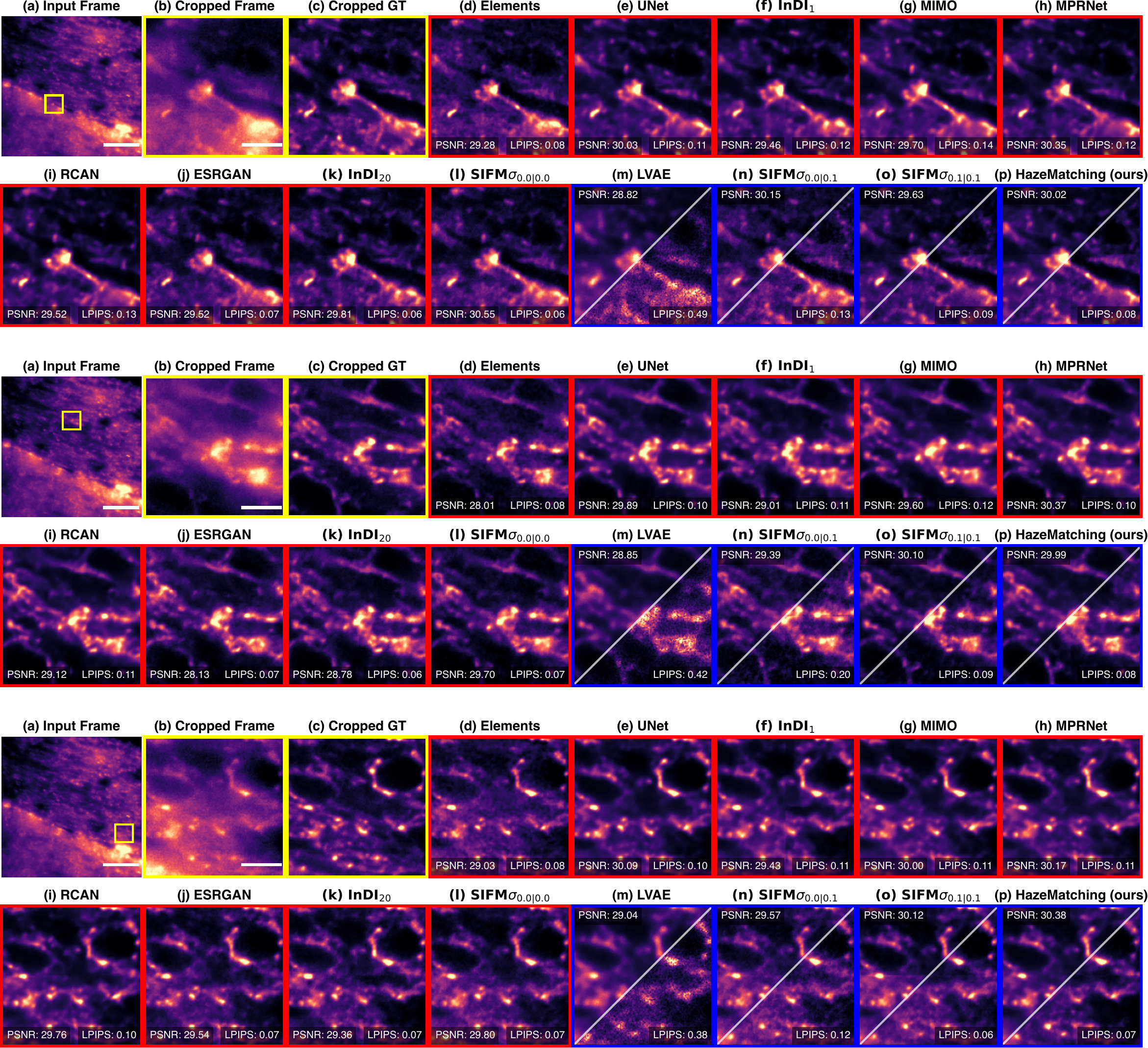}
  \caption{\textbf{Qualitative results on Organoids1 Data:} Here we present three examples showing Brain Organoids.  
  \textbf{(a)}~the full input and a selected 128$\times$128 crop (yellow box); Scalebar: 50 $\mu m$, 
  \textbf{(b)}~the selected crop; Scalebar: 10 $\mu m$, 
  \textbf{(c)}~non-hazy ground truth, 
  \textbf{(d–o)}~predictions by all baseline methods (see Section~\ref{subsec:baselines}), and 
  \textbf{(p)}~results obtained with \HazeMatching. 
  Results with red borders are predictions by point-predictors, while methods with blue borders are results by generative posterior models (see also Figure~\ref{fig:plot_main} and main text). Note that \HazeMatching consistently produces sharper and more perceptually aligned predictions (lower LPIPS) compared to both deterministic and posterior-based baselines, while maintaining comparative fidelity (PSNR). For the point-prediction methods, PSNR and LPIPS are computed on the cropped region shown in yellow. For posterior models (blue borders), we plot the MMSE estimate in the upper triangle with its PSNR and one posterior sample in the lower-triangle with its LPIPS score.
}
  \label{fig:sup_qualitative5}
\end{figure*}
}

\newcommand{\figQualitativeSuppleOrgOneThree}{
\begin{figure*}[h]
\centering
  \includegraphics[width=1.0\textwidth]{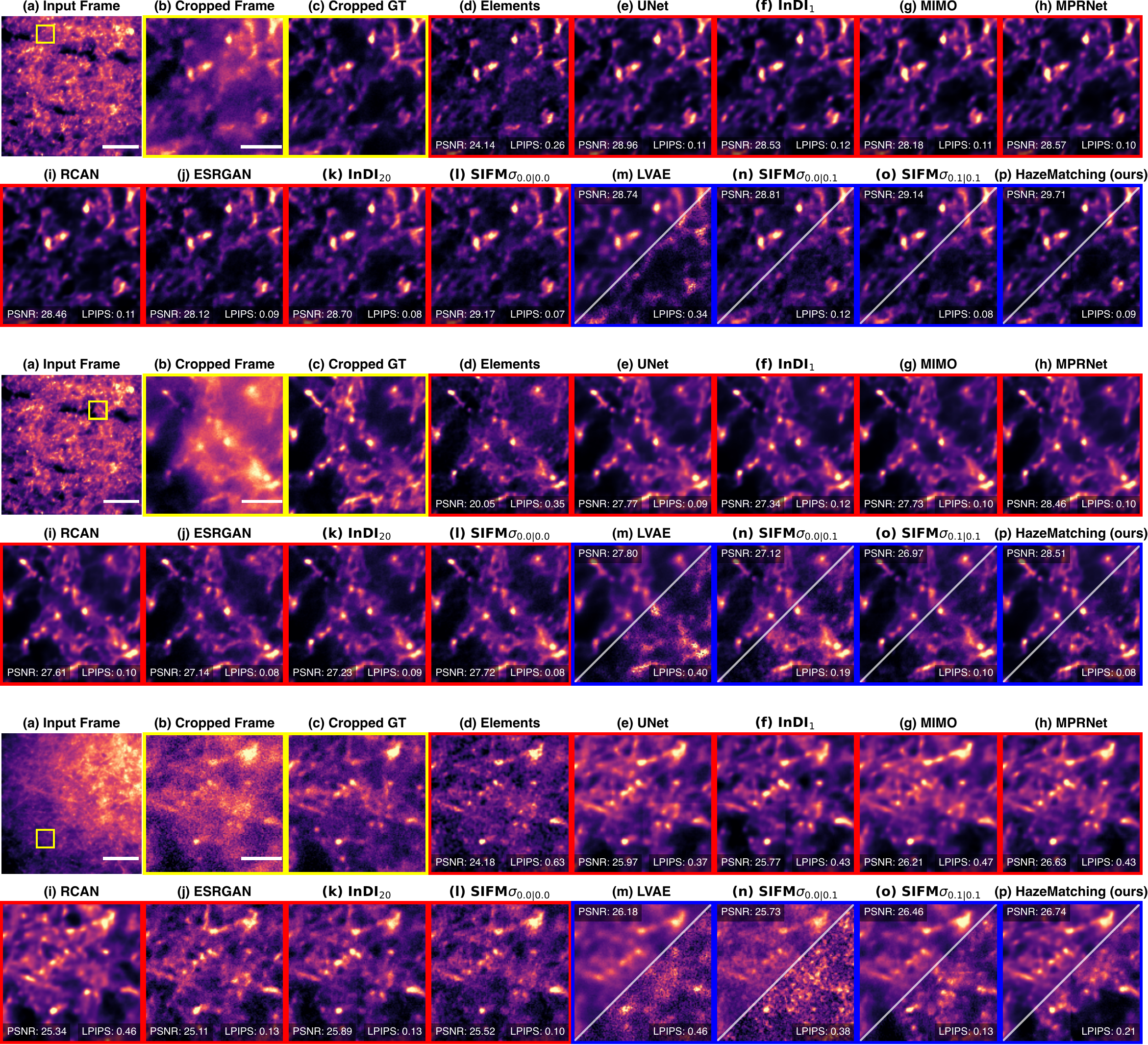}
  \caption{\textbf{Qualitative results on Organoids1 Data:} Here we present three examples showing Brain Organoids.  
  \textbf{(a)}~the full input and a selected 128$\times$128 crop (yellow box); Scalebar: 50 $\mu m$, 
  \textbf{(b)}~the selected crop; Scalebar: 10 $\mu m$, 
  \textbf{(c)}~non-hazy ground truth, 
  \textbf{(d–o)}~predictions by all baseline methods (see Section~\ref{subsec:baselines}), and 
  \textbf{(p)}~results obtained with \HazeMatching. 
  Results with red borders are predictions by point-predictors, while methods with blue borders are results by generative posterior models (see also Figure~\ref{fig:plot_main} and main text). Note that \HazeMatching consistently produces sharper and more perceptually aligned predictions (lower LPIPS) compared to both deterministic and posterior-based baselines, while maintaining comparative fidelity (PSNR). For the point-prediction methods, PSNR and LPIPS are computed on the cropped region shown in yellow. For posterior models (blue borders), we plot the MMSE estimate in the upper triangle with its PSNR and one posterior sample in the lower-triangle with its LPIPS score.
}
  \label{fig:sup_qualitative6}
\end{figure*}
}

\newcommand{\figQualitativeSuppleOrgTwoOne}{
\begin{figure*}[h]
\centering
  \includegraphics[width=1.0\textwidth]{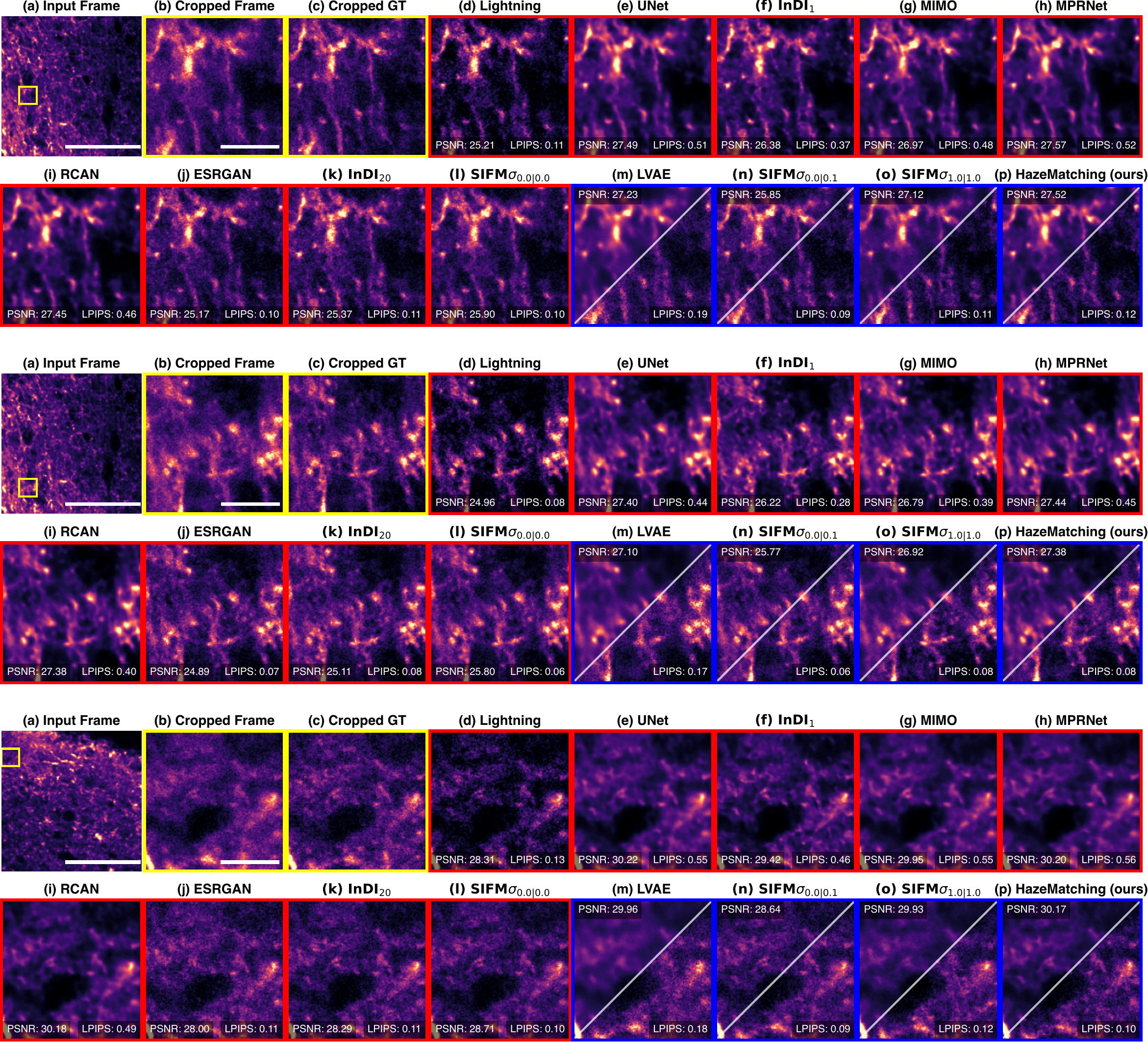}
  \caption{\textbf{Qualitative results on Organoids2 Data:} Here we present three examples showing Brain Organoids.  
  \textbf{(a)}~the full input and a selected 128$\times$128 crop (yellow box); Scalebar: 50 $\mu m$, 
  \textbf{(b)}~the selected crop; Scalebar: 5 $\mu m$, 
  \textbf{(c)}~non-hazy ground truth, 
  \textbf{(d–o)}~predictions by all baseline methods (see Section~\ref{subsec:baselines}), and 
  \textbf{(p)}~results obtained with \HazeMatching. 
  Results with red borders are predictions by point-predictors, while methods with blue borders are results by generative posterior models (see also Figure~\ref{fig:plot_main} and main text). Note that \HazeMatching consistently produces sharper and more perceptually aligned predictions (lower LPIPS) compared to both deterministic and posterior-based baselines, while maintaining comparative fidelity (PSNR). For the point-prediction methods, PSNR and LPIPS are computed on the cropped region shown in yellow. For posterior models (blue borders), we plot the MMSE estimate in the upper triangle with its PSNR and one posterior sample in the lower-triangle with its LPIPS score.
}
  \label{fig:sup_qualitative7}
\end{figure*}
}

\newcommand{\figQualitativeSuppleOrgTwoTwo}{
\begin{figure*}[h]
\centering
  \includegraphics[width=1.0\textwidth]{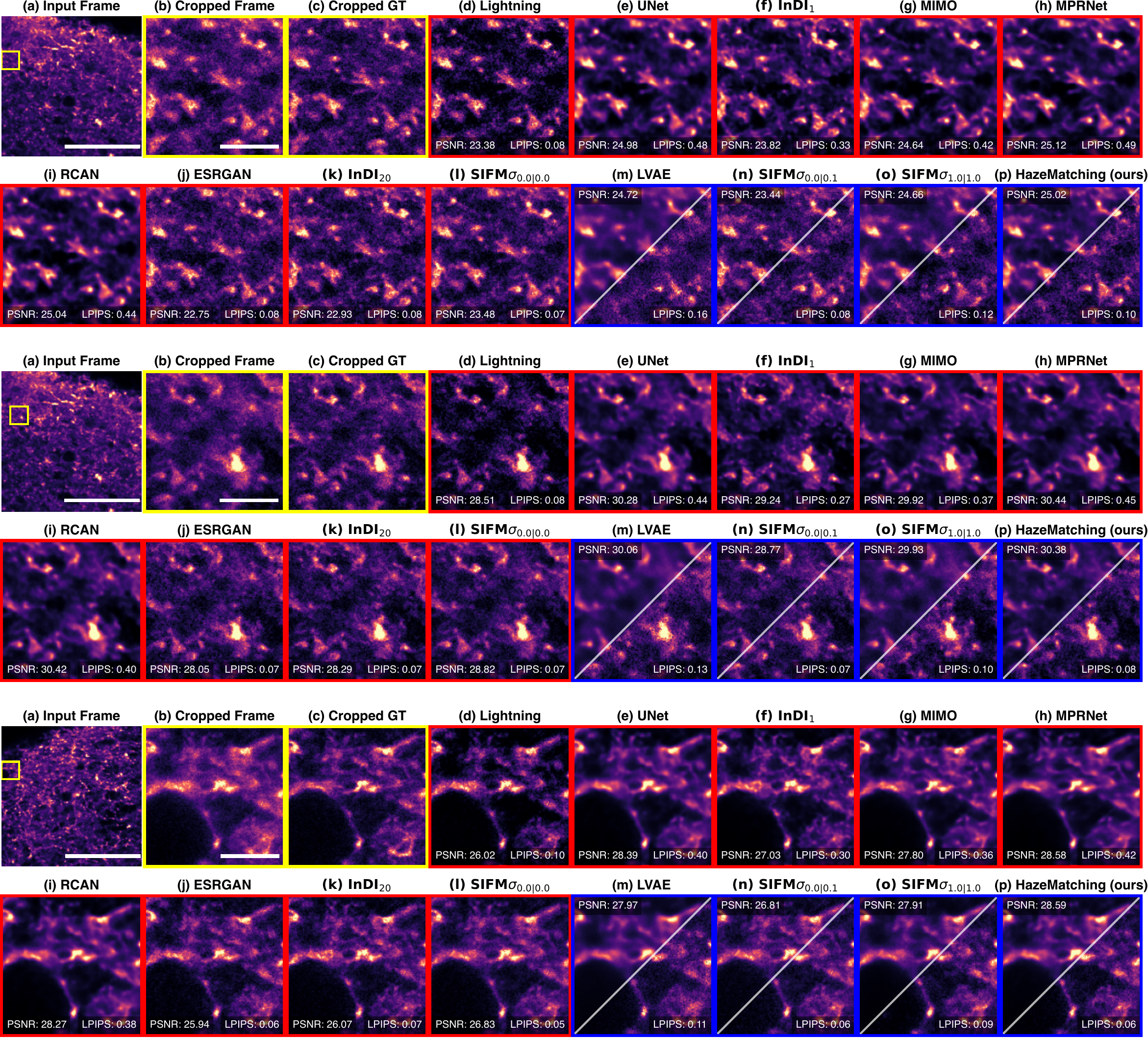}
  \caption{\textbf{Qualitative results on Organoids2 Data:} Here we present three examples showing Brain Organoids.  
  \textbf{(a)}~the full input and a selected 128$\times$128 crop (yellow box); Scalebar: 50 $\mu m$, 
  \textbf{(b)}~the selected crop; Scalebar: 5 $\mu m$, 
  \textbf{(c)}~non-hazy ground truth, 
  \textbf{(d–o)}~predictions by all baseline methods (see Section~\ref{subsec:baselines}), and 
  \textbf{(p)}~results obtained with \HazeMatching. 
  Results with red borders are predictions by point-predictors, while methods with blue borders are results by generative posterior models (see also Figure~\ref{fig:plot_main} and main text). Note that \HazeMatching consistently produces sharper and more perceptually aligned predictions (lower LPIPS) compared to both deterministic and posterior-based baselines, while maintaining comparative fidelity (PSNR). For the point-prediction methods, PSNR and LPIPS are computed on the cropped region shown in yellow. For posterior models (blue borders), we plot the MMSE estimate in the upper triangle with its PSNR and one posterior sample in the lower-triangle with its LPIPS score.
}
  \label{fig:sup_qualitative8}
\end{figure*}
}

\newcommand{\figQualitativeSuppleOrgTwoThree}{
\begin{figure*}[h]
\centering
  \includegraphics[width=1.0\textwidth]{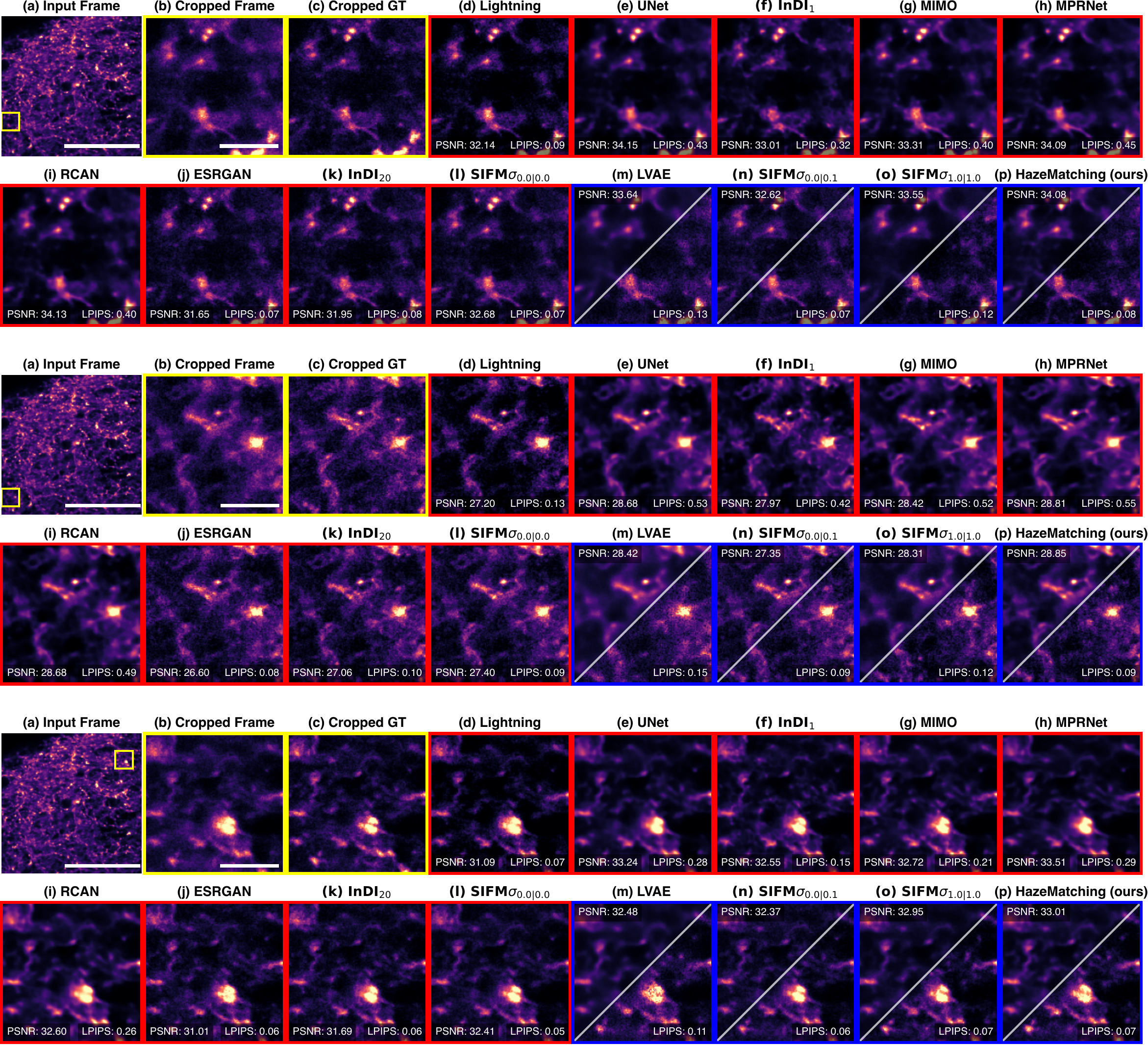}
  \caption{\textbf{Qualitative results on Organoids2 Data:} Here we present three examples showing Brain Organoids.  
  \textbf{(a)}~the full input and a selected 128$\times$128 crop (yellow box); Scalebar: 50 $\mu m$, 
  \textbf{(b)}~the selected crop; Scalebar: 5 $\mu m$, 
  \textbf{(c)}~non-hazy ground truth, 
  \textbf{(d–o)}~predictions by all baseline methods (see Section~\ref{subsec:baselines}), and 
  \textbf{(p)}~results obtained with \HazeMatching. 
  Results with red borders are predictions by point-predictors, while methods with blue borders are results by generative posterior models (see also Figure~\ref{fig:plot_main} and main text). Note that \HazeMatching consistently produces sharper and more perceptually aligned predictions (lower LPIPS) compared to both deterministic and posterior-based baselines, while maintaining comparative fidelity (PSNR). For the point-prediction methods, PSNR and LPIPS are computed on the cropped region shown in yellow. For posterior models (blue borders), we plot the MMSE estimate in the upper triangle with its PSNR and one posterior sample in the lower-triangle with its LPIPS score.
}
  \label{fig:sup_qualitative9}
\end{figure*}
}

\newcommand{\figQualitativeSuppleMicrotubuleOne}{
\begin{figure*}[h]
\centering
  \includegraphics[width=1.0\textwidth]{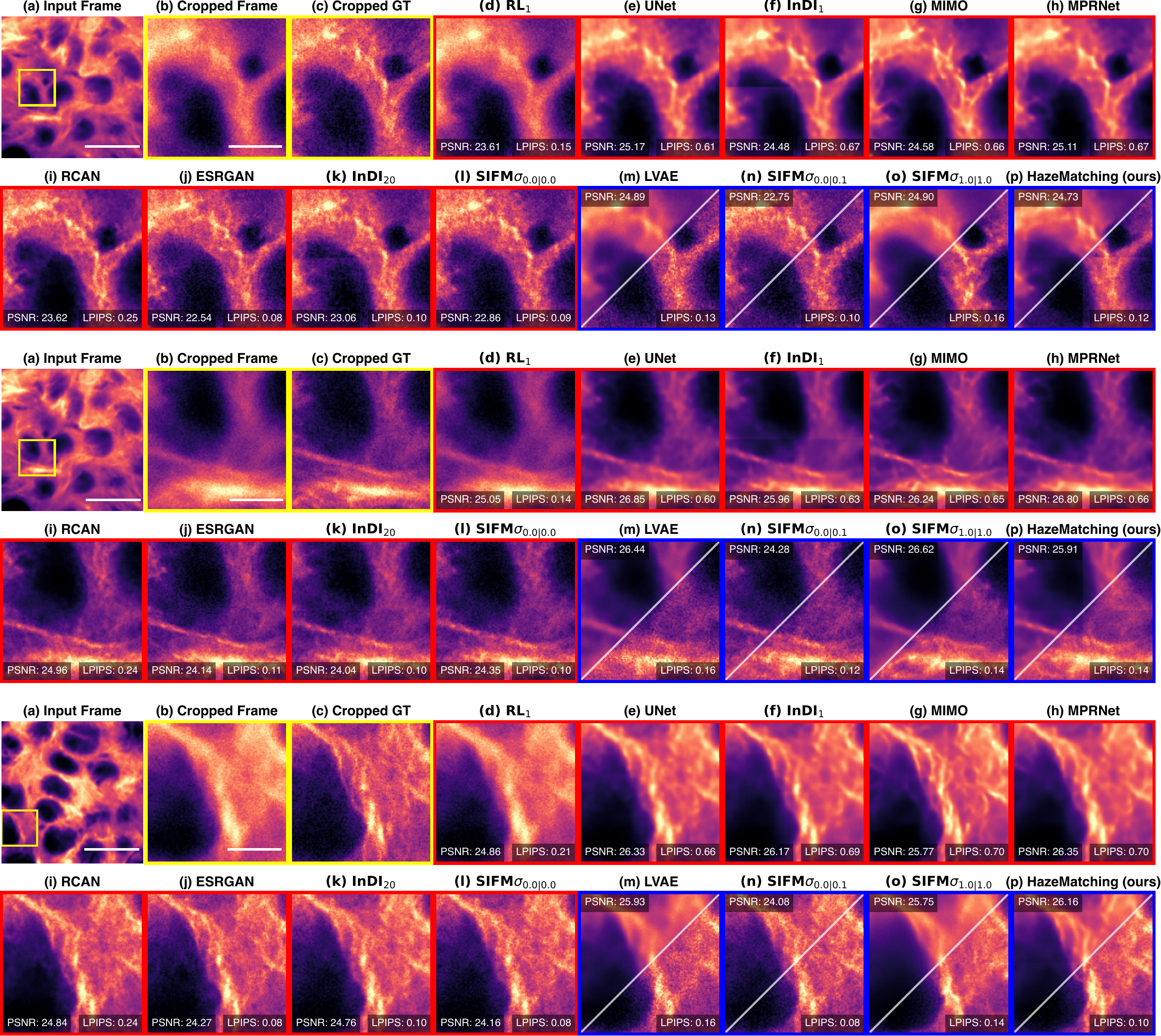}
  \caption{\textbf{Qualitative results on Microtubule Data:} Here we present three examples showing Microtubules tagged with $\alpha$-Tubulin.  
  \textbf{(a)}~the full input and a selected 128$\times$128 crop (yellow box); Scalebar: 20 $\mu m$, 
  \textbf{(b)}~the selected crop; Scalebar: 5 $\mu m$, 
  \textbf{(c)}~non-hazy ground truth, 
  \textbf{(d–o)}~predictions by all baseline methods (see Section~\ref{subsec:baselines}), and 
  \textbf{(p)}~results obtained with \HazeMatching. 
  Results with red borders are predictions by point-predictors, while methods with blue borders are results by generative posterior models (see also Figure~\ref{fig:plot_main} and main text). Note that \HazeMatching consistently produces sharper and more perceptually aligned predictions (lower LPIPS) compared to both deterministic and posterior-based baselines, while maintaining comparative fidelity (PSNR). For the point-prediction methods, PSNR and LPIPS are computed on the cropped region shown in yellow. For posterior models (blue borders), we plot the MMSE estimate in the upper triangle with its PSNR and one posterior sample in the lower-triangle with its LPIPS score.
}
  \label{fig:sup_qualitative10}
\end{figure*}
}

\newcommand{\figQualitativeSuppleMicrotubuleTwo}{
\begin{figure*}[h]
\centering
  \includegraphics[width=1.0\textwidth]{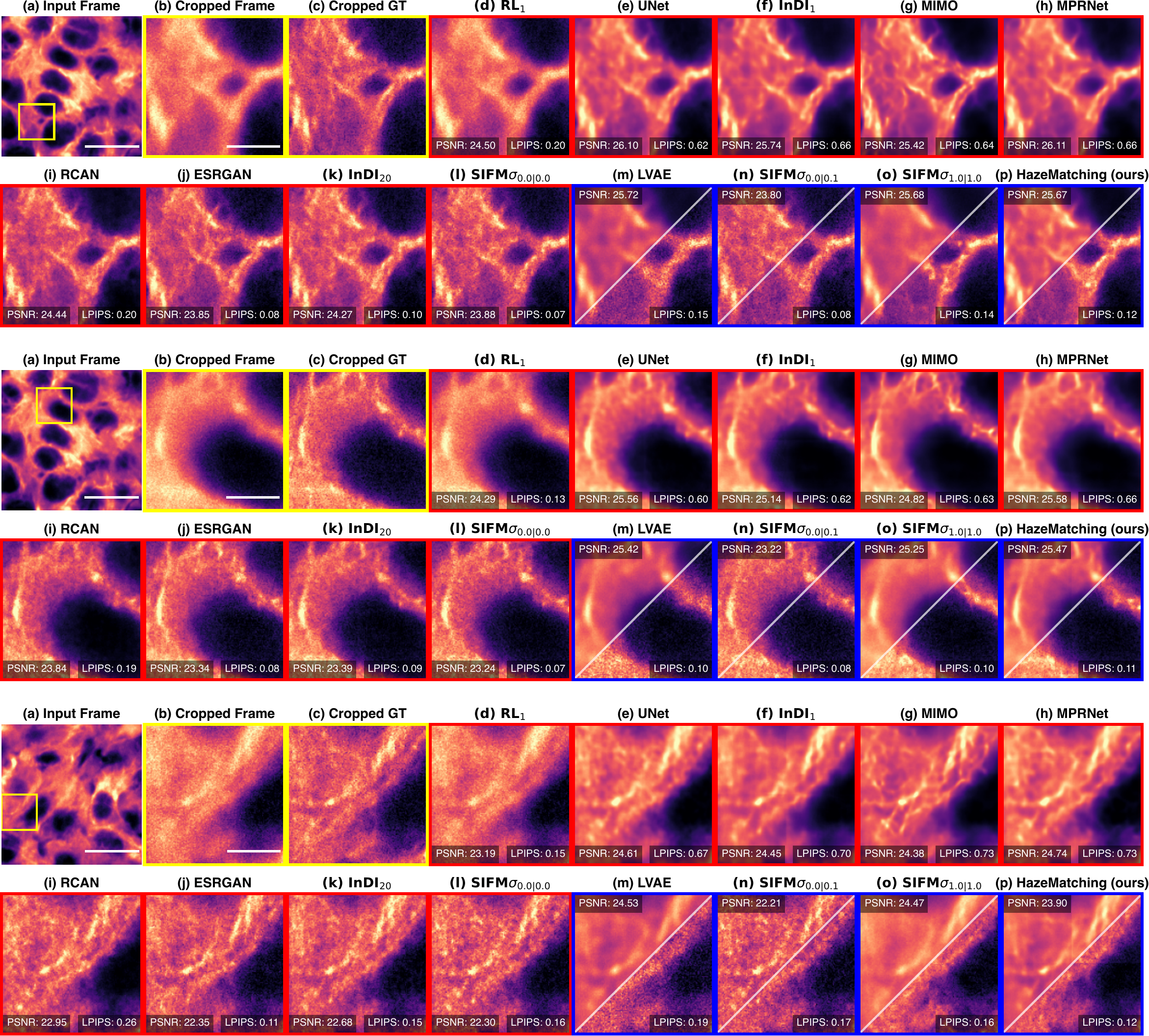}
  \caption{\textbf{Qualitative results on Microtubule Data:} Here we present three examples showing Microtubules tagged with $\alpha$-Tubulin.  
  \textbf{(a)}~the full input and a selected 128$\times$128 crop (yellow box); Scalebar: 20 $\mu m$, 
  \textbf{(b)}~the selected crop; Scalebar: 5 $\mu m$, 
  \textbf{(c)}~non-hazy ground truth, 
  \textbf{(d–o)}~predictions by all baseline methods (see Section~\ref{subsec:baselines}), and 
  \textbf{(p)}~results obtained with \HazeMatching. 
  Results with red borders are predictions by point-predictors, while methods with blue borders are results by generative posterior models (see also Figure~\ref{fig:plot_main} and main text). Note that \HazeMatching consistently produces sharper and more perceptually aligned predictions (lower LPIPS) compared to both deterministic and posterior-based baselines, while maintaining comparative fidelity (PSNR). For the point-prediction methods, PSNR and LPIPS are computed on the cropped region shown in yellow. For posterior models (blue borders), we plot the MMSE estimate in the upper triangle with its PSNR and one posterior sample in the lower-triangle with its LPIPS score.
}
  \label{fig:sup_qualitative11}
\end{figure*}
}

\newcommand{\figQualitativeSuppleMicrotubuleThree}{
\begin{figure*}[h]
\centering
  \includegraphics[width=1.0\textwidth]{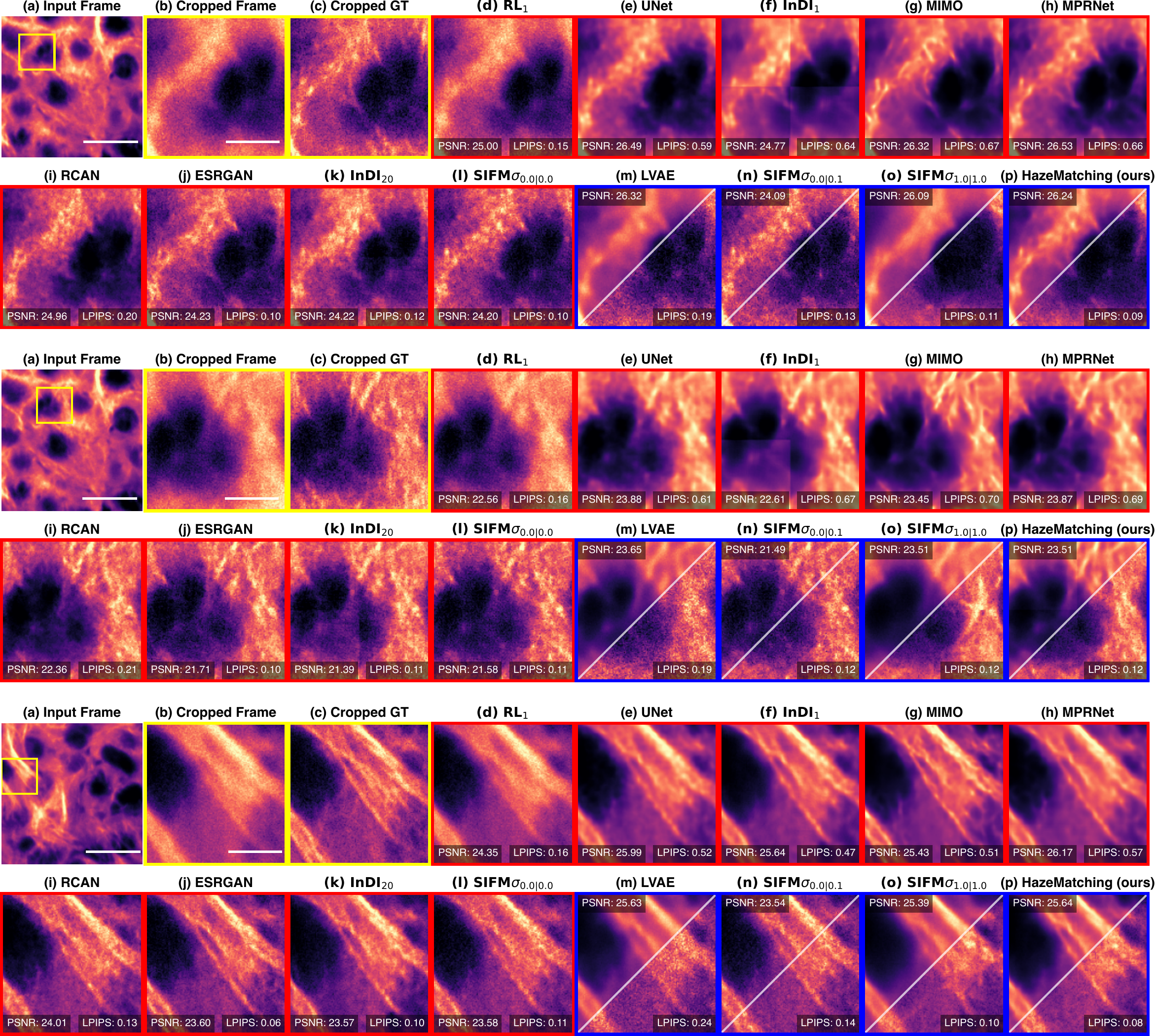}
  \caption{\textbf{Qualitative results on Microtubule Data:} Here we present three examples showing Microtubules tagged with $\alpha$-Tubulin.  
  \textbf{(a)}~the full input and a selected 128$\times$128 crop (yellow box); Scalebar: 20 $\mu m$, 
  \textbf{(b)}~the selected crop; Scalebar: 5 $\mu m$, 
  \textbf{(c)}~non-hazy ground truth, 
  \textbf{(d–o)}~predictions by all baseline methods (see Section~\ref{subsec:baselines}), and 
  \textbf{(p)}~results obtained with \HazeMatching. 
  Results with red borders are predictions by point-predictors, while methods with blue borders are results by generative posterior models (see also Figure~\ref{fig:plot_main} and main text). Note that \HazeMatching consistently produces sharper and more perceptually aligned predictions (lower LPIPS) compared to both deterministic and posterior-based baselines, while maintaining comparative fidelity (PSNR). For the point-prediction methods, PSNR and LPIPS are computed on the cropped region shown in yellow. For posterior models (blue borders), we plot the MMSE estimate in the upper triangle with its PSNR and one posterior sample in the lower-triangle with its LPIPS score.
}
  \label{fig:sup_qualitative12}
\end{figure*}
}

\newcommand{\figQualitativeSuppleNeuronOne}{
\begin{figure*}[h]
\centering
  \includegraphics[width=1.0\textwidth]{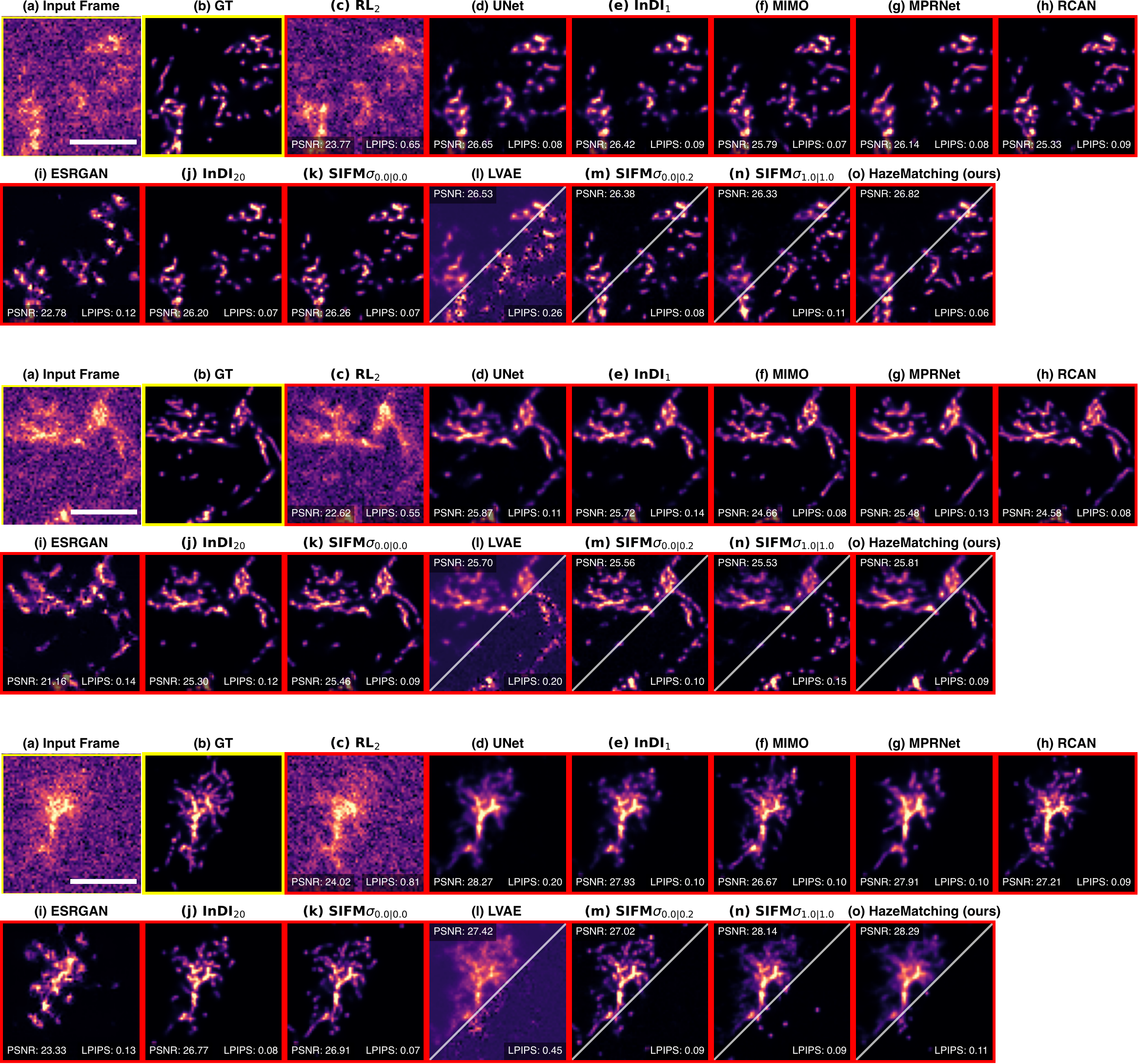}
  \caption{\textbf{Qualitative results on Neuron Data:} Here we present three examples showing Neuron data.  
  \textbf{(a)}~the full 64$\times$64 input frame (yellow box); Scalebar: 5 $\mu m$, 
  \textbf{(b)}~non-hazy ground truth, 
  \textbf{(c–n)}~predictions by all baseline methods (see Section~\ref{subsec:baselines}), and 
  \textbf{(0)}~results obtained with \HazeMatching. 
  Results with red borders are predictions by point-predictors, while methods with blue borders are results by generative posterior models (see also Figure~\ref{fig:plot_main} and main text). Note that \HazeMatching consistently produces sharper and more perceptually aligned predictions (lower LPIPS) compared to both deterministic and posterior-based baselines, while maintaining comparative fidelity (PSNR). For the point-prediction methods, PSNR and LPIPS are computed on the cropped region shown in yellow. For posterior models (blue borders), we plot the MMSE estimate in the upper triangle with its PSNR and one posterior sample in the lower-triangle with its LPIPS score.
}
  \label{fig:sup_qualitative13}
\end{figure*}
}

\newcommand{\figQualitativeSuppleNeuronTwo}{
\begin{figure*}[h]
\centering
  \includegraphics[width=1.0\textwidth]{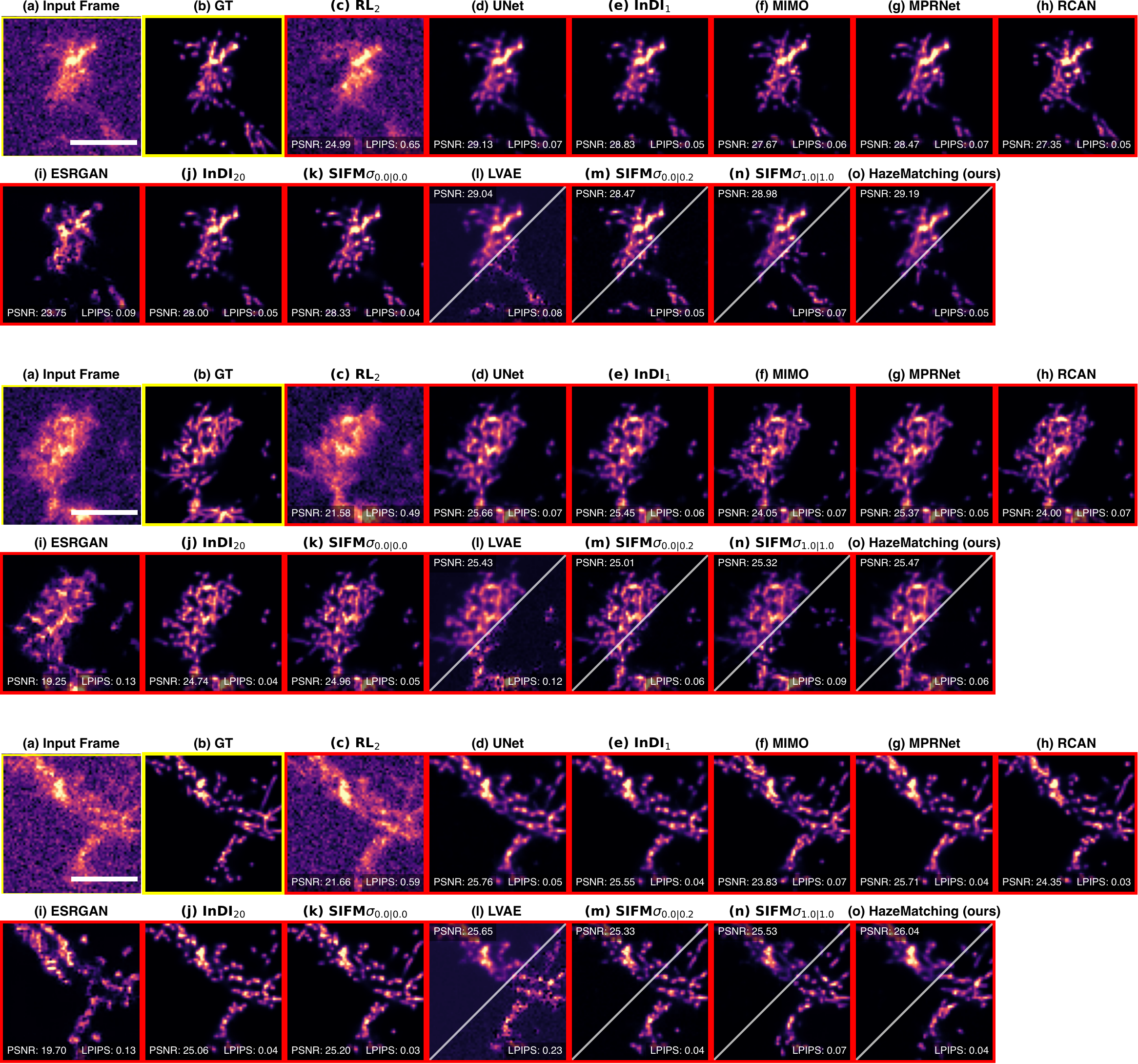}
  \caption{\textbf{Qualitative results on Neuron Data:} Here we present three examples showing Neuron data.  
  \textbf{(a)}~the full 64$\times$64 input frame (yellow box); Scalebar: 5 $\mu m$, 
  \textbf{(b)}~non-hazy ground truth, 
  \textbf{(c–n)}~predictions by all baseline methods (see Section~\ref{subsec:baselines}), and 
  \textbf{(o)}~results obtained with \HazeMatching. 
  Results with red borders are predictions by point-predictors, while methods with blue borders are results by generative posterior models (see also Figure~\ref{fig:plot_main} and main text). Note that \HazeMatching consistently produces sharper and more perceptually aligned predictions (lower LPIPS) compared to both deterministic and posterior-based baselines, while maintaining comparative fidelity (PSNR). For the point-prediction methods, PSNR and LPIPS are computed on the cropped region shown in yellow. For posterior models (blue borders), we plot the MMSE estimate in the upper triangle with its PSNR and one posterior sample in the lower-triangle with its LPIPS score.
}
  \label{fig:sup_qualitative14}
\end{figure*}
}

\newcommand{\figQualitativeSuppleNeuronThree}{
\begin{figure*}[h]
\centering
  \includegraphics[width=1.0\textwidth]{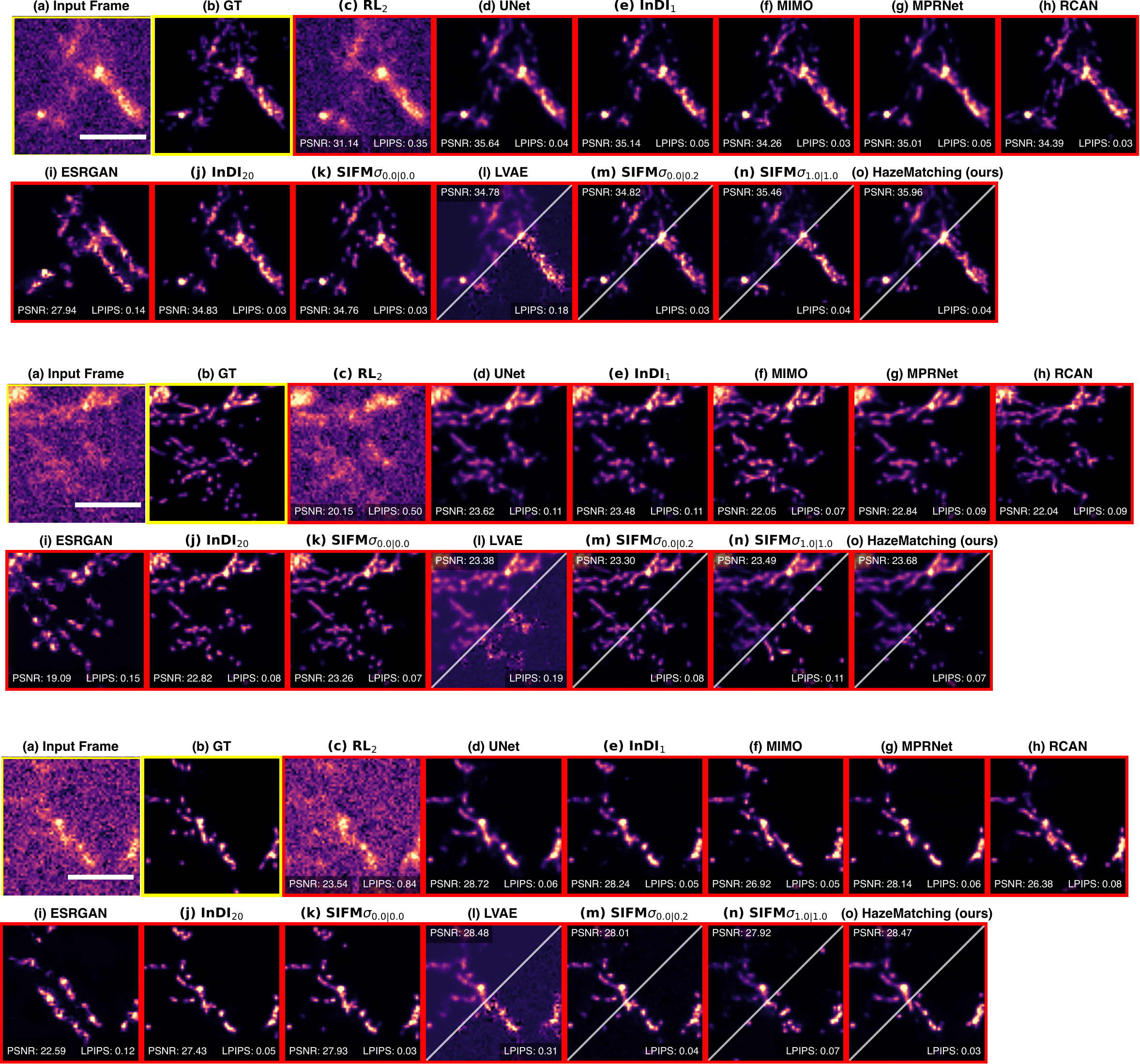}
  \caption{\textbf{Qualitative results on Neuron Data:} Here we present three examples showing Neuron data.  
  \textbf{(a)}~the full 64$\times$64 input frame (yellow box); Scalebar: 5 $\mu m$, 
  \textbf{(b)}~non-hazy ground truth, 
  \textbf{(c–n)}~predictions by all baseline methods (see Section~\ref{subsec:baselines}), and 
  \textbf{(o)}~results obtained with \HazeMatching. 
  Results with red borders are predictions by point-predictors, while methods with blue borders are results by generative posterior models (see also Figure~\ref{fig:plot_main} and main text). Note that \HazeMatching consistently produces sharper and more perceptually aligned predictions (lower LPIPS) compared to both deterministic and posterior-based baselines, while maintaining comparative fidelity (PSNR). For the point-prediction methods, PSNR and LPIPS are computed on the cropped region shown in yellow. For posterior models (blue borders), we plot the MMSE estimate in the upper triangle with its PSNR and one posterior sample in the lower-triangle with its LPIPS score.
}
  \label{fig:sup_qualitative15}
\end{figure*}
}

\newcommand{\figQualitativeZebrafishFULL}{
\begin{figure*}[h]
\centering
  \includegraphics[width=0.4\textwidth]{Figs/Qual_Zebrafish_FULL.pdf}
  \caption{\textbf{Qualitative results on Zebrafish Data:} Here we show one full-frame image from the test data.
  \textbf{(a)}~the full 1024$\times$1024 input frame; Scalebar: 50 $\mu m$, 
  \textbf{(b)}~Corresponding prediction by \HazeMatching, 
  \textbf{(c)}~the Ground Truth, Best viewed on a screen with 5$\times$ zoom.
}
  \label{fig:sup_qualitative16}
\end{figure*}
}

\newcommand{\figQualitativeOrgOneFULL}{
\begin{figure*}[h]
\centering
  \includegraphics[width=0.4\textwidth]{Figs/Qual_LIF1_FULL.pdf}
  \caption{\textbf{Qualitative results on Organoids1 Data:} Here we show one full-frame image from the test data.
  \textbf{(a)}~the full 1024$\times$1024 input frame; Scalebar: 50 $\mu m$, 
  \textbf{(b)}~Corresponding prediction by \HazeMatching, 
  \textbf{(c)}~the Ground Truth, Best viewed on a screen with 5$\times$ zoom.
}
  \label{fig:sup_qualitative17}
\end{figure*}
}

\newcommand{\figQualitativeOrgTwoFULL}{
\begin{figure*}[h]
\centering
  \includegraphics[width=0.4\textwidth]{Figs/Qual_LIF2_FULL.pdf}
  \caption{\textbf{Qualitative results on Organoids2 Data:} Here we show one full-frame image from the test data.
  \textbf{(a)}~the full 1024$\times$1024 input frame; Scalebar: 50 $\mu m$, 
  \textbf{(b)}~Corresponding prediction by \HazeMatching, 
  \textbf{(c)}~the Ground Truth, Best viewed on a screen with 5$\times$ zoom.
}
  \label{fig:sup_qualitative18}
\end{figure*}
}

\newcommand{\figQualitativeALLENFULL}{
\begin{figure*}[h]
\centering
  \includegraphics[width=0.38\textwidth]{Figs/Qual_ALLEN_FULL.pdf}
  \caption{\textbf{Qualitative results on Organoids2 Data:} Here we show one full-frame image from the test data.
  \textbf{(a)}~the full 1024$\times$1024 input frame; Scalebar: 20 $\mu m$, 
  \textbf{(b)}~Corresponding prediction by \HazeMatching, 
  \textbf{(c)}~the Ground Truth, Best viewed on a screen with 5$\times$ zoom.
}
  \label{fig:sup_qualitative19}
\end{figure*}
}
\begin{document}
\maketitle
\begin{abstract}
Fluorescence microscopy is a major driver of scientific progress in the life sciences.
Although high-end confocal microscopes are capable of filtering out-of-focus light, cheaper and more accessible microscopy modalities, such as widefield microscopy, can not, which consequently leads to hazy image data.
Computational dehazing is trying to combine the best of both worlds, leading to cheap microscopy but crisp-looking images.
The perception-distortion trade-off tells us that we can optimize either for data fidelity, e.g. low \MSE or high \PSNR, or for data realism, measured by perceptual metrics such as \LPIPS or \FID. Existing methods either prioritize fidelity at the expense of realism, or produce perceptually convincing results that lack quantitative accuracy.
In this work, we propose \HazeMatching, a novel iterative method for dehazing light microscopy images, 
which effectively balances these objectives.
Our goal was to find a balanced trade-off between the fidelity of the dehazing results and the realism of individual predictions (samples).
We achieve this by adapting the conditional flow matching framework by guiding the generative process with a hazy observation in the conditional velocity field.
We evaluate \HazeMatching on $5$ datasets, covering both synthetic and real data, assessing both distortion and perceptual quality. Our method is compared against $12$ baselines,
achieving a consistent balance between fidelity and realism on average. Additionally, with calibration analysis, we show that \HazeMatching produces well-calibrated predictions. 
Note that our method does not need an explicit degradation operator to exist, making it easily applicable on real microscopy data. All data used for training and evaluation and our code will be publicly available under a permissive license\footnote{\url{https://github.com/juglab/HazeMatching}}.
\end{abstract}
\section{Introduction}
\label{sec:intro}
Today, inverse problems in computer vision are often addressed with deep learning approaches of various sorts \cite{dl4miasurvey}, and image dehazing is one such inverse problem that has important applications in the life sciences.
When imaging biological samples with a light microscope, diffraction limited 2D images can be acquired.
The focal depth of such images can vary greatly, depending on the microscopy modality to be used: confocal microscopes have an exquisitely shallow depth of focus, leading to very crisp images subject to little haze, while widefield microscopes have a much wider depth of focus and are therefore collecting large amounts of out-of-focus light, leading to hazy images that are harder to interpret and analyze.
Confocal microscopes achieve their high image quality via optical sectioning with a physical pinhole in the light path of the microscope. 
While this is a fantastic optical trick, it has also downsides, one of which is the price of such microscopes.

\figTeaserMain

Here, we ask the question to what degree can we replace the optical ingenuity of confocal microscopes by purely computational means.
If a body of hazy widefield microscopy images is given to us, can we devise a method that can solve the inverse task of predicting how the same sample would appear if it were imaged with a confocal microscope?

Natural choices for tackling the inverse task of dehazing are supervised setups~\cite{unet, mimo, mprnet, rcan, esrgan, hvae}, diffusion-like setups~\cite{indi}, or flow matching approaches \citep{sifm}.
In all cases, we require a body of paired data, \ie pairs of images that show the same image content (the biological sample $s\in S$), once with and another time without haze.
Technically, this can be achieved by acquiring data at a confocal microscope (non-hazy images) and acquiring a second micrograph after fully opening the above-mentioned pinhole (leading to widefield-like hazy images).
Another possibility is to use synthetically generated microscopy images, using a microscopy simulator that simulates the optics of various light microscopes.
In this work, we do both -- we use synthetic data, giving us precise control and full knowledge of $s$, and we use real microscopy images to demonstrate that our method works on such data as well.

While computational dehazing enables life scientists to use cheaper, faster, and less photo-toxic microscopes while still obtaining crisp images for downstream analyzes, the perception–distortion trade-off must be considered~\citep{PDT, floriansReviewWithSulianaEtAl}. 
We can optimize either for data fidelity (\eg, low \MSE or high \PSNR) or for perceptual realism (\eg, low \LPIPS or \FID), or need to make a trade-off. 
Supervised networks trained with conservative losses such as MSE or MAE preserve fidelity by staying close to the evidence in the given input data, while generative methods like diffusion or flow matching favor realism by aligning predictions with the target domain. 
In the context of scientific imaging, fidelity to the raw data (our scientific measurements) is key, but within a set of high-fidelity solutions, we would naturally prefer the one with the highest realism.
\section{Related Work}
\label{sec:relatedwork}

Classical deconvolution methods, such as Richardson–Lucy (RL), efficiently reverse optical blur in fluorescence and confocal microscopy but amplify noise and depend on explicit point spread function (PSF) models and full 3D stacks, often introducing artifacts \cite{RL,RL_bad}. 

Learning-based point predictors, \eg a \UNet trained with an MSE loss~\cite{care, pn2v}, as well as modern convolutional restoration methods such as MIMO-UNet~\cite{mimo}, MPRNet~\cite{mprnet}, or RCAN~\cite{rcan}, tend to produce overly smooth predictions~\cite{floriansReviewWithSulianaEtAl}.
GAN-based restoration methods, such as ESRGAN~\cite{esrgan}, while able to overcome overly smooth predictions, often hallucinate structures due to their mode-seeking training objective~\cite{PDT}.

Variational approaches, such as ladder VAEs (\HVAE), and their microscopy adaptations enable uncertainty-aware sampling from learned posteriors~\cite{hvae,hdn}. 
Iterative generative methods, like diffusion models~\cite{DDPM, DDIM, DDRM} and conditional flow matching (CFM)~\cite{lipman}, progressively denoise in iterations to generate very detailed, non-smoothed predictions. Recent diffusion-based methods such as HazeDiff~\cite{hazediff} address image dehazing for natural images. 

The drawback is slow and computationally intensive inference~\cite{con_diff, cold_diff}. 
While CFM’s ODE-based transport offers faster, deterministic sampling, it is, so far, rarely used in the context of microscopy image restoration, with recent efforts like~\cite{sifm} being a notable exception.

\subsection{Overview: Conditional Flow Matching (CFM)}
\label{sec:cfm_overview}

Conditional Flow Matching (CFM) is a method for training continuous-time generative models in which a neural network learns a time-dependent vector field that transports samples from a base distribution to a target data distribution. Here, we present a brief recap of flow matching and its relation to conditional flow matching.

Let $p_0(x)$ and $p_1(x)$ be the base and target data distributions, respectively. 
The goal of Flow Matching is to learn a continuous-time transport map that evolves samples from $p_0$ into samples from $p_1$. 
This transport is represented by a family of mappings $\psi_t(x_0)$ for $t \in [0,1]$ with $\psi_t$ being the mapping function and $x_0 \sim p_0(x)$, and the evolution is governed by a time-dependent vector field $v(t, x)$, such that
\begin{equation}
    \frac{d}{dt} \psi_t(x_0) = v(t,\psi_t(x_0)), \quad \psi_0(x_0) = x_0.
\end{equation}

The density of transformed samples at time $t$, denoted by the probability density function $p_t$, is the pushforward of $p_0$ under $\psi_t$, \ie, $p_t = (\psi_t)_\# p_0$, 
where $\#$ is the pushforward operator, describing how the distribution $p_0$ changes to $p_t$ when its random variables are transformed by $\psi_t$. In~\cite{lipman}, the authors introduce a simulation-free method to train such a vector field by minimizing a regression loss, \ie
\begin{equation}
    \min_{\theta} \ \mathbb{E}_{t \sim \mathcal{U}[0,1],\ x \sim p_t} \left\| v_\theta(t,x) - v(t,x) \right\|^2,
\label{eq:fm}
\end{equation}
where $v_\theta$ is the learnable vector field, and $v(t,x)$ is the \emph{target} vector field that generates the intermediate distributions $p_t$ connecting $p_0$ to $p_1$, subject to the boundary conditions $p_{t=0} = p_0$ and $p_{t=1} = p_1$. 
Since computing $p_t$ and $v$ directly is intractable in general, the authors of~\cite{lipman} proposed a conditional formulation. 
Rather than modeling the marginals $p_t$ and the marginal vector field $v(t,x)$, one can model the \emph{conditional} path $p_t(x|x_1)$ and \emph{conditional} vector field $v(t,x|x_1)$, where $x_0 \sim p_0(x)$ and $x_1 \sim p_1(x)$ are coupled. 
In particular, one can define a deterministic interpolation path between $x_0$ and $x_1$ with $\psi_t$ as
\begin{equation}
   \psi_t(x_0|x_1) = x_t = (1 - t) x_0 + t x_1,
   \label{eq:overview_xt}
\end{equation}
and define the conditional velocity field along this path as
\begin{equation}
     v(t,x_t|x_1) = \frac{dx_t }{dt} = x_1 - x_0.
   \label{eq:overview_vt}
\end{equation}
This leads to the Conditional Flow Matching (CFM) objective
\begin{equation}
    \min_{\theta} \ \mathbb{E}_{t \sim \mathcal{U}[0,1],\ x_0 \sim p_0,\ x_1 \sim p_1} \left\| v_\theta\left(t,x_t\right) - (x_1 - x_0) \right\|^2,
\label{eq:cfm}
\end{equation}
which is the basis of our own work.

\subsection{Data-dependent realizations of CFM}

When applying Conditional Flow Matching to image restoration tasks, conditioning on the observation is essential to ensure consistency with the input and hence data fidelity. 
While recent CFM methods address related problems, they are not directly suited for image dehazing. 
For instance, in~\cite{learned_conditional}, the authors propose to learn a data-dependent coupling over $p_1(x_1)p_0(x_0|x_1)$ via a forward model $q_{\theta}(x_0|x_1)$, where $x_0 \sim \mathcal{N}(\mu_{\theta}(x_1), \Sigma_{\theta}(x_1))$. 
Although this imposes a Gaussian degradation conditioned on $x_1$, it lacks dependence on external observations (\ie the hazy measurement). 

Similarly, SIFM~\cite{sifm} proposes to define an explicit data-dependent coupling between $x_0$ and $x_1$ by introducing a degradation operator $m(x_1)$, leading to the base sample $x_0$ through $x_0 = m(x_1) + \sigma\zeta$, where $\zeta \sim \mathcal{N}(0,I)$ and $\sigma > 0$. 
This is effective when the degradation operator $m$ and the noise level $\sigma$ are known or reliably estimable. 
However, in the context of our work, $m$ is unknown, and noise properties vary widely. 
Even with access to $m(x_1)$ during training, generating multiple samples at inference is non-trivial without an accurate noise level estimate. 
In our experiments, we refer to this baseline as \texttt{SIFM\{$\sigma_{a|b}$\}}, where $a$ and $b$ denote the $\sigma$ used during training and inference, respectively, and $a,b \in \mathds{R}$. 

Another approach, PMRF~\cite{PMRF}, provides a single MMSE-optimal output via a two-step procedure (estimating posterior means $f(Y)$ given an extra MMSE estimator $f(.)$ and a degraded input $Y$, then rectifying the MMSE estimate with a conditional flow matching setup). 
Also, in \cite{PMRF}, the noise $\sigma_s\epsilon$ must be tuned carefully: if it is too small, the flow collapses; if it is too large, the MSE degrades. 
In microscopy data, the major source of noise, Poisson noise, is signal-dependent, and choosing $\sigma_s$ becomes difficult. 

In this work, we propose \textbf{\HazeMatching}, a novel framework that formulates dehazing as a guided conditional transport problem. 
We model stochastic interpolation between noisy, hazy (source) and non-hazy (target) image distributions, bridging observed haze and the underlying scene. 
We extend CFM with degradation-aware conditioning to guide the flow using hazy inputs. 
Our iterative design enables diverse, high-fidelity predictions and supports posterior sampling. 
We further show that \HazeMatching produces well-calibrated predictions.
\section{Methods}
\label{sec:methods}

\subsection{Problem definition}
\label{sec:problemdef}

The forward model that describes how hazy microscopy images are formed is described by $\mathbf{x}_M{=}H_{M}(s) + \eta(s)$, where $H_M$ is the unknown hazing function for a microscope $M$, $s$ is the true signal, \ie the imaged 3D sample, $\eta$ is a noise function that can depend on the true signal, and $\mathbf{x}_M$ is the hazy observed image.
Depending on $M$, the used microscope, the observed image $\mathbf{x}_M$ will contain different levels of signal-dependent haze.
For example, for a widefield microscope $M_0$ and a confocal microscope $M_1$, we can image two micrographs
\begin{equation} \label{eq:x0}
    \mathbf{x}_{M_0} = H_{M_0}(s) + \eta(s),
\end{equation}
\begin{equation} \label{eq:x1}
    \mathbf{x}_{M_1} = H_{M_1}(s) + \eta(s),
\end{equation}
using the same sample (imaged real-world object) $s$. 
In this way, we can acquire entire sets of images $X_0$ and $X_1$, respectively.
Let $S$ be a set of biological samples $S = \{ s^i|i = 1, 2, \dots, N \}, \quad N \in \mathbb{N}$.
An image $\mathbf{x}_M^i\in X_{j\in\{0,1\}}$ is drawn from a distribution of images $\mathbf{x}_{M_j}^i\sim p_{M_j}(\mathbf{x}|s^i)$, the distribution over all images microscope $j$ can produce for a given sample $s_j$.

With all this in mind, the task of \textit{image dehazing} is to predict
\begin{equation} \label{eq:dahazingtask}
    \hat{\mathbf{x}}_{M_1}^i = D(\mathbf{x}_{M_0}^i),
\end{equation}
with $D$ being the unknown function that solves the dehazing task, and $\hat{\mathbf{x}}_{M_1}^i$ reproducing the unknown $\mathbf{x}_{M_1}^i$ as accurately as possible.

The ambition of \HazeMatching is, given the training data $X_0$ and $X_1$, to learn an approximate $D$ that will allow microscopists to acquire images with more accessible, cheaper, and less photon-toxic widefield microscopes ($M_0$) while still recovering as much of the desired properties of data imaged with expensive confocal microscopes ($M_1$).

\subsection{Our approach}

Our goal is to learn a dehazing function $D$ that maps a hazy and noisy microscopy image to its corresponding non-hazy version, leveraging paired data acquired from two imaging modalities $M_0$ and $M_1$. 
To model this translation, we adopt the Conditional Flow Matching (CFM) framework, which constructs a continuous path from Gaussian noise to the clean image $\mathbf{x}_{M_1} \sim p_{M_1}(\mathbf{x}|s^i)$, while conditioning the learned velocity field on the degraded but observable input $\mathbf{x}_{M_0}\sim p_{M_0}(\mathbf{x}|s^i)$. 
This allows the model to learn a data-driven transport map that reconstructs high-fidelity structure informed by (conditioned on) low-quality observations.

\subsubsection{Marginal probability path for the dehazing task}
We have samples from three distributions to work with:
$(i)$~a \textit{base distribution} of pixel-wise noisy images $\mathbf{x}_0 \sim \mathcal{N}(0,I)$, which we denote as either $p_0(\mathbf{x})$ or $\mathcal{N}(0,I)$, 
$(ii)$~the \textit{source image distribution} of hazy observable images $\mathbf{x}_{M_0} \sim p_{M_0}(\mathbf{x}|s^i)$, and
$(iii)$~the non-hazy \textit{target image distribution} $\mathbf{x}_{M_1} \sim p_{M_1}(\mathbf{x}|s^i)$.

For supervised training, we draw each pair $(\mathbf{x}_{M_0}, \mathbf{x}_{M_1})$ corresponding to observations of underlying biological samples $s^i\in S$. 
The joint distribution over all variables, conditioned on the sample $s^i$, is
\begin{equation}
    p(\mathbf{x}_0, \mathbf{x}_{M_0}, \mathbf{x}_{M_1}|s^i) = p_0(\mathbf{x_0})\, p_{M_0}(\mathbf{x}^i_{M_0}|s^i)\, p_{M_1}(\mathbf{x}^i_{M_1}|s^i).
\label{eq:joint}
\end{equation}
Given this setup, we define a linear interpolation path $\mathbf{x}_t$ (as shown in equation \ref{eq:overview_xt}) between the noise sample $\mathbf{x}_0 \sim \mathcal{N}(0, I)$ and the clean target image $\mathbf{x}_{M_1}$ as
\begin{equation}
    \mathbf{x}_t = (1 - t)\mathbf{x}_0 + t \mathbf{x}_{M_1}, \quad \text{for } t \in [0, 1].
\label{eq:xt}
\end{equation}
Since $\mathbf{x}_0$ is independent Gaussian noise, the resulting marginal probability density path $p_t$ over the interpolant $\mathbf{x}_t$, conditioned on the clean image $\mathbf{x}_{M_1}$, is a Gaussian \cite{lipman} described as
\begin{equation}
\scalebox{0.94}{$
p_t(\mathbf{x}|\mathbf{x}_{M_1}) = \mathcal{N}(t \mathbf{x}_{M_1}, (1 - t)^2 I); \quad \mathbf{x}_t \sim p_t(\mathbf{x}|\mathbf{x}_{M_1}).
$}
\label{eq:pt}
\end{equation}

\subsubsection{Marginal velocity field for the dehazing task}
In the standard CFM framework (see Section \ref{sec:cfm_overview}), the conditional velocity field is defined as $v(t,\mathbf{x}_t | \mathbf{x}_{M_1})$, where $\mathbf{x}_{M_1}$ is the clean image we have available during training, making the marginal velocity field take the form
\begin{equation} \label{eq:cfm_classic} 
    v(t,\mathbf{x}_t) = \int v(t,\mathbf{x}_t \mid \mathbf{x}_{M_1}) \, p_{M_1}(\mathbf{x}_{M_1} | \mathbf{x}_t) \, d\mathbf{x}_{M_1}.
\end{equation}
However, in our framework, the conditional velocity field explicitly depends not only on the interpolant $\mathbf{x}_t$ but also on the degraded observation $\mathbf{x}_{M_0}$ and takes the form $v(t,\mathbf{x}_t,\mathbf{x}_{M_0} | \mathbf{x}_{M_1})$. 
Therefore, we extend the formulation in \cite{cfm_guide} and Equation~\ref{eq:cfm_classic}, which now becomes
\begin{equation}
    \begin{aligned}
    v(t,\mathbf{x}_t,\mathbf{x}_{M_0}) 
    &= \int v(t,\mathbf{x}_t \mid \mathbf{x}_{M_1}, \mathbf{x}_{M_0}) \\
    &\quad p_{M_1}(\mathbf{x}_{M_1} \mid \mathbf{x}_t,\mathbf{x}_{M_0}) \,
    d\mathbf{x}_{M_1}.
    \end{aligned}
\end{equation}

Hence, to solve our dehazing task through $v(t,\mathbf{x}_t,\mathbf{x}_{M_0})$, we adapt the standard CFM framework (see Equation~\ref{eq:cfm}) to become a \textit{guided} CFM framework that we will use to train a neural network $v_\theta$ that minimizes the objective
\begin{equation}
    \min_{\theta} \mathbb{E}_{(\mathbf{x}_0,\mathbf{x}_{M_0},\mathbf{x}_{M_1}), \mathbf{x}_t}\ ||v_\theta(t,\mathbf{x}_t, \mathbf{x}_{M_0}) - (\mathbf{x}_{M_1} - \mathbf{x}_0)||^2,
    \label{eq:dehazingloss}
\end{equation}
where $(\mathbf{x}_0,\mathbf{x}_{M_0},\mathbf{x}_{M_1}) \sim p(\mathbf{x}_0,\mathbf{x}_{M_0},\mathbf{x}_{M_1}|s^i)$, $\mathbf{x}_t \sim p_t(\mathbf{x}|\mathbf{x}_{M_1})$, and $\theta$ are the learnable parameters of the network. 

At inference time, we start with a given hazy observation $\mathbf{x}_{M_0} \sim p_{M_0}(\mathbf{x})$, sample an independent $\mathbf{x}_0 \sim \mathcal{N}(0,I)$ and concatenate $\mathbf{x}_0,\mathbf{x}_{M_0}$ along the channel dimension.
We then use the learned velocity field $v_{\theta}(t,\mathbf{x}_0, \mathbf{x}_{M_0})$ to predict the velocity ($d\mathbf{x}_t/dt$) iteratively for some predefined steps $t \in [0,1]$ to effectively compute
\begin{equation}
  \frac{d\mathbf{x}_t}{dt} =  v_\theta(t,\mathbf{x}_0, \mathbf{x}_{M_0}).
\end{equation}
We integrate the predicted velocities using an Euler ODE solver to predict the non-hazy image at the final integration step at $t=1$. 
Note that we can generate multiple posterior samples by repeating this process, with the stochasticity coming from sampling our base distribution $\mathbf{x}_0 \sim \mathcal{N}(0,I)$.
Further details for training and prediction procedures are shown in Algorithms~\ref{alg:Training} and~\ref{alg:Inference}, respectively.
See also Figure~\ref{fig:overview} and Supplementary Section~\ref{sup:overview}.

{\scriptsize
\begin{algorithm}
\caption{\HazeMatching: Training}\label{alg:Training}
\begin{algorithmic}[1]
\Require velocity model $v_\theta$; 
learning rate $\gamma$;
steps $T$
\Repeat
\State $s \sim \mathcal{S}$; 
$t \sim \mathcal{U}([i/T]_{i=0}^{T})$; 
$\mathbf{x}_0 \sim \mathcal{N}(0,I)$;
\State $\mathbf{x}_{M_0} \sim p_{M_0}(\mathbf{x}|s)$; $\mathbf{x}_{M_1} \sim p_{M_1}(\mathbf{x}|s)$
\State $\mathbf{x}_t=(1-t)\mathbf{x}_0+t\mathbf{x}_{M_1}$
\State $\mathcal{L}=||v_\theta(t,\mathbf{x}_t,\mathbf{x}_{M_0})-(\mathbf{x}_{M_1}-\mathbf{x}_0)||^2$
\State $\theta \leftarrow \theta-\gamma\nabla_\theta\mathcal{L}$
\Until{converged}
\State \Return $v_\theta$
\end{algorithmic}
\end{algorithm}
}


{\scriptsize
\begin{algorithm}
\caption{\HazeMatching Inference Procedure}\label{alg:Inference}
\begin{algorithmic}[1]
\Require velocity model $v_\theta$, steps $T$, input $\mathbf{x}_{M_0}$ to condition on
\State $\delta = 1/T$
\State $\mathbf{x}_0 \sim \mathcal{N}(0,I)$
\For{$t=1$ to $T$}
\State $\mathbf{x}_t = \mathbf{x}_{t-1} + \delta \cdot v_\theta(\delta\cdot(t-1),\mathbf{x}_t,\mathbf{x}_{M_0})$
\EndFor
\State \Return $\mathbf{x}_T$ 
\textcolor{gray}{\Comment{called $\hat{\mathbf{x}}_{M_1}$ in Section~\ref{sec:methods}}}
\end{algorithmic}
\end{algorithm}
}

\subsection{Evaluating the calibration of trained models}
A network is considered well-calibrated when the predicted uncertainties align with the true errors of the predictions. 
Here, we propose to assess whether the pixel-wise variability of multiple predictions (posterior samples) grows in accordance with the true error, which we assess \wrt a set of high quality ground-truth images.
The calibration procedure follows~\cite{Microsplit}, clustering pixels into bins and computing RMSE (root mean squared error) and RMV (root mean variance) per bin. 
RMV is then linearly scaled to align with RMSE. Finally, we get an RMSE vs. RMV plot.
A plot closer to $y=x$ indicates better alignment between true error (RMSE) and estimated error (RMV), indicating a better calibration.
For further details, please refer to Supplementary Section~\ref{sup:calibration}. 
\section{Experiments}
\label{sec:experiments}

In total, we evaluate our method on five datasets.
Below, we describe the three datasets shown in the main text. The additional two datasets (synthetic neurons and semi-synthetic Microtubules) can be found in Supplementary Sections~\ref{sup:microtubule} and~\ref{sup:neuron}.

\textbf{Zebrafish data:} 
We use zebrafish retina images taken with a confocal microscope and made available with~\cite{care}. 
We used \texttt{microsim}~\cite{microsim} to simulate additional and realistically looking microscopy haze. 
The simulation pipeline, along with other implementation details, is provided in Supplementary Section~\ref{sup:microsim}. 
This dataset shows fluorescently labeled nuclei, membranes, and nuclear envelopes. 

We train and evaluate all structures in a single joint model. 
The dataset consists of a total of $15$ training images, of which we reserve $3$ images for validation and test purposes.
All images are of size $1024\times1024$. 
More details can be found in Supplementary Section~\ref{sup:zebra}.

\textbf{Organoids1 and Organoids2 data}: 
These are two real microscopy datasets we have acquired for the purpose of this work. 
Organoids1 data were imaged on a spinning-disk confocal microscope, while the Organoids2 data were obtained on a point-scanning confocal setup.
Both datasets consist of pairs of hazy/non-hazy images showing the same sample ROI.
While the non-hazy data were acquired by optimizing the respective microscope for the best image quality, the hazy counterparts were acquired by removing the spinning-disk and opening the pinhole of the confocal microscope, respectively.
Both datasets consist of $15$ training images each, of which we have set aside $3$ test and $2$ validation images.
All images are of size $1024\times1024$. 
More details can be found in Supplementary Sections~\ref{sup:organoids1} and~\ref{sup:organoids2}. 

\figPlotsMain
\figCalibMain
\figQualitative

 \subsection{Training Details}
The backbone we employ in our experiments is a \UNet~\citep{diffusinOpenAI}, implemented in PyTorch~\citep{pytorch} and executed using NVIDIA V100 GPUs. 
In all training runs of \HazeMatching we used a patch size of either $64 \times 64$ or $128\times128$ and have set $T{=}20$. Details of patch sizes used during training, inference, and evaluations are provided in Supplementary Section~\ref{sup:train_setup} and~\ref{sup:eval_setup}.
The \texttt{torchCFM}~\citep{torchCFM} library is used for interpolant computation and ODE integration. 
Additional training details are provided in Supplementary Section~\ref{sup:dset_train_setup}.

\textbf{Evaluation:} 
Predictions on full-sized images use inner tiling~\cite{usplit} with the tile-size coinciding with the training patch size and a 50\% overlap.

Distortion metrics (PSNR~\citep{care} and MicroMS-SSIM~\citep{microssim}) are computed on full tiled predictions, while perceptual metrics (via LPIPS~\citep{lpips} and FID~\citep{fid}) are evaluated on non-overlapping patches tiling the full sized input images. 
See also Supplementary Section~\ref{sup:dset_train_setup} for further details.

\textbf{Baselines:}\label{subsec:baselines} We compare our results against classical Richardson–Lucy (RL) deconvolution~\citep{RL}, deterministic point-predicting models (\UNet~\citep{unet}, MIMO-Unet~\cite{mimo}, MPRNet~\cite{mprnet}, RCAN~\cite{rcan}, ESRGAN~\cite{esrgan}, InDI$_{1}$ in single step prediction mode~\citep{indi}, a transformer based model Restormer~\cite{Restormer}, and a variational \HVAE~\citep{hvae} baseline. 

Additionally, we also compare against iterative methods, \ie InDI$_{20}$ (20 step prediction mode) and multiple configurations of SIFM~\cite{sifm} that are capable of generating diverse predictions. 
For \textit{Organoids1} and \textit{Organoids2}, we additionally compare against the proprietary software \textit{Elements} from \cite{elements} and \textit{Lightning} from \cite{lightning}, respectively. 
See Supplementary Section~\ref{sup:baseline} for more details.
\section{Results}
\label{sec:results}
The main thrust of our work is to propose a method that strikes a good balance between data fidelity and perceptual distortion, \ie between being true to the lower-quality measurement and still generating predictions that are reasonably close to a target distribution of high-quality images.

Recent theoretical work~\cite{theory} establishes that the Flow Matching loss provides a deterministic upper bound on the Kullback–Leibler (KL) divergence between the true and predicted data distributions. 
This implies that minimizing the flow-matching loss leads to better perceptual performance. 
On the other hand, since we condition our flow matching approach on a real microscopy image and can compute an MMSE estimate by averaging multiple posterior samples, \HazeMatching also achieves good fidelity scores.

\textbf{Perception–distortion trade-off:} 
In the context of microscopy data analysis, it is of central importance that predicted structures are not hallucinated to satisfy perceptual consistency with a given target distribution, but instead coincide with the biological structures that the original sample we have put into the microscope indeed contains. 
Fidelity measures, such as PSNR, therefore prefer blurry predictions over realistic-looking but misplaced image details.
Unfortunately, methods that lead to very good PSNR values generate overly smooth predictions, motivating our work on \HazeMatching, where we do aim for high PSNR values, \ie excellent data fidelity. 
However, among all predictions of comparable fidelity, we want to choose the one with the best perceptual quality, \ie results that show structures as clearly and unambiguously as possible.

To this end, we plot LPIPS vs.\ PSNR and FID vs.\ PSNR in Figure~\ref{fig:plot_main}, jointly assessing content fidelity (PSNR) and perceptual quality (LPIPS, FID). 
We prefer models that have high PSNR and low LPIPS/FID, which, in our plot, are located closer to the origin. 
\HazeMatching consistently achieves a favorable balance, improving on deterministic baselines in perceptual quality while remaining competitive with the best fidelity-optimized generative posterior baselines. 
Posterior models (blue markers) generate diverse and realistic looking outputs; point predictors (red) yield single predictions. 
For posterior models, we report the PSNR of an approximate MMSE (average of $50$ samples) and the average of the per-sample LPIPS and FID scores.
More details can be found in Supplementary Sections~\ref{sup:rl_iters} and~\ref{sup:sifm_configs}.

Choosing how we want to trade perceptual quality against data fidelity within our proposed framework is possible by adjusting the number of integration steps $T$. 
In Supplementary Section~\ref{sup:pdt_control}, we show this effect.

\textbf{Calibration analysis:} We assess how \HazeMatching deals with data uncertainty by conducting a calibration analysis on predicted variance maps. 
The calibration procedure we employ is taken and adapted from~\cite{Microsplit}.
Accordingly, in Figure~\ref{fig:calib_main} we plot RMSE vs.\ RMV (root mean variance) for the three main datasets ($y{=}x$ shown as a dashed line).
Note that our trained models are well calibrated, suggesting that we can use the variability in posterior samples as a surrogate for the true prediction error. 
To better demonstrate sample efficiency for calibration, we measured the calibration slope $\alpha$ and MMSE PSNR as a function of posterior sample count $k$ for all the datasets (see Supplementary Section~\ref{sup:calibration_k}).

\textbf{Qualitative comparisons:} In Figure~\ref{fig:qualitative}, we show representative results for the three main datasets, with PSNR and LPIPS reported on a highlighted crop. 
Red and blue outlines indicate whether a model is a point-predictor or a posterior model, respectively. 
Deterministic models tend to predict overly smooth results, while posterior models, especially \HazeMatching, produce sharper, more realistic looking outputs. 
Note that \HazeMatching is consistently located in the bottom left corner of the evaluation plots, thus striking a consistently good balance between data fidelity and perceptual quality.
See also Supplementary Section~\ref{sup:qualitative} for more qualitative results.

\textbf{Additional Results:} As mentioned before, in Supplementary Section~\ref{sup:results}, we present extended evaluations on the two additional synthetic datasets—\textit{Neuron}~(\ref{sup:neuron_results}) and \textit{Microtubule}~(\ref{sup:microtubule_results}).
In Supplementary Section~\ref{sup:pdt_microssim}, we highlight the trade-offs between fine structure (via MicroMS-SSIM) and perceptual quality (LPIPS and FID).
In Supplementary Section~\ref{sup:full_tables}, we provide full quantitative tables for PSNR, LPIPS, FID, MicroMS-SSIM, MS-SSIM~\cite{msssim}, and two alternative perceptual quality scores, FSIM~\cite{fsim} and GMSD~\cite{gmsd}.
In Supplementary Section~\ref{sup:time}, we present an inference runtime comparison between \HazeMatching and the baseline methods. 
We also show an ablation for our choice of the concatenation-based conditioning mechanism in Supplementary Section~\ref{sup:condition_ablation}.
Finally, in Supplementary Section~\ref{sup:qualitative}, we show additional qualitative results.

\textbf{Practical usage guideline for posterior sampling:}
To benefit from posterior samples in \HazeMatching, we recommend evaluating both the MMSE estimate (obtained by averaging multiple samples) and the calibrated uncertainty map (posterior variance map scaled by the learned calibration factor $\alpha$). The uncertainty map can be used to identify reliable regions across individual posterior samples, enabling the rejection of uncertain structures. Well-calibrated uncertainty, therefore, provides a principled way for biologists to assess the trustworthiness of predicted structures. 
In regions of consistently high uncertainty, we recommend either rejecting this part of the data for downstream processing or, if strictly needed, relying on the MMSE estimate only.

\textbf{Limitations:}
Since \HazeMatching requires paired training data, its application is limited to use-cases where such data exists.
Additionally, to generate multiple posterior samples, \HazeMatching requires multiple predictions, each of which requires multiple forward passes.
This increases the inference time for MMSE predictions, making this approach computationally more intensive than some of its competitors and, today, likely not as easily applicable to online live-cell imaging applications as other approaches. 
\section{Discussion and Conclusion}
\label{sec:discussion}
\HazeMatching addresses a key challenge in microscopy, namely the prediction of dehazed images.
Unlike deterministic point-predicting models that lead to high fidelity but overly smooth predictions~\cite{floriansReviewWithSulianaEtAl}, \HazeMatching also preserves perceptual quality to the degree possible.
Hence, \HazeMatching balances distortions with perceptual performance better and more consistently than other methods (see Figure~\ref{fig:plot_main}).
In particular, our experiments with real and simulated data show that \HazeMatching often achieves a favorable perception–distortion trade-off that is consistently good across all conducted experiments. 

Our results show that \HazeMatching consistently achieves the best data fidelity among all tested posterior models (shown in blue in Figure~\ref{fig:qualitative}) and closely matches the performance of even the best deterministic baselines (\eg, the MIMO-UNet), while leading to considerably improved perceptual quality predictions.

In addition to these properties, and unlike point predictors, \HazeMatching learns an implicit posterior of possible solutions from which we can generate multiple samples conditioned on the same hazy input.
The data fidelity scores can be further improved by averaging multiple such samples, leading to a smoother yet higher fidelity MMSE prediction.
Overall, \HazeMatching proposes an iterative image dehazing method that extends the CFM framework and provides access to multiple samples that deviate from each other in accordance with the uncertainty in the data (as demonstrated by our calibration experiments).

It will be interesting to explore whether the advantages of \HazeMatching extend beyond image dehazing to other ill-posed inverse problems in microscopy, with computational super-resolution being an especially promising candidate. 
A broader adoption of our approach will ultimately depend on the value that the improved reconstructions and posterior samples provide to downstream analysis pipelines. 
If these uncertainty-aware predictions meaningfully enhance tasks such as segmentation, tracking, or quantitative feature extraction, the additional computational cost inherent to iterative generative methods may be well justified. 
Future work will therefore examine the integration of \HazeMatching into larger microscopy workflows to assess these practical benefits.
\subsection*{Acknowledgements}
\label{sec:acknowledgements}

We thank Francesca Casagrande, Alessandra Fasciani, Jacopo Zasso, Ilaria Laface, Dario Ricca, and Eugenia Cammarota for their valuable contributions to this work. We also acknowledge the support of Talley Lambert (Harvard Medical School) and Vera Galinova in setting up the \texttt{microsim} pipeline and some baselines, as well as the entire Jug Group for insightful discussions.
This work was supported by the European Union through the Horizon Europe program (IMAGINE project, grant agreement 101094250-IMAGINE and AI4Life project, grant agreement 101057970-AI4LIFE) and the generous core funding of Human Technopole.

{
    \small
    \bibliographystyle{ieeenat_fullname}
    \bibliography{main}

@String(CVPR= {IEEE Conf. Comput. Vis. Pattern Recog.})

@String(ICCV= {Int. Conf. Comput. Vis.})

@String(ECCV= {Eur. Conf. Comput. Vis.})

@String(CVPR  = {CVPR})

@String(ICCV  = {ICCV})

@String(ECCV  = {ECCV})

@article{RL_bad,
author = {Liu, Yiming and Panezai, Spozmai and Wang, Yutong and Stallinga, Sjoerd},
year = {2025},
month = {01},
title = {Noise amplification and ill-convergence of Richardson-Lucy deconvolution},
volume = {16},
journal = {Nature Communications},
doi = {10.1038/s41467-025-56241-x}
}

@unknown{pn2v,
author = {Krull, Alexander and Vicar, Tomas and Jug, Florian},
year = {2019},
month = {06},
title = {Probabilistic Noise2Void: Unsupervised Content-Aware Denoising},
doi = {10.48550/arXiv.1906.00651}
}

@inproceedings{
hdn,
title={Interpretable Unsupervised Diversity Denoising and Artefact Removal},
author={Mangal Prakash and Mauricio Delbracio and Peyman Milanfar and Florian Jug},
booktitle={International Conference on Learning Representations},
year={2022},
url={https://openreview.net/forum?id=DfMqlB0PXjM}
}

@inproceedings{DDPM,
 author = {Ho, Jonathan and Jain, Ajay and Abbeel, Pieter},
 booktitle = {Advances in Neural Information Processing Systems},
 editor = {H. Larochelle and M. Ranzato and R. Hadsell and M.F. Balcan and H. Lin},
 pages = {6840--6851},
 publisher = {Curran Associates, Inc.},
 title = {Denoising Diffusion Probabilistic Models},
 url = {https://proceedings.neurips.cc/paper_files/paper/2020/file/4c5bcfec8584af0d967f1ab10179ca4b-Paper.pdf},
 volume = {33},
 year = {2020}
}

@inproceedings{
DDIM,
title={Denoising Diffusion Implicit Models},
author={Jiaming Song and Chenlin Meng and Stefano Ermon},
booktitle={International Conference on Learning Representations},
year={2021},
url={https://openreview.net/forum?id=St1giarCHLP}
}

@misc{DDRM,
      title={Denoising Diffusion Restoration Models}, 
      author={Bahjat Kawar and Michael Elad and Stefano Ermon and Jiaming Song},
      year={2022},
      eprint={2201.11793},
      archivePrefix={arXiv},
      primaryClass={eess.IV},
      url={https://arxiv.org/abs/2201.11793}, 
}

@misc{cold_diff,
      title={Cold Diffusion: Inverting Arbitrary Image Transforms Without Noise}, 
      author={Arpit Bansal and Eitan Borgnia and Hong-Min Chu and Jie S. Li and Hamid Kazemi and Furong Huang and Micah Goldblum and Jonas Geiping and Tom Goldstein},
      year={2022},
      eprint={2208.09392},
      archivePrefix={arXiv},
      primaryClass={cs.CV},
      url={https://arxiv.org/abs/2208.09392}, 
}

@article {CARE,
	author = {Weigert, Martin and Schmidt, Uwe and Boothe, Tobias and M{\"u}ller, Andreas and Dibrov, Alexandr and Jain, Akanksha and Wilhelm, Benjamin and Schmidt, Deborah and Broaddus, Coleman and Culley, Si{\^a}n and Rocha-Martins, Mauricio and Segovia-Miranda, Fabi{\'a}n and Norden, Caren and Henriques, Ricardo and Zerial, Marino and Solimena, Michele and Rink, Jochen and Tomancak, Pavel and Royer, Loic and Jug, Florian and Myers, Eugene W.},
	title = {Content-Aware Image Restoration: Pushing the Limits of Fluorescence Microscopy},
	elocation-id = {236463},
	year = {2018},
	doi = {10.1101/236463},
	publisher = {Cold Spring Harbor Laboratory},
	URL = {https://www.biorxiv.org/content/early/2018/07/03/236463},
	eprint = {https://www.biorxiv.org/content/early/2018/07/03/236463.full.pdf},
	journal = {bioRxiv}
}

@inproceedings{sifm,
title = {Stochastic Interpolants with Data-Dependent Couplings},
author = {Albergo, Michael Samuel and Goldstein, Mark and Boffi, Nicholas Matthew and Ranganath, Rajesh and Vanden-Eijnden, Eric},
booktitle = {Proceedings of the 41st International Conference on Machine Learning},
pages = {921--937},
year = {2024},
editor = {Salakhutdinov, Ruslan and Kolter, Zico and Heller, Katherine and Weller, Adrian and Oliver, Nuria and Scarlett, Jonathan and Berkenkamp, Felix},
volume = {235},
series = {Proceedings of Machine Learning Research},
month = {21--27 Jul},
publisher = {PMLR},
url = {https://proceedings.mlr.press/v235/albergo24a.html}
}

@article{learned_conditional,
  title={Minimizing Trajectory Curvature of ODE-based Generative Models},
  author={Lee, Sangyun and Kim, Beomsu and Ye, Jong Chul},
  journal={arXiv preprint arXiv:2301.12003},
  year={2023}
}

@inproceedings{PDT,
  author       = {Yochai Blau and
                  Tomer Michaeli},
  title        = {The Perception-Distortion Tradeoff},
  booktitle    = {2018 {IEEE} Conference on Computer Vision and Pattern Recognition,
                  {CVPR} 2018, Salt Lake City, UT, USA, June 18-22, 2018},
  pages        = {6228--6237},
  publisher    = {Computer Vision Foundation / {IEEE} Computer Society},
  year         = {2018},
  url          = {http://openaccess.thecvf.com/content\_cvpr\_2018/html/Blau\_The\_Perception-Distortion\_Tradeoff\_CVPR\_2018\_paper.html},
  doi          = {10.1109/CVPR.2018.00652},
  bibsource    = {dblp computer science bibliography, https://dblp.org}
}

@article{floriansReviewWithSulianaEtAl,
author = {Shroff, Hari and Testa, Ilaria and Jug, Florian and Manley, Suliana},
year = {2024},
month = {02},
title = {Live-cell imaging powered by computation},
volume = {25},
journal = {Nature Reviews Molecular Cell Biology},
doi = {10.1038/s41580-024-00702-6}
}

@InProceedings{usplit,
  author    = {Ashesh, Ashesh and Krull, Alexander and Di Sante, Moises and Pasqualini, Francesco and Jug, Florian},
  title     = {uSplit: Image Decomposition for Fluorescence Microscopy},
  booktitle = {Proceedings of the IEEE/CVF International Conference on Computer Vision (ICCV)},
  month     = {October},
  year      = {2023},
  pages     = {21219--21229},
  url       = {https://openaccess.thecvf.com/content/ICCV2023/html/Ashesh_uSplit_Image_Decomposition_for_Fluorescence_Microscopy_ICCV_2023_paper.html}
}

@article {Microsplit,
	author = {Ashesh, Ashesh and Carrara, Federico and Zubarev, Igor and Galinova, Vera and Croft, Melisande and Pezzotti, Melissa and Gong, Daozheng and Casagrande, Francesca and Colombo, Elisa and Giussani, Stefania and Restelli, Elena and Cammarota, Eugenia and Battagliotti, Juan Manuel and Klena, Nikolai and Di Sante, Moises and Pigino, Gaia and Taverna, Elena and Harschnitz, Oliver and Maghelli, Nicola and Scherer, Norbert and Dalle Nogare, Damian Edward and Deschamps, Joran and Pasqualini, Francesco and Jug, Florian},
	title = {MicroSplit: Semantic Unmixing of Fluorescent Microscopy Data},
	elocation-id = {2025.02.10.637323},
	year = {2025},
	doi = {10.1101/2025.02.10.637323},
	publisher = {Cold Spring Harbor Laboratory},
	URL = {https://www.biorxiv.org/content/early/2025/02/11/2025.02.10.637323},
	eprint = {https://www.biorxiv.org/content/early/2025/02/11/2025.02.10.637323.full.pdf},
	journal = {bioRxiv}
}

@inproceedings{pytorch,
  title={Automatic differentiation in PyTorch},
  author={Paszke, Adam and Gross, Sam and Chintala, Soumith and Chanan, Gregory and Yang, Edward and DeVito, Zachary and Lin, Zeming and Desmaison, Alban and Antiga, Luca and Lerer, Adam},
  booktitle={NIPS-W},
  year={2017}
}

@misc{torchCFM,
  author       = {Alexander Y. Tong},
  title        = {TorchCFM: A Conditional Flow Matching Library},
  year         = {2024},
  publisher    = {GitHub},
  url          = {https://github.com/atong01/conditional-flow-matching},
  note         = {Accessed: 2025-05-04}
}

@ARTICLE{RL,
       author = {{Lucy}, L.~B.},
        title = "{An iterative technique for the rectification of observed distributions}",
      journal = {The Astronomical Journal},
         year = 1974,
       volume = {79},
        pages = {745},
          doi = {10.1086/111605},
       adsurl = {https://ui.adsabs.harvard.edu/abs/1974AJ.....79..745L}
}

@inproceedings{unet,
  author       = {Olaf Ronneberger and
                  Philipp Fischer and
                  Thomas Brox},
  editor       = {Nassir Navab and
                  Joachim Hornegger and
                  William M. Wells III and
                  Alejandro F. Frangi},
  title        = {U-Net: Convolutional Networks for Biomedical Image Segmentation},
  booktitle    = {Medical Image Computing and Computer-Assisted Intervention - {MICCAI}
                  2015 - 18th International Conference Munich, Germany, October 5 -
                  9, 2015, Proceedings, Part {III}},
  series       = {Lecture Notes in Computer Science},
  volume       = {9351},
  pages        = {234--241},
  publisher    = {Springer},
  year         = {2015},
  url          = {https://doi.org/10.1007/978-3-319-24574-4\_28},
  doi          = {10.1007/978-3-319-24574-4\_28},
  bibsource    = {dblp computer science bibliography, https://dblp.org}
}

@article{indi,
  author       = {Mauricio Delbracio and
                  Peyman Milanfar},
  title        = {Inversion by Direct Iteration: An Alternative to Denoising Diffusion
                  for Image Restoration},
  journal      = {Trans. Mach. Learn. Res.},
  volume       = {2023},
  year         = {2023},
  url          = {https://openreview.net/forum?id=VmyFF5lL3F},
  bibsource    = {dblp computer science bibliography, https://dblp.org}
}

@inproceedings{hvae,
  author       = {Casper Kaae S{\o}nderby and
                  Tapani Raiko and
                  Lars Maal{\o}e and
                  S{\o}ren Kaae S{\o}nderby and
                  Ole Winther},
  editor       = {Daniel D. Lee and
                  Masashi Sugiyama and
                  Ulrike von Luxburg and
                  Isabelle Guyon and
                  Roman Garnett},
  title        = {Ladder Variational Autoencoders},
  booktitle    = {Advances in Neural Information Processing Systems 29: Annual Conference
                  on Neural Information Processing Systems 2016, December 5-10, 2016,
                  Barcelona, Spain},
  pages        = {3738--3746},
  year         = {2016},
  url          = {https://proceedings.neurips.cc/paper/2016/hash/6ae07dcb33ec3b7c814df797cbda0f87-Abstract.html},
  bibsource    = {dblp computer science bibliography, https://dblp.org}
}

@misc{cfm_guide,
      title={Flow Matching Guide and Code}, 
      author={Yaron Lipman and Marton Havasi and Peter Holderrieth and Neta Shaul and Matt Le and Brian Karrer and Ricky T. Q. Chen and David Lopez-Paz and Heli Ben-Hamu and Itai Gat},
      year={2024},
      eprint={2412.06264},
      archivePrefix={arXiv},
      primaryClass={cs.LG},
      url={https://arxiv.org/abs/2412.06264}, 
}

@inproceedings{
lipman,
title={Flow Matching for Generative Modeling},
author={Yaron Lipman and Ricky T. Q. Chen and Heli Ben-Hamu and Maximilian Nickel and Matthew Le},
booktitle={The Eleventh International Conference on Learning Representations },
year={2023},
url={https://openreview.net/forum?id=PqvMRDCJT9t}
}

@misc{microsim,
  author       = {Talley Lambert},
  title        = {{microsim}: A physics-based confocal and widefield microscopy simulator},
  year         = {2022},
}

@misc{Neuron,
  author       = {{Allen Institute for Brain Science}},
  title        = {{Allen Cell Types Database}},
  url          = {https://celltypes.brain-map.org/},
  note         = {Accessed: 2025-05-04}
}

@misc{Tubulin,
  author       = {{Allen Institute for Cell Science}},
  title        = {{Allen Cell Explorer}},
  url          = {https://www.allencell.org/},
  note         = {Accessed: 2025-05-04}
}

@inproceedings{n2v,
  title={Noise2void-learning denoising from single noisy images},
  author={Krull, Alexander and Buchholz, Tim-Oliver and Jug, Florian},
  booktitle={Proceedings of the IEEE Conference on Computer Vision and Pattern Recognition},
  pages={2129--2137},
  year={2019}
}

@misc{diffusinOpenAI,
      title={Diffusion Models Beat GANs on Image Synthesis}, 
      author={Prafulla Dhariwal and Alex Nichol},
      year={2021},
      eprint={2105.05233},
      archivePrefix={arXiv},
      primaryClass={cs.LG},
      url={https://arxiv.org/abs/2105.05233}, 
}

@misc{microssim,
      title={MicroSSIM: Improved Structural Similarity for Comparing Microscopy Data}, 
      author={Ashesh Ashesh and Joran Deschamps and Florian Jug},
      year={2024},
      eprint={2408.08747},
      archivePrefix={arXiv},
      primaryClass={eess.IV},
      url={https://arxiv.org/abs/2408.08747}, 
}

@inproceedings{lpips,
  title={The Unreasonable Effectiveness of Deep Features as a Perceptual Metric},
  author={Zhang, Richard and Isola, Phillip and Efros, Alexei A and Shechtman, Eli and Wang, Oliver},
  booktitle={Proceedings of the IEEE Conference on Computer Vision and Pattern Recognition (CVPR)},
  year={2018},
  pages={586--595}
}

@inproceedings{fid,
  title={GANs Trained by a Two Time-Scale Update Rule Converge to a Local Nash Equilibrium},
  author={Heusel, Martin and Ramsauer, Hubert and Unterthiner, Thomas and Nessler, Bernhard and Hochreiter, Sepp},
  booktitle={Advances in Neural Information Processing Systems (NeurIPS)},
  year={2017},
  pages={6626--6637}
}

@article{fsim,
  title={FSIM: A Feature Similarity Index for Image Quality Assessment},
  author={Zhang, Lin and Zhang, Lei and Mou, Xuanqin and Zhang, David},
  journal={IEEE Transactions on Image Processing},
  volume={20},
  number={8},
  pages={2378--2386},
  year={2011},
  publisher={IEEE}
}

@article{gmsd,
  title={Gradient Magnitude Similarity Deviation: A Highly Efficient Perceptual Image Quality Index},
  author={Xue, Wufeng and Zhang, Lei and Mou, Xuanqin and Bovik, Alan C.},
  journal={IEEE Transactions on Image Processing},
  volume={23},
  number={2},
  pages={684--695},
  year={2014},
  publisher={IEEE}
}

@article{msssim,
  title={Image Quality Assessment: From Error Visibility to Structural Similarity},
  author={Wang, Zhou and Bovik, Alan C. and Sheikh, Hamid R. and Simoncelli, Eero P.},
  journal={IEEE Transactions on Image Processing},
  volume={13},
  number={4},
  pages={600--612},
  year={2004},
  publisher={IEEE}
}

@misc{elements,
  author       = {{Nikon Instruments Inc.}},
  title        = {{NIS-Elements Imaging Software}}
}

@misc{lightning,
  author       = {{Leica Microsystems}},
  title        = {{Obtain Maximum Information from Your Specimen with LIGHTNING}}
}

@inproceedings{calibration,
  title     = {On Calibration of Modern Neural Networks},
  author    = {Chuan Guo and Geoff Pleiss and Yu Sun and Kilian Q. Weinberger},
  booktitle = {Proceedings of the 34th International Conference on Machine Learning},
  pages     = {1321--1330},
  year      = {2017},
  volume    = {70},
  series    = {Proceedings of Machine Learning Research},
  publisher = {PMLR},
  url       = {https://proceedings.mlr.press/v70/guo17a.html}
}

@article{dl4miasurvey,
  title={A survey on applications of deep learning in microscopy image analysis},
  author={Liu, Zhichao and Jin, Luhong and Chen, Jincheng and Fang, Qiuyu and Ablameyko, Sergey and Yin, Zhaozheng and Xu, Yingke},
  journal={Computers in Biology and Medicine},
  volume={134},
  pages={104523},
  year={2021},
  publisher={Elsevier},
  doi={10.1016/j.compbiomed.2021.104523}
}

@article{con_diff,
title={Image super-resolution via iterative refinement},
author={Saharia, Chitwan and Ho, Jonathan and Chan, William and Salimans, Tim and Fleet, David J and Norouzi, Mohammad},
journal={arXiv:2104.07636},
year={2021}
}

@inproceedings{
    PMRF,
    title={Posterior-Mean Rectified Flow: Towards Minimum {MSE} Photo-Realistic Image Restoration},
    author={Guy Ohayon and Tomer Michaeli and Michael Elad},
    booktitle={The Thirteenth International Conference on Learning Representations},
    year={2025},
    url={https://openreview.net/forum?id=hPOt3yUXii}
}

@misc{theory,
      title={On Flow Matching KL Divergence}, 
      author={Maojiang Su and Jerry Yao-Chieh Hu and Sophia Pi and Han Liu},
      year={2025},
      eprint={2511.05480},
      archivePrefix={arXiv},
      primaryClass={cs.LG},
      url={https://arxiv.org/abs/2511.05480}, 
}

@inproceedings{esrgan,
  title     = {ESRGAN: Enhanced Super-Resolution Generative Adversarial Networks},
  author    = {Wang, Xintao and Yu, Ke and Wu, Shixiang and Gu, Jinjin and Liu, Yihao and Dong, Chao and Qiao, Yu and Loy, Chen Change},
  booktitle = {Proc. ECCV Workshops},
  year      = {2018},
  pages     = {63--79},
  url       = {https://doi.org/10.1007/978-3-030-11021-5_5}
}

@inproceedings{rcan,
  title     = {Image Super-Resolution Using Very Deep Residual Channel Attention Networks},
  author    = {Yulun Zhang and Kunpeng Li and Kai Li and Lichen Wang and Bineng Zhong and Yun Fu},
  booktitle = {Proceedings of the European Conference on Computer Vision (ECCV)},
  year      = {2018},
  pages     = {286--301}
}

@INPROCEEDINGS{mimo,
  author={Cho, Sung-Jin and Ji, Seo-Won and Hong, Jun-Pyo and Jung, Seung-Won and Ko, Sung-Jea},
  booktitle={2021 IEEE/CVF International Conference on Computer Vision (ICCV)}, 
  title={Rethinking Coarse-to-Fine Approach in Single Image Deblurring}, 
  year={2021},
  volume={},
  number={},
  pages={4621-4630},
  doi={10.1109/ICCV48922.2021.00460}}

@INPROCEEDINGS{mprnet,
  author={Zamir, Syed Waqas and Arora, Aditya and Khan, Salman and Hayat, Munawar and Khan, Fahad Shahbaz and Yang, Ming-Hsuan and Shao, Ling},
  booktitle={2021 IEEE/CVF Conference on Computer Vision and Pattern Recognition (CVPR)}, 
  title={Multi-Stage Progressive Image Restoration}, 
  year={2021},
  volume={},
  number={},
  pages={14816-14826},
  doi={10.1109/CVPR46437.2021.01458}}

@article{hazediff,
  title   = {Hazediff: A training-free diffusion-based image dehazing method with pixel-level feature injection},
  author  = {Lin, X. and Li, Z. and Huang, D. and Feng, W. and An, X. and Sun, L. and others},
  journal = {PLOS ONE},
  volume  = {20},
  number  = {10},
  pages   = {e0329759},
  year    = {2025},
  doi     = {10.1371/journal.pone.0329759}
}

@inproceedings{Restormer,
    title={Restormer: Efficient Transformer for High-Resolution Image Restoration}, 
    author={Syed Waqas Zamir and Aditya Arora and Salman Khan and Munawar Hayat 
            and Fahad Shahbaz Khan and Ming-Hsuan Yang},
    booktitle={CVPR},
    year={2022}
}
}

\newpage
\onecolumn 
\begin{center}
    {\LARGE \textbf{\HazeMatching: Dehazing Light Microscopy Images with \\ Guided Conditional Flow Matching}}
\end{center}

\vspace{0.5em}

\begin{center}
    {\Large \textbf{\textit{Supplementary Material}}}
\end{center}

\setcounter{figure}{0} 
\renewcommand{\thefigure}{S\arabic{figure}} 

\setcounter{table}{0} 
\renewcommand{\thetable}{S\arabic{table}} 

\renewcommand{\theequation}{s.\arabic{equation}}
\setcounter{equation}{0}

\appendix
\section{Overview of the \HazeMatching}
\label{sup:overview}
\figTraining

\section{Uncertainty and Calibration}
\label{sup:calibration}
In this section, we describe our uncertainty estimation and calibration procedure~\cite{calibration}.
We use pixel-wise variability across multiple predictions of \HazeMatching and assess if this value scales with the true prediction error, based on a body of ground-truth data.
Under the hood, we adopt a binning-based calibration procedure inspired by~\cite{Microsplit} where we generate multiple predictions for each test image and analyze the relationship between predicted standard deviation and the observed true prediction error. 

For each test image $\mathbf{x} \in \mathds{R}^{H\times W}$, we generate $k = 50$ samples such that for each pixel $p$ in the test image $\mathbf{x}$, we have $k$ predictions. We then use these $k$ predictions to compute the pixel-wise standard deviation $\sigma(p)$ for the pixel location $p$. Next, we sort $\sigma(p)$ and bin them over $l=50$ equally sized bins $B: \{B_1, B_2 ... B_l\}$. 
We then compute the \textit{Root Mean Variance} (RMV) and \textit{Root Mean Squared Error} (RMSE) for each $B_j$ as
\begin{equation}
\text{RMV}(j) = \sqrt{ \frac{1}{|B_j|} \sum_{p \in B_j} \sigma(p)^2 }, \qquad  
\end{equation}
\begin{equation}
\text{RMSE}(j) = \sqrt{ \frac{1}{|B_j|} \sum_{p \in B_j} \left( \hat{y}(p) - y(p) \right)^2 },
\end{equation}
where $\hat{y}$ is the predicted mean (MMSE estimate) and $y$ is the ground truth. Then using our validation set, we then fit a linear relationship between RMSE and RMV as 
\begin{equation}
    \min_{\alpha, \beta} ||\text{RMSE} - (\alpha\cdot \text{RMV} + \beta)||^2,
\end{equation}
where $\alpha$ is a learnable scaling factor and $\beta$ is the learned offset. To ensure a meaningful mapping, we constrain $\alpha > 0$. To assess the calibration quality on the test set, we generate multiple predictions for each input and compute the pixel-wise standard deviation $\sigma$ across these samples. The corresponding estimate of the pixel-wise RMSE is then obtained using the learned linear mapping $\text{RMSE} = \alpha \cdot \sigma + \beta$. Note that this procedure does not alter the original predictions but instead learns a mapping that best predicts the measured error. All the calibrations were performed on the normalized data.

\subsection{Sample efficiency for calibration}
\label{sup:calibration_k}

In Table~\ref{tab:calibration_k_all}, we report the calibration slope $\alpha$ and the PSNR of the MMSE prediction as a function of the number of posterior samples $k$ across the five datasets.
Across all datasets, $\alpha$ increases steadily with $k$, reflecting improved variance calibration as the posterior sample grows. PSNR also rises consistently but saturates early. By $k\approx 20$ the gains become marginal, and beyond $k=40$ all datasets exhibit changes below $0.03$dB. In this high–$k$ regime, $\alpha$ varies only slightly between $k=40$ and $k=50$ (Zebrafish: $+2.6\%$, Organoids1: $-0.4\%$, Organoids2: $-2.5\%$, Microtubule: $+4.5\%$), indicating small fluctuations rather than systematic trends, with only exception being the Neuron data with a change of $+25.9\%$. We therefore use $k=50$ throughout the manuscript.

\begin{table}[h]
\centering
\setlength{\tabcolsep}{4pt} 
\begin{tabular}{c
                cc
                cc
                cc
                cc
                cc}
\toprule
\multirow{2}{*}{$k$} &
\multicolumn{2}{c}{\textbf{Zebrafish}} &
\multicolumn{2}{c}{\textbf{Organoids1}} &
\multicolumn{2}{c}{\textbf{Organoids2}} &
\multicolumn{2}{c}{\textbf{Microtubule}} &
\multicolumn{2}{c}{\textbf{Neuron}} \\
\cmidrule(lr){2-3} \cmidrule(lr){4-5} \cmidrule(lr){6-7} \cmidrule(lr){8-9} \cmidrule(lr){10-11}
 & $\alpha$ & PSNR$\uparrow$ & $\alpha$ & PSNR$\uparrow$ & $\alpha$ & PSNR$\uparrow$ & $\alpha$ & PSNR$\uparrow$ & $\alpha$ & PSNR$\uparrow$ \\
\midrule
5   & 1.0931 & 27.57 & 1.2554 & 36.58 & 0.6676  & 34.43 & 0.3924 & 27.50 & 0.5677 & 28.39  \\
10  & 1.3012 & 27.68 & 1.8286 & 36.71 & 0.9729  & 34.74 & 0.6153 & 27.71 & 0.6295 & 28.69  \\
20  & 1.5117 & 27.74 & 2.0911 & 36.78 & 1.1892  & 34.92 & 0.9049 & 27.80 & 0.8641 & 28.86  \\
30  & 1.5575 & 27.76 & 2.0880 & 36.81 & 1.2655  & 34.97 & 1.0013 & 27.84 & 0.7876 & 28.91  \\
40  & 1.6602 & 27.78 & 2.4993 & 36.82 & 1.3992  & 35.00 & 1.1114 & 27.85 & 0.7804 & 28.94  \\
50  & 1.7033 & 27.78 & 2.4903 & 36.83 & 1.3646  & 35.02 & 1.1671 & 27.87 & 0.9825 & 28.96  \\
\bottomrule
\end{tabular}
\caption{Calibration slope $\alpha$ and PSNR of the MMSE prediction for varying posterior sample count $k$ across the five datasets.}
\label{tab:calibration_k_all}
\end{table}

\subsection{Calibration factor validity}
\label{sup:calibration_val_val}

To verify that the learned calibration parameters ($\alpha$ and $\beta$) are meaningful, we apply the \texttt{scale} and \texttt{offset} computed on the validation set back onto the same validation set (rather than onto the test set used in the main results). This produces curves that align more closely with the identity line, confirming that the calibration behaves as expected as shown in Figure~\ref{fig:calib_val_val}.
\figCalibValVal
\newpage
\section{Datasets}
\label{sup:datasets}

\subsection{Microsim: Widefield-Confocal simulator}
\label{sup:microsim}

Conceptually, a confocal microscope can be viewed as a widefield microscope augmented with a pinhole aperture placed in a plane conjugate to the focal plane of the specimen. In widefield microscopy, the entire specimen is illuminated, and emitted fluorescence from all depths is collected, leading to images with significant background blur due to out-of-focus light. By introducing a pinhole at the conjugate image plane, the confocal setup selectively allows only in-focus light to reach the detector, effectively rejecting out-of-focus fluorescence and enhancing image clarity. By manipulating the pinhole aperture, one can simulate different imaging conditions:
\begin{itemize}
    \item \textbf{Pinhole Restricted (Confocal Mode):} The pinhole restricts detection to in-focus light, producing images with reduced background haze.
    \item \textbf{Pinhole Open (Widefield Mode):} Without the pinhole, both in-focus and out-of-focus light are detected, resulting in images with increased background haze.
\end{itemize}
This toggling mechanism facilitates the generation of paired datasets comprising "clean" (confocal) and "hazy" (widefield) images. Such datasets are invaluable for training and evaluating image restoration algorithms, particularly in the context of microscopy image dehazing.

We use \texttt{microsim}~\cite{microsim} to generate paired data for the \textit{Zebrafish}, \textit{Microtubule}, and \textit{Neuron} datasets. For \textit{Zebrafish} and \textit{Microtubule}, we start from clean confocal images and remove residual pixel-independent noise using \texttt{Noise2Void}~\cite{n2v}. We then simulate widefield-like counterparts using realistic physical pixel sizes and a confocal PSF with an open pinhole, incorporating microscopy-specific noise such as shot noise. This results in pairs with independent but statistically identical noise, preventing shortcut learning. For the \textit{Neuron} dataset, we skip denoising since we begin from segmentation labels, which are noise-free. During training, we treat the confocal image as the clean target and the simulated widefield image as the hazy input.

\textbf{Pixel size:} In the \texttt{microsim} library, the \texttt{scale} parameter defines the physical size of each voxel along the respective spatial axes, typically specified in micrometers. For instance, a scale of \texttt{(0.02, 0.01, 0.01)} corresponds to voxel dimensions of 20\,nm in the Z-axis and 10\,nm in both Y and X axes. This parameter is crucial for ensuring spatial realism in simulations, as it maps the discrete voxel grid to real-world physical dimensions. Accurate specification of the \texttt{scale} allows for realistic modeling of optical phenomena and ensures that simulated images correspond to the physical dimensions observed in actual microscopy data. To ensure spatial realism, we assign a physical pixel size (in nanometers) based on metadata from the original confocal datasets, further refined via expert visual inspection. This calibration enables accurate structure quantification and realistic widefield simulation. Detailed values for pixel sizes and pinhole diameters are provided in each of the dataset description below.

\subsection{Zebrafish Data}
\label{sup:zebra}

We use the confocal \textit{Zebrafish} dataset introduced in~\cite{care}, which contains three distinct structures: nuclei, nuclear membrane, and nuclear envelope. After removing pixel-independent noise, haze is simulated as described in Section~\ref{sup:microsim}, using a pinhole diameter of 30 Airy Units (AU) to emulate widefield degradation. For spatial realism, the \texttt{scale} parameter is set to \texttt{(0.4, 0.2, 0.2)} for Z, Y and X axes. As the three structures differ in intensity statistics, we normalize each structure independently using the mean and standard deviation computed from the training set. A single \HazeMatching model is trained across all structures. The original 3D confocal volume has shape $52 \times 1024 \times 1024$. After simulation, we extract the central plane, which exhibits the most haze. However for evaluating our method against the classical Richardson Lucy (RL) method, we used this entire stack because RL requires the input to be a 3D stack. We use a crop size of $128 \times 128$ during training and validation. The dataset contains 15 training images and 3 images each for validation and testing, all of size $1024 \times 1024$. We generated a total of 3000 patches for training. Pixel size is 0.2 $\mu m$.

\subsection{Organoids1}
\label{sup:organoids1}

This is a real microscopy dataset acquired using a spinning-disk confocal system, which is particularly relevant to our work due to its ability to toggle between confocal and wide-field modes simply by removing the spinning disk from the optical path. This enables the acquisition of non-hazy (confocal) and hazy (widefield) image pairs under nearly identical conditions, making it ideal for evaluating dehazing algorithms. The dataset was obtained using a spinning disk confocal system, consisting of a CrestOptics V3 Light scan-head (configured with 50$\mu m$ pinhole) mounted on a \texttt{Nikon Ti2-E} inverted microscope equipped with a motorized stage and a Photometrics Prime 95B 25mm camera (pixel size 11$\mu m$). The samples were acquired with a PLAN APO Lambda S 40x/1.25NA silicon immersion objective in combination with a 1.5x optovar lens (final magnification 60x) using Celesta Lumencor solid-state lasers as light source. For each acquisition, a specific filter set was used: a penta-band excitation filter (MXR00543-CELESTA-DAPI/FITC/TRITC/Cy5/Cy7-Full Multiband Penta), a penta-band dichroic filter (MXR00543-CELESTA-DAPI/FITC/TRITC/Cy5/Cy7-Full Multiband Penta) and a FITC band-pass emission filter (Semrock FF01-511/20-25). For each Field of View (FOV), two single-plane images were acquired: a confocal and a widefield image. The widefield images were acquired without the Nipkow disk using the 477nm laser line at 1\% and 80ms of exposure time, while the confocal images were captured with the Nipkow disk along the light path using the 477nm laser line at 45\% and 500ms of exposure time. All images were acquired using a 1024$\times$1024 pixel format and a pixel size of 190nm. We normalize each image using the mean and standard deviation computed from the training set. We use a crop size of $128 \times 128$ during training and validation. The dataset contains 15 training images and 2 images for validation and 3 images for testing, all of size $1024 \times 1024$. We generated a total of 3000 patches for training. Pixel size is 0.190 $\mu m$.

\subsection{Organoids2}
\label{sup:organoids2}

This human brain organoid dataset was acquired on a Leica Stellaris 8 point-scanning confocal system mounted on a DMI8 inverted microscope with a motorized stage, using a 405nm diode laser line in combination with a white light laser (WLL; 440nm -790nm) as excitation sources. The images were acquired with a HC PL APO 63x/1.40NA oil immersion objective using the 488nm laser line (WLL) at 5\% and 500nm – 600nm as emission collecting range. The pixel format of the 16-bit images was 1024$\times$1024 and the pixel size was 90nm, while the pixel dwell time was 887ns. For each FOV, two single-plane images were acquired: one using the fully open pinhole corresponding to 6.28 Airy Units (AU) and 600$\mu m$ as widefield-like image and the other one using 1 AU pinhole size corresponding to 95.52$\mu m$ as confocal image. By acquiring matched image pairs with the pinhole closed and fully open, this setup allows for controlled confocal and widefield-like imaging from the same sample, making it well-suited for dehazing evaluation. We normalize each image using the mean and standard deviation computed from the training set. We use a crop size of $128 \times 128$ during training and validation. The dataset contains 15 training images and 2 images for validation and 3 images for testing, all of size $1024 \times 1024$. We generated a total of 3000 patches for training. Pixel size is 0.090 $\mu m$.

\subsection{Microtubule data} 
\label{sup:microtubule}

We use confocal images of \textit{Microtubules} tagged with $\alpha$-Tubulin from the Allen Cell dataset~\cite{Tubulin}. Haze is simulated as described in Section~\ref{sup:microsim}, using a pinhole diameter of 45 Airy Units (AU) to emulate widefield degradation. For spatial realism, the \texttt{scale} parameter is set to \texttt{(0.55, 0.169, 0.169)} for Z, Y and X axes. We normalize each image using the mean and standard deviation computed from the training set. A crop size of $128 \times 128$ is used for training and validation. Similar to the Zebrafish data, these are real confocal images, they already contain pixel-independent noise (e.g., Gaussian and Poisson). To prevent the simulator from applying haze on top of existing noise—thus resulting in similar noise statistics for both clean and degraded images—we first denoise them using \texttt{Noise2Void}~\cite{n2v}, which removes only pixel-independent noise. The denoised image is then passed into the simulator to generate a hazy version with added realistic microscopy noise. Importantly, for training, we use the original confocal image as the clean target and the simulated hazy version as input. As the simulator requires a 3D volume, we use a stack of shape $52 \times 624 \times 624$ and select the middle slice, which contains the most haze, for training and evaluation. However, for evaluating our method against the classical Richardson Lucy (RL) method, we used this entire stack because RL requires the input to be a 3D stack. The dataset includes 52 training images of size $624 \times 624$ and 5 images each for validation and testing of size $512 \times 512$. We generated a total of 3016 patches for training. Pixel size is 0.1 $\mu m$.

\subsection{Neuron data} 
\label{sup:neuron}

This is a fully simulated dataset constructed from scratch using 3D segmentation labels from the Allen Cell Atlas~\cite{Neuron}, which contains 667 labeled volumes. To create realistic biological variability, we randomly sample 8--12 of these volumes chosen u.a.r, applying random rotations (sampled uniformly from $[-\pi, \pi]$) and isotropic scaling (sampled uniformly from $[0.8, 1.2]$). The transformed volumes are placed within a fixed spatial volume of size $256 \times 512 \times 512$ (Z, Y, X) and convolved with a point spread function (PSF) using \texttt{microsim} to simulate the imaging process. The \texttt{scale} parameter is set to \texttt{(0.04, 0.02, 0.02)}. For the confocal image, we use a pinhole diameter of 0.5 AU and for the widefield image we use a pinhole diameter of 5.0 AU. Detector noise is then added to produce realistic image degradation. For realistic simulation, we also isotropically downscale the simulated image by a factor of 8. Although the simulation generates full 3D stacks, we use only the middle slice for training and 2D evaluation, as it exhibits the strongest haze. However, for comparison with the classical Richardson Lucy (RL) deconvolution method, we use the entire 3D volume since RL operates on volumetric data. The final dataset consists of 5800 training images and 100 images each for validation and testing, all of size $64 \times 64$. Since this dataset is fully simulated, we can generate paired inputs and targets: the noisy, hazy image is used as input, and the clean (non-hazy, noise-free) image as target. This allows evaluation of both dehazing and denoising performance. Owing to the simplicity of the synthetic structures, this dataset exhibits less inherent data uncertainty compared to the other datasets used in this work. Pixel size is 0.16 $\mu m$.

\subsection{Organoids sample preparation}
\label{sup:organoids_sampleprep}

We introduced two real datasets that were acquired in-house -- \textit{Organoids1} and \textit{Organoids2}. Human brain organoids were fixed at 117 div, OCT embedded, frozen and prepared for sectioning. They were cut at 20$\mu m$ and 50$\mu m$ thickness using a cryostat. The cryosections were permeabilized with 0.5\% Triton X-100 in PBS for 30 min. Blocking was performed in blocking solution (5\% Normal Donkey Serum and 0.25\% Triton X-100 in PBS) for 30 min. All sections were stained using phalloidin conjugated with AlexaFluor-488 (A12379, ThermoFischer Scientific), to label actin filaments. Phalloidin was incubated in blocking solution for 1 hour. After three washings, the sections were then mounted in Prolong Glass (P36980, ThermoFisher Scientific). All the steps of the immunofluorescence procedure were performed at room temperature.

\section{Experimental setup}
\label{sup:dset_train_setup}

\subsection{Training setup}
\label{sup:train_setup}

For training, we use a patch size of $128 \times 128$ for \textit{Microtubule}, \textit{Zebrafish}, \textit{Organoids1}, and \textit{Organoids2} datasets, with 3016 patches for \textit{Microtubule} and 3000 patches for each of the other three datasets. For the \textit{Neuron} dataset, we use the original patch size of $64 \times 64$ for both training and evaluation with 5800 training patches. For evaluation, we adopt a rolling tile strategy: predictions are made on $128 \times 128$ tiles ($64 \times 64$ for \textit{Neuron}) and only the central $64 \times 64$ region ($32 \times 32$ for \textit{Neuron}) is retained. These are then stitched together to reconstruct the full image. All models are trained with $T{=}20$ integration steps. We implement \HazeMatching using a \UNet architecture as in~\cite{diffusinOpenAI}, and compute the interpolants and solve the forward ODE using the \texttt{torchCFM} package~\cite{torchCFM}. Training is performed in PyTorch~\cite{pytorch} on a half NVIDIA V100 GPU, with a batch size of 16, a learning rate of $10^{-4}$, and the Adam optimizer.

\subsection{Evaluation metrics and procedure}
\label{sup:eval_setup}

Distortion is evaluated using PSNR (a variant of the PSNR developed in (\cite{care} for light microscopy images) and MicroMS-SSIM (a light microscopy-specific variant of MS-SSIM introduced in (\cite{microssim}). For perceptual quality, we report LPIPS (\cite{lpips}) and FID (\cite{fid}). LPIPS measures perceptual similarity via deep feature embeddings (from AlexNet), while FID computes the Fréchet distance between feature statistics of real and generated images. 

Evaluation uses full-frame images for PSNR and MicroMS-SSIM. Perceptual metrics (LPIPS, FID, FSIM, GMSD) are computed on non-overlapping $64\times 64$ patches ($32\times 32$ for \textit{Neuron}) to eliminate stitching–related biases. Due to limited test samples, we estimate the clean image distribution using non-overlapping patches from the training sets and compute FID against predictions on the test set. This ensures robust evaluation without introducing data leakage, as training patches are used only for distribution estimation, not performance evaluation. A more detailed note on this can be found in supplementary Section \ref{sup:note_fid}.  Additionally, we present extended evaluations on the two additional synthetic datasets—\textit{Neuron}~(\ref{sup:neuron_results}) and \textit{Microtubule}~(\ref{sup:microtubule_results}). In Section~\ref{sup:pdt_microssim}, we highlight trade-offs between fine structure (via MicroMS-SSIM) and perceptual quality (LPIPS and FID). Next, in Section~\ref{sup:full_tables} we provide full quantitative tables for PSNR, LPIPS, FID, MicroMS-SSIM, MS-SSIM~\cite{msssim}, and two alternative perceptual quality scores FSIM~\cite{fsim}, and GMSD~\cite{gmsd}. Finally, in Section~\ref{sup:qualitative}, we show additional qualitative results to offer a more comprehensive view of all baseline and \HazeMatching. Note that all evaluations were conducted in the denormalized space, except for the \textit{Neuron} dataset. For this dataset, evaluations were performed in the normalized space. In the case of Richardson–Lucy (RL) deconvolution, the \textit{Neuron} data was normalized using min-max normalization to the $[0,1]$ range instead of standard mean–standard deviation normalization. This unusual deviation from the default denormalization strategy was a practical decision, as the data—being fully simulated, exhibited very high intensity ranges due to the use of non-noisy ground truth segmentation volumes. For LPIPS, we normalize each image independently to the range $[-1, 1]$ using its own minimum and maximum intensity values, whereas for FID, we apply the same per-image normalization to the range $[0, 1]$.

\subsubsection{On FID evaluation protocol} 
\label{sup:note_fid}
Due to the limited number of test patches in our datasets, directly computing the FID between predicted and real test samples can yield unstable results. To address this, we follow a practical approach: we aggregate non-overalpping $64\times64$ patches (or $32\times32$) from the training sets for each of the datasets to construct a representative empirical distribution of real clean images. Importantly, this does \textit{not} mean we evaluate model performance on training data; rather, the training patches are used solely to estimate the underlying distribution of real clean images. The predictions on the test set are then compared against this aggregated distribution to compute FID. This strategy allows us to maintain a robust evaluation while ensuring that the test set remains completely unseen during training and inference.

\subsubsection{Effect of Patch Size at Inference}
\label{sup:patch_size}

\textcolor{black}{While it is true that we train on $128\times 128$ sized patches, the data itself is composed of much larger micrographs. 
The $128\times 128$ constraint is imposed only by our data loader during training to enable efficient batching and GPU utilization. 
Since \HazeMatching is fully convolutional, it can, in principle, process inputs of arbitrary spatial size, subject only to GPU memory limitations. 
When full micrographs are too large to fit in memory, we use a tiling strategy during inference (see also Supplementary Section~\ref{sup:dset_train_setup}).}

\textcolor{black}{To assess whether larger patch sizes improve performance, we evaluated our trained model on the \textbf{Organoids1} real spinning disk dataset using patch sizes of $128$, $256$, and $512$ pixels at inference. 
The results, summarized in Table~\ref{tab:patch_size}, show no meaningful improvement in PSNR when increasing the patch size beyond $128$ pixels.}

\begin{table}[h]
\centering
\setlength{\tabcolsep}{6pt}
\begin{tabular}{lccc}
\toprule
\textbf{Model} & \textbf{Patch Size} & \textbf{PSNR}$\uparrow$ ($\mu$) & \textbf{PSNR} ($\sigma$) \\
\midrule
\HazeMatching  & $128$ & 36.83 & 5.894 \\
\HazeMatching  & $256$ & 36.85 & 5.964 \\
\HazeMatching  & $512$ & 36.76 & 6.217 \\
\bottomrule
\end{tabular}
\caption{Effect of patch size at inference on \textbf{Organoids1} real spinning disk data.}
\label{tab:patch_size}
\end{table}

\textcolor{black}{These results indicate that, for our model, larger patch sizes at inference do not yield measurable improvements in restoration quality. 
This suggests that \HazeMatching's performance is not constrained by the spatial context provided during training, and the network can be reliably applied to full-resolution images via tiling when needed.}
\section{Baselines}
\label{sup:baseline}

We include the Richardson–Lucy (RL) deconvolution algorithm~\citep{RL}, a classical iterative method that operates on 3D image stacks and requires knowledge of the point spread function (PSF). We evaluate RL with varying numbers of iterations and select the best-performing result based on peak signal-to-noise ratio (PSNR). We denote this as $\text{RL}_m$, where $m$ indicates the number of iterations. In the main paper we show the iteration that has the best PSNR. Detailed results across iterations are provided in the supplementary material (Figure~\ref{fig:plot_rl_iters}). We use the implementation provided by \texttt{DeconvolutionLab2}\footnote{DeconvolutionLab2: An Open-Source Software for Deconvolution Microscopy. Sage et. al, Methods—Image Processing for Biologists, 2017} Fiji plugin, adapted to our 3D microscopy data. 

We evaluate seven point-predictors: a \UNet~\citep{unet}, InDI$_1$~\citep{indi}, MIMO-UNet~\cite{mimo}, MPRNet~\cite{mprnet}, RCAN~\cite{rcan}, Restormer~\cite{Restormer}, and ESRGAN~\cite{esrgan} all of which produce a single prediction per input. For all of the baselines, we use a batch size of 16 except RCAN on Neuron data, where we use a batch size of 12, and Restormer, where we use a batch size of 8. We use a learning rate of $10^{-4}$ for all the models except Restormer where we follow the official implementation. In ESRGAN training, the VGG-based discriminator requires a minimum input resolution of \(128 \times 128\). Since the \textit{Neuron} dataset provides only \(64 \times 64\) patches, we upsample each predicted patch to \(128 \times 128\) via nearest-neighbour interpolation before feeding it to the discriminator. Inference remains unchanged. We use the open-source \footnote{\texttt{BasicSR}: Image and Video Restoration Toolbox} implementation for RCAN and ESRGAN, and the official implementations for MIMO-UNet and MPRNet. 

We also evaluate the iterative variant InDI$_{20}$, which runs the prediction for 20 refinement steps. For InDI, we adopt the training settings specified for the \texttt{defocus deblurring} task in their original paper~\citep{indi}. As the official implementation of InDI is not available, we re-implement it following the setup for the \texttt{defocus deblurring} task using code available at \cite{con_diff}. We use a sampling schedule for the \texttt{defocus deblurring} task, train with an $L_1$ loss, a learning rate of $10^{-4}$, batch size of 16, and the Adam optimizer. 

We also compare with \HVAE~\citep{hvae}, a ladder VAE model capable of generating diverse predictions from a single input in one forward pass. 
We use a learned top-level prior initialized to zero. Training is done with a batch size of 16, a learning rate of $10^{-4}$, and 5 levels of hierarchy (reduced to 4 for the \textit{Neuron} dataset due to its smaller size). 

We present several configurations of SIFM (\cite{sifm}): SIFM{$\sigma_{a=0.0|b\geq0.0}$} and SIFM{$\sigma_{a>0.0|b=a}$}. These approaches can generate a single output (SIFM{$\sigma_{a=0.0|b=0.0}$}) or can generate multiple predictions in an iterative manner (like \HazeMatching). Note that for this approach we test different values of $a$ and $b$ for SIFM baselines but report only the configuration with best PSNR. We report the PSNR of all the configurations in Figure~\ref{fig:plot_sifm_configs}. As the official code is not available, we use the same \HazeMatching code with change in the loss function according to\cite{sifm}. 

For the real dataset \textit{Organoids1}, we also compare against the proprietary software \textit{Elements} from \cite{elements} and for the real dataset \textit{Organoids2}, we compare against the proprietary software \textit{Lightning} from \cite{lightning}. The \textit{Elements} AR 5.42.02 (Nikon) was used for deconvolution. It was performed using the NIS-integrated deconvolution tool. The deconvolution parameters were set as follows: deconvolution type: 2D; modality: widefield; numerical aperture: 1.25; immersion refractive index: 1.406; calibration: 0.1902$\mu m$; iterations: 70-120. The \textit{Lightning} software was from  Leica Application Suite X (LAS X, Leica Microsystems) version 4.7.0.28176 and its integrated deconvolution module. The parameters used for the post-processing of the widefield images using Lightning are the following: strategy: adaptive; type: confocal; number of iterations: automatic; optimization: 0; contrast enhancement: automatic; cutoff: automatic; regularization parameter: 0; smoothing: none; excitation wavelength: 488nm; emission wavelength: 515nm; pinhole: 6.28 AU; normalization: range; objective numerical aperture: 1.40; immersion refractive index: 1.518; magnification: 63; z-offset: 0$\mu m$; coverslip refractive index 1.523; coverslip thickness: 170 um; mounting medium: Prolong Glass; mounting medium refractive index: 1.520.
\newpage
\section{Results}
\label{sup:results}

\subsection{Neuron data results}
\label{sup:neuron_results}
\figPlotsMainNeuron
\figCalibNeuron

\newpage
\subsection{Microtubule data results}
\label{sup:microtubule_results}

\figPlotsMainMicrotubule
\figCalibMicrotubule

\newpage
\subsection{MicroMS-SSIM vs LPIPS/FID}
\label{sup:pdt_microssim}
\figPlotsSSIM

\newpage
\subsection{Richardson Lucy iterations}
\label{sup:rl_iters}
\figPlotsRL

\subsection{SIFM configurations}
\label{sup:sifm_configs}
\figPlotsSIFM

\newpage
\subsection{Full quantitative results}
\label{sup:full_tables}

Here we report PSNR, LPIPS, FID, MicroMS-SSIM, MS-SSIM, as well as two additional perceptual quality metrics: FSIM and GMSD. These metrics jointly capture pixel-wise fidelity, structural preservation, and perceptual realism. MicroMS-SSIM emphasizes fine structural detail, while LPIPS and FID evaluate perceptual quality based on learned features. FSIM evaluates perceptual similarity based on low-level features such as phase congruency and gradient magnitude, while GMSD quantifies image quality by measuring gradient magnitude similarity and is particularly sensitive to local distortions. Standard deviations are reported wherever applicable (mini-row at the bottom), computed across the test set for each dataset. Note that we consider the average of the samples for our perceptual metrics. In this case, for MMSE/point predictors only have one samples. Nevertheless for completeness we show the perceptual quality of the MMSE estimates for all the methods.

\begin{table*}[htbp]
\centering
\renewcommand{\arraystretch}{1.2}
\small
\vspace{0.5em}
\scalebox{0.55}{
\begin{tabular}{c|l|c|c|c||c|c|c|c||c|c|c|cc}

\multirow{2}{*}{\centering \textbf{Dataset}} & \multicolumn{1}{c|}{\multirow{2}{*}{\textbf{Methods}}}
 & \multicolumn{3}{c||}{\textbf{Distortion on MMSE/Point-Prediction}} & \multicolumn{4}{c||}{\textbf{Perception on MMSE/Point-Prediction}} & \multicolumn{4}{c}{\textbf{Average Perception on Samples}} \\ \cline{3-13}

&  & \textbf{PSNR $\uparrow$} & \textbf{MS-SSIM $\uparrow$} & \textbf{MicroMS-SSIM $\uparrow$} & \textbf{FSIM $\uparrow$} & \textbf{LPIPS $\downarrow$} & \textbf{FID $\downarrow$} & \textbf{GMSD $\downarrow$} & \textbf{FSIM $\uparrow$} & \textbf{LPIPS $\downarrow$} & \textbf{FID $\downarrow$} & \textbf{GMSD $\downarrow$}\\ 
\hline \hline

\multirow{18}{*}{\centering \textbf{Zebrafish}}

& RL$_{3}$ & \makecell{26.56 \\ 4.105} & \makecell{0.992 \\ 0.0018} & \makecell{0.858 \\ 0.0368} & \makecell{0.788 \\ 0.0351} & \makecell{0.249} & \makecell{0.430} & \makecell{0.166 \\ 0.0289} &&&&  \\ \cline{2-9}
& UNet & \makecell{27.85 \\ 3.546} & \makecell{0.902 \\ 0.0138} & \makecell{0.901 \\ 0.0119} & \makecell{0.790 \\ 0.0646} & \makecell{0.383} & \makecell{9.146} & \makecell{0.151 \\ 0.0306} &&&& \\ \cline{2-9}
& InDI$_{1}$ & \makecell{27.28 \\ 3.833} & \makecell{0.885 \\ 0.0201} & \makecell{0.886 \\ 0.0178} & \makecell{0.748 \\ 0.0907} & \makecell{0.455} & \makecell{12.326} & \makecell{0.165 \\ 0.0322} &&&&  \\ \cline{2-9}

& MIMO-UNet & \makecell{27.74 \\ 3.538} & \makecell{0.891 \\ 0.0237} & \makecell{0.895 \\ 0.0128} & \makecell{0.773 \\ 0.0800} & \makecell{0.322} & \makecell{9.235} & \makecell{0.155 \\ 0.0312} &&&&  \\ \cline{2-9}
& MPRNet & \makecell{28.08 \\ 3.590} & \makecell{0.904 \\ 0.0152} & \makecell{0.903 \\ 0.0122} & \makecell{0.769 \\ 0.0844} & \makecell{0.482} & \makecell{11.587} & \makecell{0.151 \\ 0.0310} &&&&   \\ \cline{2-9}
& RCAN & \makecell{27.70 \\ 3.535} & \makecell{0.892 \\ 0.0221} & \makecell{0.896 \\ 0.0135} & \makecell{0.792 \\ 0.0591} & \makecell{0.246} & \makecell{4.957} & \makecell{0.151 \\ 0.0335} &&&&   \\ \cline{2-9}
& ESRGAN & \makecell{27.07 \\ 3.841} & \makecell{0.878 \\ 0.0197} & \makecell{0.881 \\ 0.0234} & \makecell{0.811 \\ 0.0353} & \makecell{0.128} & \makecell{0.857} & \makecell{0.151 \\ 0.0305} &&&&   \\ \cline{2-9}

& Restormer & \makecell{28.38 \\ 3.506} & \makecell{0.912 \\ 0.0159} & \makecell{0.911 \\ 0.0110} & \makecell{0.772 \\ 0.0814} & \makecell{0.417} & \makecell{10.148} & \makecell{0.150 \\ 0.0344} &&&&  \\ \cline{2-9}

& InDI$_{20}$ & \makecell{26.35 \\ 4.158} & \makecell{0.867 \\ 0.0208} & \makecell{0.850 \\ 0.0398} & \makecell{0.787 \\ 0.0326} & \makecell{0.300} & \makecell{0.633} & \makecell{0.157 \\ 0.0260} &&&& \\ \cline{2-9}
& SIFM{$\sigma_{0.0|0.0}$} & \makecell{27.42 \\ 3.583} & \makecell{0.888 \\ 0.0191} & \makecell{0.893 \\ 0.0151} & \makecell{0.823 \\ 0.0377} & \makecell{0.119} & \makecell{0.917} & \makecell{0.140 \\ 0.0318} &&&&  \\ \cline{2-13}

& LVAE & \makecell{27.76 \\ 3.729} & \makecell{0.903 \\ 0.0126} & \makecell{0.899 \\ 0.0143} & \makecell{0.772 \\ 0.0759} & \makecell{0.329} & \makecell{6.490} & \makecell{0.149 \\ 0.0329} & \makecell{0.7911 \\ 0.0323} & \makecell{0.318 \\ 0.0210} & \makecell{0.631 \\ 0.2790} & \makecell{0.161 \\ 0.0361}  \\ \cline{2-13}
& SIFM{$\sigma_{0.0|0.1}$} & \makecell{27.36 \\ 3.651} & \makecell{0.885 \\ 0.0150} & \makecell{0.889 \\ 0.0187} & \makecell{0.822 \\ 0.0352} & \makecell{0.123} & \makecell{0.668} & \makecell{0.140 \\ 0.0302} & \makecell{0.8079 \\ 0.0352} & \makecell{0.152 \\ 0.0228} & \makecell{1.128 \\ 1.0379} & \makecell{0.148 \\ 0.0273}  \\ \cline{2-13}
& SIFM{$\sigma_{0.2|0.2}$} & \makecell{27.72 \\ 3.858} & \makecell{0.903 \\ 0.0121} & \makecell{0.898 \\ 0.0157} & \makecell{0.815 \\ 0.0468} & \makecell{0.179} & \makecell{3.054} & \makecell{0.142 \\ 0.0316} & \makecell{0.8093 \\ 0.0361} & \makecell{0.156 \\ 0.0146} & \makecell{1.372 \\ 1.3301} & \makecell{0.151 \\ 0.0288}  \\ \cline{2-13}

\hhline{~|============|~}
& \textbf{\HazeMatching (ours)} & \makecell{27.78 \\ 3.658} & \makecell{0.899 \\ 0.0200} & \makecell{0.897 \\ 0.0109} & \makecell{0.764 \\ 0.0849} & \makecell{0.439} & \makecell{9.996} & \makecell{0.154 \\ 0.0315} & \makecell{0.8061 \\ 0.0422} & \makecell{0.145 \\ 0.0182} & \makecell{2.165 \\ 1.7052} & \makecell{0.156 \\ 0.0331}

\end{tabular}
}
\caption{\textbf{Full quantitative comparison for the Zebrafish Dataset}}
\label{tab:CareSupple}
\end{table*}
\begin{table*}[htbp]
\centering
\renewcommand{\arraystretch}{1.2}
\small
\vspace{0.5em}
\scalebox{0.55}{
\begin{tabular}{c|l|c|c|c||c|c|c|c||c|c|c|cc}

\multirow{2}{*}{\centering \textbf{Dataset}} & \multicolumn{1}{c|}{\multirow{2}{*}{\textbf{Methods}}}
 & \multicolumn{3}{c||}{\textbf{Distortion on MMSE/Point-Prediction}} & \multicolumn{4}{c||}{\textbf{Perception on MMSE/Point-Prediction}} & \multicolumn{4}{c}{\textbf{Average Perception on Samples}} \\ \cline{3-13}
&  & \textbf{PSNR $\uparrow$} & \textbf{MS-SSIM $\uparrow$} & \textbf{MicroMS-SSIM $\uparrow$} & \textbf{FSIM $\uparrow$} & \textbf{LPIPS $\downarrow$} & \textbf{FID $\downarrow$} & \textbf{GMSD $\downarrow$} & \textbf{FSIM $\uparrow$} & \textbf{LPIPS $\downarrow$} & \textbf{FID $\downarrow$} & \textbf{GMSD $\downarrow$}\\ 
\hline \hline

\multirow{18}{*}{\centering \textbf{Organoids1}}

& Elements & \makecell{33.20 \\ 5.645} & \makecell{0.876 \\ 0.1215} & \makecell{0.785 \\ 0.0793} & \makecell{0.866 \\ 0.0336} & \makecell{0.336} & \makecell{0.697} & \makecell{0.103 \\ 0.0273} &&&&  \\ \cline{2-9}
& UNet & \makecell{37.07 \\ 6.023} & \makecell{0.993 \\ 0.0045} & \makecell{0.953 \\ 0.0283} & \makecell{0.865 \\ 0.0510} & \makecell{0.211} & \makecell{2.200} & \makecell{0.119 \\ 0.0287} &&&&  \\ \cline{2-9}
& InDI$_{1}$ & \makecell{36.05 \\ 6.164} & \makecell{0.990 \\ 0.0054} & \makecell{0.936 \\ 0.0360} & \makecell{0.849 \\ 0.0554} & \makecell{0.244} & \makecell{2.590} & \makecell{0.131 \\ 0.0284} &&&& \\ \cline{2-9}

& MIMO-UNet & \makecell{36.84 \\ 6.044} & \makecell{0.993 \\ 0.0051} & \makecell{0.951 \\ 0.0278} & \makecell{0.857 \\ 0.0574} & \makecell{0.253} & \makecell{2.800} & \makecell{0.122 \\ 0.0276} &&&&   \\ \cline{2-9}
& MPRNet & \makecell{37.39 \\ 6.073} & \makecell{0.993 \\ 0.0040} & \makecell{0.955 \\ 0.0273} & \makecell{0.865 \\ 0.0558} & \makecell{0.237} & \makecell{2.716} & \makecell{0.115 \\ 0.0281} &&&&   \\ \cline{2-9}
& RCAN & \makecell{36.55 \\ 6.136} & \makecell{0.992 \\ 0.0049} & \makecell{0.942 \\ 0.0362} & \makecell{0.847 \\ 0.0598} & \makecell{0.243} & \makecell{2.656} & \makecell{0.134 \\ 0.0307} &&&&  \\ \cline{2-9}
& ESRGAN & \makecell{36.16 \\ 6.080} & \makecell{0.991 \\ 0.0057} & \makecell{0.941 \\ 0.0324} & \makecell{0.852 \\ 0.0406} & \makecell{0.101} & \makecell{0.038} & \makecell{0.131 \\ 0.0268} &&&&  \\ \cline{2-9}

& Restormer & \makecell{37.52 \\ 5.870} & \makecell{0.994 \\ 0.0043} & \makecell{0.957 \\ 0.0265} & \makecell{0.865 \\ 0.0584} & \makecell{0.220} & \makecell{2.303} & \makecell{0.113 \\ 0.0290} &&&&  \\ \cline{2-9}

& InDI$_{20}$ & \makecell{33.52 \\ 5.511} & \makecell{0.977 \\ 0.0163} & \makecell{0.890 \\ 0.0509} & \makecell{0.822 \\ 0.0316} & \makecell{0.317} & \makecell{3.010} & \makecell{0.140 \\ 0.0187} &&&&  \\ \cline{2-9}
& SIFM{$\sigma_{0.0|0.0}$} & \makecell{36.57 \\ 6.526} & \makecell{0.991 \\ 0.0047} & \makecell{0.945 \\ 0.0301} & \makecell{0.869 \\ 0.0437} & \makecell{0.090} & \makecell{0.047} & \makecell{0.115 \\ 0.0299} &&&&  \\ \cline{2-13}

& LVAE & \makecell{33.99 \\ 1.763} & \makecell{0.991 \\ 0.0050} & \makecell{0.951 \\ 0.0253} & \makecell{0.851 \\ 0.0471} & \makecell{0.197} & \makecell{0.491} & \makecell{0.123 \\ 0.0253} & \makecell{0.8031 \\ 0.0355} & \makecell{0.427 \\ 0.0247} & \makecell{3.988 \\ 3.1306} & \makecell{0.156 \\ 0.0320}  \\ \cline{2-13}
& SIFM{$\sigma_{0.0|0.1}$} & \makecell{36.21 \\ 6.392} & \makecell{0.987 \\ 0.0044} & \makecell{0.942 \\ 0.0287} & \makecell{0.865 \\ 0.0421} & \makecell{0.100} & \makecell{0.100} & \makecell{0.117 \\ 0.0292} & \makecell{0.8011 \\ 0.0668} & \makecell{0.226 \\ 0.0877} & \makecell{2.825 \\ 3.6920} & \makecell{0.147 \\ 0.0382}  \\ \cline{2-13}
& SIFM{$\sigma_{0.1|0.1}$} & \makecell{36.49 \\ 6.061} & \makecell{0.991 \\ 0.0051} & \makecell{0.947 \\ 0.0290} & \makecell{0.866 \\ 0.0542} & \makecell{0.155} & \makecell{1.070} & \makecell{0.113 \\ 0.0291} & \makecell{0.8472 \\ 0.0516} & \makecell{0.121 \\ 0.0531} & \makecell{0.851 \\ 0.1649} & \makecell{0.134 \\ 0.0369}  \\ \cline{2-13}
\hhline{~|============|~}
& \textbf{\HazeMatching (ours)} & \makecell{36.83 \\ 5.894} & \makecell{0.992 \\ 0.0042} & \makecell{0.954 \\ 0.0256} & \makecell{0.867 \\ 0.0533} & \makecell{0.236} & \makecell{2.650} & \makecell{0.114 \\ 0.0282} & \makecell{0.8449 \\ 0.0487} & \makecell{0.140 \\ 0.0755} & \makecell{0.905 \\ 0.5233} & \makecell{0.136 \\ 0.0288}

\end{tabular}
}
\caption{\textbf{Full quantitative comparison for the Organoids1 Dataset}}
\label{tab:Organoids1Supple}
\end{table*}

\begin{table*}[htbp]
\centering
\renewcommand{\arraystretch}{1.2}
\small
\vspace{0.5em}
\scalebox{0.55}{
\begin{tabular}{c|l|c|c|c||c|c|c|c||c|c|c|cc}

\multirow{2}{*}{\centering \textbf{Dataset}} & \multicolumn{1}{c|}{\multirow{2}{*}{\textbf{Methods}}}
 & \multicolumn{3}{c||}{\textbf{Distortion on MMSE/Point-Prediction}} & \multicolumn{4}{c||}{\textbf{Perception on MMSE/Point-Prediction}} & \multicolumn{4}{c}{\textbf{Average Perception on Samples}} \\ \cline{3-13}
&  & \textbf{PSNR $\uparrow$} & \textbf{MS-SSIM $\uparrow$} & \textbf{MicroMS-SSIM $\uparrow$} & \textbf{FSIM $\uparrow$} & \textbf{LPIPS $\downarrow$} & \textbf{FID $\downarrow$} & \textbf{GMSD $\downarrow$} & \textbf{FSIM $\uparrow$} & \textbf{LPIPS $\downarrow$} & \textbf{FID $\downarrow$} & \textbf{GMSD $\downarrow$}\\ 
\hline \hline

\multirow{18}{*}{\centering \textbf{Organoids2}}

& Lightning & \makecell{32.38 \\ 1.358} & \makecell{0.935 \\ 0.0041} & \makecell{0.929 \\ 0.0102} & \makecell{0.824 \\ 0.0349} & \makecell{0.095} & \makecell{0.380} & \makecell{0.113 \\ 0.0248} &&&& \\ \cline{2-9}
& UNet & \makecell{35.08 \\ 1.159} & \makecell{0.977 \\ 0.0007} & \makecell{0.967 \\ 0.0057} & \makecell{0.825 \\ 0.0566} & \makecell{0.483} & \makecell{7.386} & \makecell{0.094 \\ 0.0302} &&&&  \\ \cline{2-9}
& InDI$_{1}$ & \makecell{34.00 \\ 1.057} & \makecell{0.967 \\ 0.0005} & \makecell{0.955 \\ 0.0076} & \makecell{0.813 \\ 0.0545} & \makecell{0.366} & \makecell{6.316} & \makecell{0.113 \\ 0.0290} &&&& \\ \cline{2-9}

& MIMO-UNet & \makecell{34.64 \\ 1.107} & \makecell{0.974 \\ 0.0010} & \makecell{0.963 \\ 0.0062} & \makecell{0.812 \\ 0.0625} & \makecell{0.452} & \makecell{7.138} & \makecell{0.100 \\ 0.0286} &&&&  \\ \cline{2-9}
& MPRNet & \makecell{35.14 \\ 1.151} & \makecell{0.978 \\ 0.0011} & \makecell{0.968 \\ 0.0056} & \makecell{0.819 \\ 0.0613} & \makecell{0.495} & \makecell{6.959} & \makecell{0.092 \\ 0.0306} &&&&   \\ \cline{2-9}
& RCAN & \makecell{34.92 \\ 1.142} & \makecell{0.977 \\ 0.0009} & \makecell{0.965 \\ 0.0054} & \makecell{0.825 \\ 0.0538} & \makecell{0.440} & \makecell{5.871} & \makecell{0.092 \\ 0.0288} &&&&  \\ \cline{2-9}
& ESRGAN & \makecell{32.75 \\ 1.144} & \makecell{0.961 \\ 0.0018} & \makecell{0.948 \\ 0.0107} & \makecell{0.831 \\ 0.0347} & \makecell{0.089} & \makecell{0.401} & \makecell{0.109 \\ 0.0273} &&&&  \\ \cline{2-9}

& Restormer & \makecell{35.18 \\ 1.162} & \makecell{0.978 \\ 0.0014} & \makecell{0.969 \\ 0.0053} & \makecell{0.818 \\ 0.0605} & \makecell{0.514} & \makecell{7.367} & \makecell{0.093 \\ 0.0309} &&&&  \\ \cline{2-9}

& InDI$_{20}$ & \makecell{33.28 \\ 1.039} & \makecell{0.965 \\ 0.0007} & \makecell{0.955 \\ 0.0073} & \makecell{0.840 \\ 0.0316} & \makecell{0.109} & \makecell{0.552} & \makecell{0.097 \\ 0.0232} &&&& \\ \cline{2-9}
& SIFM{$\sigma_{0.0|0.0}$} & \makecell{33.57 \\ 1.223} & \makecell{0.970 \\ 0.0015} & \makecell{0.960 \\ 0.0080} & \makecell{0.847 \\ 0.0358} & \makecell{0.077} & \makecell{0.164} & \makecell{0.093 \\ 0.0276} &&&&  \\ \cline{2-13}

& LVAE & \makecell{34.73 \\ 1.066} & \makecell{0.975 \\ 0.0013} & \makecell{0.965 \\ 0.0057} & \makecell{0.799 \\ 0.0586} & \makecell{0.448} & \makecell{4.592} & \makecell{0.098 \\ 0.0287} & \makecell{0.8205 \\ 0.0341} & \makecell{0.149 \\ 0.0172} & \makecell{1.013 \\ 0.5901} & \makecell{0.120 \\ 0.0292}  \\ \cline{2-13}
& SIFM{$\sigma_{0.0|0.1}$} & \makecell{33.53 \\ 1.229} & \makecell{0.970 \\ 0.0015} & \makecell{0.960 \\ 0.0080} & \makecell{0.847 \\ 0.0356} & \makecell{0.076} & \makecell{0.196} & \makecell{0.093 \\ 0.0278} & \makecell{0.8414 \\ 0.0389} & \makecell{0.084 \\ 0.0098} & \makecell{0.690 \\ 0.3667} & \makecell{0.098 \\ 0.0307}  \\ \cline{2-13}
& SIFM{$\sigma_{1.0|1.0}$} & \makecell{34.57 \\ 1.113} & \makecell{0.974 \\ 0.0012} & \makecell{0.963 \\ 0.0060} & \makecell{0.800 \\ 0.0576} & \makecell{0.329} & \makecell{3.507} & \makecell{0.099 \\ 0.0291} & \makecell{0.8117 \\ 0.0420} & \makecell{0.117 \\ 0.0117} & \makecell{0.354 \\ 0.0948} & \makecell{0.138 \\ 0.0294}  \\ \cline{2-13}
\hhline{~|============|~}
& \textbf{\HazeMatching (ours)} & \makecell{35.02 \\ 1.113} & \makecell{0.977 \\ 0.0004} & \makecell{0.967 \\ 0.0058} & \makecell{0.825 \\ 0.0542} & \makecell{0.435} & \makecell{5.517} & \makecell{0.093 \\ 0.0290} & \makecell{0.8300 \\ 0.0355} & \makecell{0.088 \\ 0.0085} & \makecell{0.385 \\ 0.1600} & \makecell{0.113 \\ 0.0271} 

\end{tabular}
}
\caption{\textbf{Full quantitative comparison for the Organoids2 Dataset}}
\label{tab:Organoids2Supple}
\end{table*}

\begin{table*}[htbp]
\centering
\renewcommand{\arraystretch}{1.2}
\small
\vspace{0.5em}
\scalebox{0.55}{
\begin{tabular}{c|l|c|c|c||c|c|c|c||c|c|c|cc}

\multirow{2}{*}{\centering \textbf{Dataset}} & \multicolumn{1}{c|}{\multirow{2}{*}{\textbf{Methods}}}
 & \multicolumn{3}{c||}{\textbf{Distortion on MMSE/Point-Prediction}} & \multicolumn{4}{c||}{\textbf{Perception on MMSE/Point-Prediction}} & \multicolumn{4}{c}{\textbf{Average Perception on Samples}} \\ \cline{3-13}

&  & \textbf{PSNR $\uparrow$} & \textbf{MS-SSIM $\uparrow$} & \textbf{MicroMS-SSIM $\uparrow$} & \textbf{FSIM $\uparrow$} & \textbf{LPIPS $\downarrow$} & \textbf{FID $\downarrow$} & \textbf{GMSD $\downarrow$} & \textbf{FSIM $\uparrow$} & \textbf{LPIPS $\downarrow$} & \textbf{FID $\downarrow$} & \textbf{GMSD $\downarrow$}\\ 
\hline \hline

\multirow{14}{*}{\centering \textbf{Microtubule}} 
& RL$_{1}$ & \makecell{27.16 \\ 0.749} & \makecell{0.994 \\ 0.0005} & \makecell{0.867 \\ 0.0126} & \makecell{0.812 \\ 0.0283} & \makecell{0.175} & \makecell{1.456} & \makecell{0.131 \\ 0.0207} &&&& \\ \cline{2-9}
& U-Net & \makecell{28.53 \\ 0.729} & \makecell{0.906 \\ 0.0089} & \makecell{0.898 \\ 0.0087} & \makecell{0.779 \\ 0.0367} & \makecell{0.645} & \makecell{8.209} & \makecell{0.131 \\ 0.0190} &&&&  \\ \cline{2-9}
& InDI$_{1}$ & \makecell{27.58 \\ 0.657} & \makecell{0.897 \\ 0.0091} & \makecell{0.884 \\ 0.0064} & \makecell{0.759 \\ 0.0425} & \makecell{0.658} & \makecell{7.441} & \makecell{0.133 \\ 0.0188} &&&&  \\ \cline{2-9}

& MIMO-UNet & \makecell{27.60 \\ 0.659} & \makecell{0.896 \\ 0.0082} & \makecell{0.884 \\ 0.0077} & \makecell{0.752 \\ 0.0519} & \makecell{0.705} & \makecell{8.570} & \makecell{0.141 \\ 0.0183} &&&&  \\ \cline{2-9}
& MPRNet & \makecell{28.67 \\ 0.782} & \makecell{0.908 \\ 0.0087} & \makecell{0.901 \\ 0.0082} & \makecell{0.770 \\ 0.0402} & \makecell{0.712} & \makecell{8.679} & \makecell{0.131 \\ 0.0192} &&&&  \\ \cline{2-9}
& RCAN & \makecell{26.92 \\ 0.830} & \makecell{0.869 \\ 0.0114} & \makecell{0.857 \\ 0.0157} & \makecell{0.806 \\ 0.0220} & \makecell{0.230} & \makecell{0.254} & \makecell{0.146 \\ 0.0196} &&&& \\ \cline{2-9}
& ESRGAN & \makecell{26.40 \\ 0.772} & \makecell{0.860 \\ 0.0103} & \makecell{0.848 \\ 0.0168} & \makecell{0.818 \\ 0.0194} & \makecell{0.110} & \makecell{0.989} & \makecell{0.142 \\ 0.0203} &&&&  \\ \cline{2-9}

& Restormer & \makecell{28.62 \\ 0.755} & \makecell{0.909 \\ 0.0080} & \makecell{0.902 \\ 0.0078} & \makecell{0.764 \\ 0.0454} & \makecell{0.712} & \makecell{8.626} & \makecell{0.130 \\ 0.0184} &&&&\\ \cline{2-9}

& InDI$_{20}$ & \makecell{27.23 \\ 0.745} & \makecell{0.876 \\ 0.0095} & \makecell{0.863 \\ 0.0140} & \makecell{0.814 \\ 0.0174} & \makecell{0.178} & \makecell{1.473} & \makecell{0.131 \\ 0.0199} &&&&  \\ \cline{2-9}
& SIFM{$\sigma_{0.0|0.0}$} & \makecell{26.43 \\ 0.734} & \makecell{0.872 \\ 0.0104} & \makecell{0.861 \\ 0.0158} & \makecell{0.822 \\ 0.0194} & \makecell{0.114} & \makecell{1.243} & \makecell{0.133 \\ 0.0225} &&&&  \\ \cline{2-13}
& LVAE & \makecell{28.38 \\ 0.797} & \makecell{0.900 \\ 0.0090} & \makecell{0.894 \\ 0.0082} & \makecell{0.759 \\ 0.0382} & \makecell{0.458} & \makecell{2.927} & \makecell{0.130 \\ 0.0174} & \makecell{0.8030 \\ 0.0184} & \makecell{0.195 \\ 0.0327} & \makecell{3.325 \\ 1.0124} & \makecell{0.150 \\ 0.0203}  \\ \cline{2-13}
& SIFM{$\sigma_{0.0|0.1}$} & \makecell{26.34 \\ 0.722} & \makecell{0.870 \\ 0.0108} & \makecell{0.858 \\ 0.0161} & \makecell{0.822 \\ 0.0191} & \makecell{0.108} & \makecell{1.357} & \makecell{0.132 \\ 0.0220} & \makecell{0.8159 \\ 0.0188} & \makecell{0.127 \\ 0.0203} & \makecell{2.164 \\ 0.6513} & \makecell{0.139 \\ 0.0216}  \\ \cline{2-13}
& SIFM{$\sigma_{1.0|1.0}$} & \makecell{28.17 \\ 0.744} & \makecell{0.886 \\ 0.0103} & \makecell{0.882 \\ 0.0082} & \makecell{0.734 \\ 0.0436} & \makecell{0.487} & \makecell{4.272} & \makecell{0.137 \\ 0.0188} & \makecell{0.7996 \\ 0.0222} & \makecell{0.162 \\ 0.0201} & \makecell{0.711 \\ 0.3178} & \makecell{0.165 \\ 0.0231}  \\ \cline{2-13}
\hhline{~|============|~}
& \textbf{\HazeMatching (ours)} & \makecell{27.87 \\ 0.755} & \makecell{0.904 \\ 0.0085} & \makecell{0.891 \\ 0.0073} & \makecell{0.758 \\ 0.0433} & \makecell{0.607} & \makecell{6.304} & \makecell{0.129 \\ 0.0181} & \makecell{0.8151 \\ 0.0200} & \makecell{0.127 \\ 0.0139} & \makecell{0.710 \\ 0.2743} & \makecell{0.142 \\ 0.0217}

\end{tabular}
}
\caption{\textbf{Full quantitative comparison for the Microtubule Dataset}}
\label{tab:AllenSupple}
\end{table*}

\begin{table*}[h]
\centering
\renewcommand{\arraystretch}{1.2}
\small
\vspace{0.5em}
\scalebox{0.55}{
\begin{tabular}{c|l|c|c|c||c|c|c|c||c|c|c|cc}

\multirow{2}{*}{\centering \textbf{Dataset}} & \multicolumn{1}{c|}{\multirow{2}{*}{\textbf{Methods}}}
 & \multicolumn{3}{c||}{\textbf{Distortion on MMSE/Point-Prediction}} & \multicolumn{4}{c||}{\textbf{Perception on MMSE/Point-Prediction}} & \multicolumn{4}{c}{\textbf{Average Perception on Samples}} \\ \cline{3-13}
&  & \textbf{PSNR $\uparrow$} & \textbf{MS-SSIM $\uparrow$} & \textbf{MicroMS-SSIM $\uparrow$} & \textbf{FSIM $\uparrow$} & \textbf{LPIPS $\downarrow$} & \textbf{FID $\downarrow$} & \textbf{GMSD $\downarrow$} & \textbf{FSIM $\uparrow$} & \textbf{LPIPS $\downarrow$} & \textbf{FID $\downarrow$} & \textbf{GMSD $\downarrow$}\\ 
\hline \hline

\multirow{18}{*}{\centering \textbf{Neuron}}

& RL$_{2}$ & \makecell{24.44 \\ 2.335} & \makecell{0.493 \\ 0.0953} & \makecell{0.562 \\ 0.1081} & \makecell{0.458 \\ 0.1399} & \makecell{0.264} & \makecell{15.090} & \makecell{0.310 \\ 0.0225} &&&& \\ \cline{2-9}
& UNet & \makecell{29.01 \\ 2.956} & \makecell{0.980 \\ 0.0206} & \makecell{0.936 \\ 0.0532} & \makecell{0.877 \\ 0.0869} & \makecell{0.032} & \makecell{0.559} & \makecell{0.108 \\ 0.0587} &&&& \\ \cline{2-9}
& InDI$_{1}$ & \makecell{28.73 \\ 2.914} & \makecell{0.973 \\ 0.0331} & \makecell{0.929 \\ 0.0616} & \makecell{0.890 \\ 0.0736} & \makecell{0.029} & \makecell{0.438} & \makecell{0.110 \\ 0.0598} &&&&  \\ \cline{2-9}

& MIMO-UNet & \makecell{27.50 \\ 2.847} & \makecell{0.974 \\ 0.0232} & \makecell{0.923 \\ 0.0612} & \makecell{0.875 \\ 0.0719} & \makecell{0.027} & \makecell{0.395} & \makecell{0.124 \\ 0.0600} &&&& \\ \cline{2-9}
& MPRNet & \makecell{28.52 \\ 2.969} & \makecell{0.976 \\ 0.0273} & \makecell{0.933 \\ 0.0578} & \makecell{0.888 \\ 0.0927} & \makecell{0.033} & \makecell{0.495} & \makecell{0.118 \\ 0.0660} &&&& \\ \cline{2-9}
& RCAN & \makecell{27.43 \\ 2.893} & \makecell{0.975 \\ 0.0222} & \makecell{0.925 \\ 0.0622} & \makecell{0.881 \\ 0.0723} & \makecell{0.025} & \makecell{0.385} & \makecell{0.124 \\ 0.0606} &&&&  \\ \cline{2-9}
& ESRGAN & \makecell{22.70 \\ 2.080} & \makecell{0.889 \\ 0.0455} & \makecell{0.629 \\ 0.0926} & \makecell{0.680 \\ 0.0987} & \makecell{0.057} & \makecell{2.505} & \makecell{0.270 \\ 0.0447} &&&& \\ \cline{2-9}

& Restormer & \makecell{29.01 \\ 2.984} & \makecell{0.978 \\ 0.0265} & \makecell{0.940 \\ 0.0558} & \makecell{0.903 \\ 0.0703} & \makecell{0.030} & \makecell{0.487} & \makecell{0.107 \\ 0.0634} &&&&  \\ \cline{2-9}

& InDI$_{20}$ & \makecell{25.28 \\ 2.454} & \makecell{0.948 \\ 0.0451} & \makecell{0.594 \\ 0.0983} & \makecell{0.468 \\ 0.1521} & \makecell{0.290} & \makecell{15.493} & \makecell{0.311 \\ 0.0238} &&&& \\ \cline{2-9}
& SIFM{$\sigma_{0.0|0.0}$} & \makecell{28.46 \\ 3.010} & \makecell{0.973 \\ 0.0330} & \makecell{0.925 \\ 0.0669} & \makecell{0.889 \\ 0.0774} & \makecell{0.025} & \makecell{0.412} & \makecell{0.115 \\ 0.0671} &&&& \\ \cline{2-13}
& LVAE & \makecell{28.62 \\ 2.823} & \makecell{0.977 \\ 0.0217} & \makecell{0.931 \\ 0.0591} & \makecell{0.870 \\ 0.0930} & \makecell{0.032} & \makecell{0.561} & \makecell{0.114 \\ 0.0622} & \makecell{0.7224 \\ 0.1061} & \makecell{0.123 \\ 0.0874} & \makecell{8.677 \\ 3.0876} & \makecell{0.310 \\ 0.0448}  \\ \cline{2-13}
& SIFM{$\sigma_{0.0|0.2}$} & \makecell{28.53 \\ 2.992} & \makecell{0.974 \\ 0.0303} & \makecell{0.907 \\ 0.0682} & \makecell{0.856 \\ 0.0944} & \makecell{0.029} & \makecell{0.559} & \makecell{0.118 \\ 0.0632} & \makecell{0.8490 \\ 0.0971} & \makecell{0.032 \\ 0.0386} & \makecell{0.792 \\ 1.1414} & \makecell{0.129 \\ 0.0658}  \\ \cline{2-13}
& SIFM{$\sigma_{1.0|1.0}$} & \makecell{28.65 \\ 2.935} & \makecell{0.971 \\ 0.0358} & \makecell{0.914 \\ 0.0732} & \makecell{0.856 \\ 0.0969} & \makecell{0.036} & \makecell{0.425} & \makecell{0.113 \\ 0.0593} & \makecell{0.8141 \\ 0.0877} & \makecell{0.040 \\ 0.0271} & \makecell{0.450 \\ 0.5625} & \makecell{0.191 \\ 0.0688}  \\ \cline{2-13}
\hhline{~|============|~}
& \textbf{\HazeMatching (ours)} & \makecell{28.96 \\ 2.961} & \makecell{0.978 \\ 0.0183} & \makecell{0.933 \\ 0.0541} & \makecell{0.887 \\ 0.0711} & \makecell{0.031} & \makecell{0.464} & \makecell{0.107 \\ 0.0563} & \makecell{0.8493 \\ 0.0856} & \makecell{0.033 \\ 0.0299} & \makecell{0.554 \\ 0.8043} & \makecell{0.147 \\ 0.0653} 

\end{tabular}
}
\caption{\textbf{Full quantitative comparison for the Neuron Dataset}}
\label{tab:NeuronSupple}
\end{table*}

\newpage
\subsection{Inference runtime comparison}
\label{sup:time}

Below in Table~\ref{tab:runtime_comparison}, we report wall-clock inference times (averaged over 5 runs on an NVIDIA V100) on an image of size $1024 \times 1024$, alongside model sizes and sampling capabilities.

\begin{table}[h]
\centering
\begin{tabular}{llllll}
\toprule
\textbf{Model} & \textbf{Mean (5) [sec.]} & \textbf{Std (5)} & \textbf{Per Forward Pass [sec.]} & \textbf{Params (M)} \\
\midrule
RL           & 13.6  & 3.16   & N/A     & N/A \\
Elements     & 3.233 & 0.136  & N/A     & N/A \\
Lightning    & 1.0 (est.) & N/A & N/A     & N/A \\
UNet         & 0.2944 & 0.0393 & N/A     & 13.39 \\

MIMO-UNet & 0.6358 & 0.0032 & N/A & 6.80 \\
MPRNet & 5.4713 & 0.0046 & N/A & 20.12 \\
RCAN & 8.6271 & 0.0293 & N/A & 15.29 \\
ESRGAN & 1.1839 & 0.0566 & N/A & 16.70 \\
Restormer &4.5067 &0.0255 & N/A & 26.12 \\
InDI$_1$     & 7.1888 & 0.0930 & N/A     & 2.44 \\
InDI$_{20}$  & 143.3886 & 0.5544 & 7.16943 & 2.44 \\

\HVAE        & 1.0118 & 0.4468 & N/A     & 5.67 \\
SIFM            & 4.3120 & 0.0633 & 0.2156  & 3.70 \\
\HazeMatching   & 4.7498 & 0.0049 & 0.23749 & 3.70 \\

\bottomrule
\end{tabular}
\caption{\textbf{Inference runtime (mean and std. over 5 runs)}}
\label{tab:runtime_comparison}
\end{table}

\paragraph{Parameter Efficiency}
\textcolor{black}{At \textbf{3.7M} parameters, \HazeMatching is more compact than UNet (13.4M) and still supports iterative refinement. Given that iterative generative models are generally computationally intensive, \HazeMatching delivers a \textbf{practical runtime} ($\sim 4.7$~s for $1024\times1024$ images) while enabling \textbf{multi-sample posterior inference}, outperforming other iterative methods by over an order of magnitude in speed.}

\clearpage
\newpage
\subsection{Flexibility in Fidelity--Realism tradeoff (Organoids1 Dataset)}
\label{sup:pdt_control}

\paragraph{Integration Steps ($T$):} 
\textcolor{black}{Increasing the number of integration steps generally yields smoother flows and improved perceptual realism:}

\begin{table}[h]
\centering
\begin{tabular}{cccccc}
\toprule
\textbf{$T$ Step} & \textbf{LPIPS} $\downarrow$ ($\mu$) & \textbf{LPIPS} ($\sigma$) & \textbf{PSNR} $\uparrow$ ($\mu$) & \textbf{PSNR} ($\sigma$) & \textbf{Time (s)} \\
\midrule
2   & 0.1419 & 0.0572 & 36.73 & 6.099 & 0.26  \\
3   & 0.1843 & 0.0935 & 36.60 & 6.034 & 0.48  \\
4   & 0.1938 & 0.1096 & 36.39 & 6.046 & 0.68  \\
5   & 0.1939 & 0.1155 & 36.24 & 6.063 & 0.89  \\
10  & 0.1753 & 0.1092 & 35.92 & 6.101 & 1.96  \\
15  & 0.1570 & 0.0929 & 35.80 & 6.134 & 3.04 \\
30  & 0.1158 & 0.0436 & 35.63 & 6.181 & 6.05 \\
50  & 0.1124 & 0.0383 & 35.57 & 6.200 & 9.92 \\
80  & 0.1117 & 0.0367 & 35.55 & 6.204 & 15.75 \\
100 & 0.1114 & 0.0362 & 35.55 & 6.202 & 19.67  \\
150 & 0.1116 & 0.0359 & 35.53 & 6.206 & 29.42 \\
200 & 0.1115 & 0.0359 & 35.52 & 6.209 & 39.20 \\
250 & 0.1115 & 0.0359 & 35.52 & 6.211 & 48.99 \\
\bottomrule
\end{tabular}
\caption{Effect of integration steps $T$ on perceptual (LPIPS) and distortion (PSNR) metrics and runtime per $1024\times1024$ sample. Note that time shown is for 1 sample.}
\label{tab:int_steps}
\end{table}

\paragraph{Sample Averaging:}
\textcolor{black}{Averaging more stochastic samples reduces variance and increases PSNR (fidelity). Results shown for $T = 20$:}

\begin{table}[h]
\centering
\begin{tabular}{ccc}
\toprule
\textbf{\# Samples} & \textbf{PSNR} $\uparrow$ ($\mu$) & \textbf{PSNR} ($\sigma$) \\
\midrule
01 & 35.74 & 6.153 \\
02 & 36.23 & 6.042 \\
05 & 36.58 & 5.985 \\
10 & 36.71 & 5.947 \\
20 & 36.78 & 5.916 \\
30 & 36.81 & 5.900 \\
40 & 36.82 & 5.897 \\
50 & 36.83 & 5.894 \\
\bottomrule
\end{tabular}
\caption{Effect of number of posterior samples on PSNR (mean and std.) for $T=20$.}
\label{tab:sample_avg}
\end{table}

\subsection{Choice of conditioning mechanism}
\label{sup:condition_ablation}

Following prior work in diffusion models, we adopt \textbf{concatenation} of the noisy input and conditioning observation, which has become standard for inverse problems~\cite{con_diff}. Additionally, we perform ablation with addition of the condition to the input on \textit{Organoids1} (Table~\ref{tab:cond_ablation}).

\begin{table}[h]
\centering
\begin{tabular}{lcc}
\toprule
\textbf{Conditioning} & \textbf{PSNR} $\uparrow$ & \textbf{LPIPS} $\downarrow$ \\
\midrule
Concatenation (ours)   & \textbf{36.83} & \textbf{0.140} \\
Element-wise add & 33.66 & 0.217 \\
\bottomrule
\end{tabular}
\caption{Ablation of conditioning mechanisms with \HazeMatching on \textit{Organoids1}.}
\label{tab:cond_ablation}
\end{table}

\subsection{Comparison with conditional diffusion}
\label{sup:conditional_diffusion}

Here, we compare PSNR (of the MMSE) and LPIPS (averaged over samples) between \HazeMatching and a conditional diffusion model~\citep{con_diff}. For each dataset, we evaluate both methods using two posterior samples generated from a single test image. The conditional diffusion model is run with 1200 backward steps, while \HazeMatching uses $T=20$ steps. The quantitative results are summarized in Table~\ref{tab:compare_cdiff}, along with the wall-clock time required to generate a single posterior sample for each dataset. The diffusion model takes considerably longer per sample, making it less practical to use.

\begin{table}[h]
\centering
\begin{tabular}{lcccccl}
\toprule
\textbf{Dataset} 
& \multicolumn{3}{c}{\textbf{\HazeMatching}} 
& \multicolumn{3}{c}{\textbf{Cond. Diffusion~\citep{con_diff}}} \\
\cmidrule(lr){2-4} \cmidrule(lr){5-7}
& PSNR $\uparrow$ & LPIPS $\downarrow$ & Time (s) 
& PSNR $\uparrow$ & LPIPS $\downarrow$ & Time (s) \\
\midrule
Zebrafish (1024$\times$1024)   &31.43  &0.170  &4.15  &30.34  &0.144  &10067.45  \\
Organoids1 (1024$\times$1024)  &34.42  &0.094  &4.15  &32.33  &0.280  &10244.28  \\
Organoids2 (1024$\times$1024)  &32.97  &0.096  &4.15  &30.07  &0.415  &10203.20  \\
Microtubule (512$\times$512) &25.82  &0.145  &1.24  &23.96  &0.128  &2766.09  \\
Neuron (64$\times$64)      &26.83  &0.027  &0.24  &26.17  &0.027  &25.24  \\
\bottomrule
\end{tabular}
\caption{Comparison of \HazeMatching and conditional diffusion across five datasets (1 test image, 2 posterior samples) in terms of PSNR, LPIPS, and wall-clock time per posterior sample.}
\label{tab:compare_cdiff}
\end{table}

\subsection{Effect of different dynamic range regimes}
\label{sup:dynamic_range}

We evaluate the robustness of the model to dynamic range variations by constructing two additional datasets from the original Organoids1 data: an ``\textit{OK}'' variant obtained by scaling the intensity of the input images by $0.7$ and adding a constant offset of $12$, and a more extreme ``\textit{Bad}'' version obtained by scaling the intensity by $0.2$ with the same offset of $12$. By training with these datasets, we obtain the \textit{OK Model} and the \textit{Bad Model}, respectively. Note that the \textit{Good Model} is the original \HazeMatching model trained on the actual Organoids1 data.

All models are then evaluated across all data variants, and their performance is reported in Table~\ref{tab:dynamic_range}. Across these settings, PSNR exhibits only a slight decrease under dynamic range mismatch, with a drop of less than $\sim$1 dB even in the most severe case, indicating that reconstruction fidelity is largely preserved. In contrast, LPIPS remains effectively unchanged across all combinations of training and testing conditions, suggesting that perceptual similarity is invariant to such intensity transformations. Overall, these results demonstrate that the model is robust to moderate and even strong dynamic range shifts, with only minimal impact on pixel-wise accuracy.

Due to the data normalization prior to feeding input patches, these results might be, at most, moderately surprising to some readers. 
We attribute this section to the curiosity of one of our reviewers.

\begin{table}[h]
\centering
\small
\begin{tabular}{llccc}
\toprule
\textbf{Metric} & \textbf{Data} & \textbf{Good Model} & \textbf{OK Model} & \textbf{Bad Model} \\
\midrule
\multirow{3}{*}{PSNR $\uparrow$} 
& Good (100\%) & 36.83 & 35.48 & 35.22 \\
& OK (70\%)   & 35.72 & 35.47 & 35.22 \\
& Bad (20\%) & 35.73 & 35.47 & 35.22 \\
\midrule
\multirow{3}{*}{LPIPS $\downarrow$} 
& Good (100\%) & 0.140 & 0.135 & 0.138 \\
& OK (70\%)   & 0.140 & 0.135 & 0.138 \\
& Bad (20\%)  & 0.140 & 0.135 & 0.138 \\
\bottomrule
\end{tabular}
\caption{Effect of dynamic range variations on model performance. The OK and Bad datasets are generated by scaling the intensity (0.7 and 0.2, respectively) and adding a constant offset of 12.}
\label{tab:dynamic_range}
\end{table}

\clearpage
\subsection{More qualitative results}
\label{sup:qualitative}

\figQualitativeSuppleZebrafishtwo 
\figQualitativeSuppleZebrafishthree
\figQualitativeSuppleZebrafishOne

\figQualitativeSuppleOrgOneOne 
\figQualitativeSuppleOrgOneTwo
\figQualitativeSuppleOrgOneThree

\figQualitativeSuppleOrgTwoOne 
\figQualitativeSuppleOrgTwoTwo
\figQualitativeSuppleOrgTwoThree

\figQualitativeSuppleMicrotubuleOne 
\figQualitativeSuppleMicrotubuleTwo
\figQualitativeSuppleMicrotubuleThree

\figQualitativeSuppleNeuronOne 
\figQualitativeSuppleNeuronTwo
\figQualitativeSuppleNeuronThree

\figQualitativeZebrafishFULL
\figQualitativeOrgOneFULL
\figQualitativeOrgTwoFULL
\figQualitativeALLENFULL

\end{document}